\documentclass{aa}
\usepackage{graphicx}
\usepackage{txfonts}
\usepackage{natbib}
\usepackage[colorlinks=true, pdfstartview=FitV, linkcolor=blue,citecolor=blue, urlcolor=blue]{hyperref}
\usepackage{subfigure}

\begin{document} 

\title{Cosmology with Galaxy Cluster Properties using Machine Learning}
\author{Lanlan Qiu \inst{1,2},
Nicola R. Napolitano \inst{1,2,3} \thanks{E-mail: napolitano@mail.sysu.edu.cn},
Stefano Borgani \inst{4,5,6,7,8},
Fucheng Zhong\inst{1,2}\thanks{E-mail: zhongfch@mail2.sysu.edu.cn},\\
Xiaodong Li\inst{1,2},
Mario Radovich\inst{9},
Weipeng Lin\inst{1,2}, 
Klaus Dolag\inst{10,11},
Crescenzo Tortora\inst{12},
Yang Wang\inst{2,13}, \\
Rhea-Silvia Remus\inst{9},
Sirui Wu \inst{1,2},
Giuseppe Longo\inst{3}
}
\authorrunning{Qiu, L., et al.}
\institute{School of Physics and Astronomy, Sun Yat-sen University Zhuhai Campus, 2 Daxue Road, Tangjia, Zhuhai 519082, P.R. China
\and
CSST Science Center for Guangdong-Hong Kong-Macau Great Bay Area, Zhuhai 519082, P.R. China
\and
Department of Physics E. Pancini, University Federico II, Via Cinthia 6, 80126-I, Naples, Italy
\and
Astronomy Unit, Department of Physics, University of Trieste, via Tiepolo 11, I-34131 Trieste, Italy 
\and 
INAF-Osservatorio Astronomico di Trieste, via G. B. Tiepolo 11, I-34143 Trieste, Italy
\and
IFPU, Institute for Fundamental Physics of the Universe, Via Beirut 2, 34014 Trieste, Italy
\and
INFN, Instituto Nazionale di Fisica Nucleare, Via Valerio 2, I-34127, Trieste, Italy
\and
ICSC - Italian Research Center on High Performance Computing, Big Data, and Quantum Computing
\and
INAF - Osservatorio Astronomico di Padova, via dell'Osservatorio 5, 35122 Padova, Italy
\and
Universitäts-Sternwarte, Fakultät für Physik, Ludwig-Maximilians-Universität München, Scheinerstr.1, 81679 München, Germany
\and
Max-Planck-Institut für Astrophysik, Karl-Schwarzschild-Straße 1, 85741 Garching, Germany
\and 
INAF -- Osservatorio Astronomico di Capodimonte, Salita Moiariello 16, 80131 - Napoli, Italy
\and
Peng Cheng Laboratory, No.2, Xingke 1st Street, Shenzhen, 518000, P.R. China
}

\date{\today}

\abstract
{Galaxy clusters are the largest gravitating structures in the universe and their mass assembly is sensitive to the underlying cosmology. Their mass function, baryon fraction, and mass distribution have been used to infer cosmological parameters, despite the presence of systematics. 
However, the complexity of the scaling relations among galaxy cluster properties has never been fully exploited, limiting their potential as a cosmological probe.}
{We propose the first Machine Learning (ML) method using galaxy cluster properties from hydrodynamical simulations in different cosmologies to predict cosmological parameters combining a series of canonical cluster observables, like gas mass, gas bolometric luminosity, gas
temperature, stellar mass, cluster radius, total mass, and velocity dispersion at different redshifts.}
{The machine learning model is trained on mock ``measurements'' of these observable quantities from Magneticum multi-cosmology simulations to derive unbiased constraints on a set of cosmological parameters. These include the mass density parameter, $\Omega_m$, the power spectrum normalization, $\sigma_8$, the baryonic density parameter, $\Omega_b$, and the reduced Hubble constant, $h_0$.}
{We test the ML model on catalogs of a few hundred clusters taken, in turn, from each simulation and find that the ML model can correctly predict the cosmology they have been picked from. The cumulative accuracy depends on the cosmology, ranging from 21\% to 75\%. We demonstrate that this is sufficient to derive unbiased constraints on the main cosmological parameters with errors of the order of {$\sim14$\% for $\Omega_m$, $\sim8$\% for $\sigma_8$, $\sim6$\% for $\Omega_b$, and $\sim3 \%$ for $h_0$.}}
{This proof-of-concept analysis, yet based on a limited variety of multi-cosmology simulations, shows that machine learning can efficiently map the correlations in the multi-dimensional space of the observed quantities to the cosmological parameter space and narrow down the probability that a given sample belongs to a given cosmological parameter combination. More large-volume, mid-resolution, multi-cosmology hydro-simulations need to be produced to expand the applicability to a wider cosmological parameter range. However, this first test is exceptionally promising, as it shows that these ML tools can be applied to cluster samples from multi-wavelength observations from surveys like Rubin/LSST, CSST, {\it Euclid}, Roman in optical and near-infrared bands, and eROSITA in X-rays, to constrain cosmology and the effect of the baryonic feedback.}

\maketitle
%
\section{Introduction}
\label{sec:intro}
According to the hierarchical clustering scenario, galaxy clusters are the largest and the most massive collapsed objects in the universe, typically residing in the nodes of the cosmic web. The virial mass of a typical rich cluster is about $10^{14}-10^{15}\rm{M_{\odot}}$, consisting of approximately $2\%$ galaxies, $12\%$ hot gas, and $86\%$ dark matter. Due to their spatial distribution in the universe and specific mass composition, they have been widely investigated, both as an effective cosmological probe and a natural astrophysical laboratory \citep{2011ARA&A..49..409A,2012ARA&A..50..353K,2022A&A...659A..88L,2022A&A...665A.100L,2022MNRAS.511.1484I}.

With respect to their cosmological application, cluster masses, and in particular, the cluster mass function, can be used to constrain both the universe mean matter density $\Omega_m$ and the density fluctuation amplitude $\sigma_8$. 
However, their constraining capacity is inevitably limited by the difficulty of deriving accurate mass estimates from observations \citep{2019SSRv..215...25P}. The most precise mass estimates come from weak gravitational lensing. 
This has been widely exploited to calibrate mass estimations from other methods, but the cluster triaxiality and projection effects of lensing measurements limit the precision of individual cluster mass to about $5\%$ (e.g. \citealt{2015MNRAS.449..685H,2016ApJ...821..116U,2017MNRAS.465.1454H,2017MNRAS.469.4899M,2017MNRAS.465.3361H,2023arXiv230200687E}). 
Besides, weak lensing is also observationally difficult to perform, and yet today there is a rather limited statistics of clusters having accurate weak lensing mass (e.g. \citealt{2011MNRAS.416.3187S,2015MNRAS.450.3665S,2020ApJ...890..148U,2021A&A...653A..19Giocoli}). 
Other direct mass estimates are obtained through the virial theorem, i.e. by measuring the velocity field of galaxy members (e.g. \citealt{2020ApJ...901...90A}), or via Jeans analysis (e.g. \citealt{2006MNRAS.367.1463L,2013MNRAS.436.2639F,2013A&A...558A...1Biviano13,2014A&A...566A..68M}). 
However, the application of the virial theorem and Jeans analysis is also limited by the difficulty of measuring a large number of redshifts in individual clusters and the presence of systematics like outliers and underlying modeling assumptions, that are hard to control. 
Customarily, to overcome at least the observational difficulties, cluster masses are widely estimated indirectly by various means. 
For instance, some multi-band integrated observables of galaxy clusters are generally expected to scale with cluster masses and be used as mass proxies. 
Typical observables may come from the X-ray emission (e.g. \citealt{2001Natur.409...39B,2009ApJ...692.1060V,2010MNRAS.406.1759M,2022arXiv220712429C}), optical richness (e.g. \citealt{1999ApJ...527..561B,2016ApJS..224....1R,2019MNRAS.485..498M,2020PhRvD.102b3509A}), and millimeter-wave thermal Sunyaev-Zel'dovich signal (e.g. \citealt{2015ApJS..216...27B,2016A&A...594A..24P,2019ApJ...878...55B,2021ApJS..253....3H}). 
However, the scaling relations connecting these quantities with mass are generally very noisy and not bias-free \citep{2016MNRAS.463.3582M,2019MNRAS.483.2871D,2022A&A...661A...7B}. 
In general, cluster masses based on various methods tend to be rather scattered, leaving the constraints based on these systems under-exploited, despite the large potential \citep{2020ApJ...901...90A,2022A&A...659A..88L}.

Recent studies have shown the potential of using AI-based methods to cluster science, e.g. for mass estimation using tools trained on simulations. 
These studies have used a variety of cluster features, like the velocity distribution of the cluster members (\citealt{2015ApJ...803...50N}), the velocity distribution along with mock X-ray and weak-lensing analyses (\citealt{2019MNRAS.484.1526A}), richness, velocity distribution, and other simulated multi-wavelength measurements \citep{2020MNRAS.491.1575C}),
or directly emulating the richness-mass relation \citep{2023A&A...675A..77R}.
Other studies have also considered the cluster phase space distribution (e.g. \citealt{Ho_2019, 2020MNRAS.499.1985K, 2021MNRAS.501.4080K}), and stellar mass, X-ray flux, or the Compton $y$ parameter (e.g. \citealt{2020MNRAS.499.3445Y, 2022NatAs...6.1325D}). 
These simulation-based AI schemes have been found very promising as alternatives to classical methods of cluster mass estimation.

Despite these many efforts to enhance the cosmological application of galaxy clusters by improving the accuracy of mass estimates, very little has been done to exploit the potential of all other direct observables connected to the baryonic components, that, being tightly correlated with masses, can also keep significant cosmological information. 
The one-to-one correlations among some typical observables, such as stellar mass, gas mass, and X-ray flux, i.e. the so-called scaling relations, represent a viable approach to constrain cosmology (\citealt{2020MNRAS.494.3728S}). 
In principle, to fully exploit the cosmological potential of the cluster properties, one could combine the information encoded in all of the existing scaling relations among various mass-related quantities.
Machine Learning (ML) is the ideal tool to extract valuable scientific information and execute joint analysis out of such a multi-dimensional feature space and help establish internal links between these features and their environmental information. To be linked to cosmology and baryonic physics, these need to be trained using realistic mock data samples for which the ground truths are given.
Cosmological simulations can provide such training samples as they have currently reached a rather advanced technological and theoretical level to predict the effect of cosmology (and feedback) on the baryonic + dark scaling relations over different scales, from galaxies to clusters (see e.g. \citealt{2018ARA&A..56..435Wechsler} for a review). 
Modern hydrodynamical simulations can capture most of this physics with fair accuracy and study the effect of the complex baryon processes over the dark matter distribution (e.g. \citealt{2004MNRAS.348.1078B,2009MNRAS.399..497D,2012MNRAS.423.2279C,2014MNRAS.444.1518V,2017MNRAS.464.3742Remus17,2018MNRAS.473.4077P}), though, they mostly focus on one single cosmological model.

On the other hand, multi-cosmology hydro-dynamical simulations would be of paramount importance to combine cosmology and baryonic physics and possibly solve the degeneracies coming from the interplay of the dark and baryonic components (\citealt{2018ARA&A..56..435Wechsler,2022ApJ...929..132V}). An effective strategy is to fully explore the multi-dimension parameter space where, on one side, one can change the cosmology, meaning the cosmological parameters and the DM flavors, and, on the other side, one can explore different galaxy formation models, including the stellar initial mass function, the duration, power, and location of star formation, the stellar feedback including the supernova explosions, the AGN effect, etc. 

By combining Machine Learning and multi-cosmology hydrodynamical simulations, we have the possibility 
to build a new effective model to predict the cosmology and the formation scenario from catalogs of astronomical observables.
Among the first attempts to collect predictions from a different combination of cosmology and baryonic physics scenarios, the CAMELS project$\footnote{\url{https://www.camel-simulations.org/}}$ \citep{2021ApJ...915...71V} is designed for galaxy scales while Magneticum project$\footnote{\url{http://magneticum.org/}}$ (\citealt{2020MNRAS.494.3728S}) is tailored for galaxy cluster scales. 
The bottleneck of these applications is the availability of sufficiently large volume simulations with enough mass resolution to investigate the widest range of the systems under exams. For galaxy scales, simulation samples are sufficient to directly test the application to mock galaxy samples (e.g. \citealt{2022ApJ...929..132V,2023arXiv230912048C,2023ApJ...954..125E}). For cluster scales, on the other hand, there are still limited multi-cosmological samples to use. One way to expand the simulation library can be the adoption of emulators or generative models, that have been already used to reproduce cosmological statistics such as galaxy clustering (e.g. \citealt{2022arXiv221003203S}), galaxy power spectrum (e.g. \citealt{2022PhRvD.105h3517K}) and halo mass function (e.g. \citealt{2020ApJ...901....5B}).

In this first article, we start by testing the predictive power encoded in the galaxy clusters' multi-wavelength and spectroscopic data of next-generation surveys to constrain the cosmology testing a suite of machine learning tools on Magneticum multi-cosmology simulations \citep{2020MNRAS.494.3728S}.
We postpone the constraints of the feedback in this analysis because of the limited variety of feedback models currently available for these simulations.
The observables available in simulations are gas mass, gas bolometric luminosity, gas temperature, stellar mass, cluster size, total mass, and velocity dispersion at different redshifts. 
In particular, we aim to demonstrate that machine learning can be trained on multi-cosmology simulations to recognize the correct universe a given cluster catalog belongs to. Then, by defining the probability for each cluster of being drawn by a cosmology with a series of cosmological parameters, we will derive the posterior probability distribution of any given cosmological parameter. 
Albeit we make this proof-of-concept experiment realistic enough, by including observationally motivated measurement errors, this remains a ``toy model'' approach. To move to real data applications, it will need a more methodical derivation of fiducial observables from simulations, in order to minimize the systematics due to the ``observational realism''. The inclusions of these aspects, as well as the study of the impact of other sources of systematics that can be introduced by simulation set-ups (e.g. resolution, numerical methods, etc.), are beyond the scope of this paper and will be only touched here but fully addressed in the second phase of the project, where we will investigate the application to real cluster catalogs.

This paper is organized as follows. Sect. \ref{sec:data} introduces the data we use to check this idea and the algorithm for getting the pre-processed data and preparing training and test samples. Sect. \ref{sec:method} illustrates all the machine learning algorithms and evaluation metrics involved to quantify the constraining power of each experiment. Sect. \ref{sec:result} lists all the results about the proper classifier, the classification of cosmological models, and the cosmological parameter inferences. 
In Sect. \ref{sec:discussion}, we discuss the robustness of our results and some sources of systematics. Finally, we draw conclusions and outline future perspectives in Sect. \ref{sec:conclusion}.

\begin{table*}[h]
\centering
\caption{Cosmological parameter values for 13 cosmological models.}
\resizebox{0.9\textwidth}{!}{
\setlength{\tabcolsep}{5pt}
\begin{tabular}{c c c c c c c c c c c c c c}
\hline
\hline
\noalign{\smallskip}
& M1 & M2 & M3 & M4 & M5 & M6 & M7 & M8 & M9 & M10 & M11 & M12 & M13\\
\hline
\noalign{\smallskip}
$\Omega_m$ & 0.200 & 0.204 & 0.222 & 0.232 & 0.268 & 0.272 & 0.301 & 0.304 & 0.342 & 0.363 & 0.400 & 0.406 & 0.428 \\
$\sigma_8$ & 0.850 & 0.739 & 0.793 & 0.687 & 0.721 & 0.809 & 0.824 & 0.886 & 0.834 & 0.884 & 0.650 & 0.867 & 0.830 \\
$h_0$ & 0.730 & 0.689 & 0.676 & 0.670 & 0.699 & 0.704 & 0.707 & 0.740 & 0.708 & 0.729 & 0.675 & 0.712 & 0.732 \\
$\Omega_b$ & 0.0415 & 0.0437 & 0.0421 & 0.413 & 0.0449 & 0.0456 & 0.0460 & 0.0504 & 0.0462 & 0.0490 & 0.0485 & 0.0466 & 0.0492\\
\noalign{\smallskip}
\hline
\end{tabular}
\label{tab:para_map}}
\end{table*}

\begin{figure*}
\includegraphics[scale=0.392]{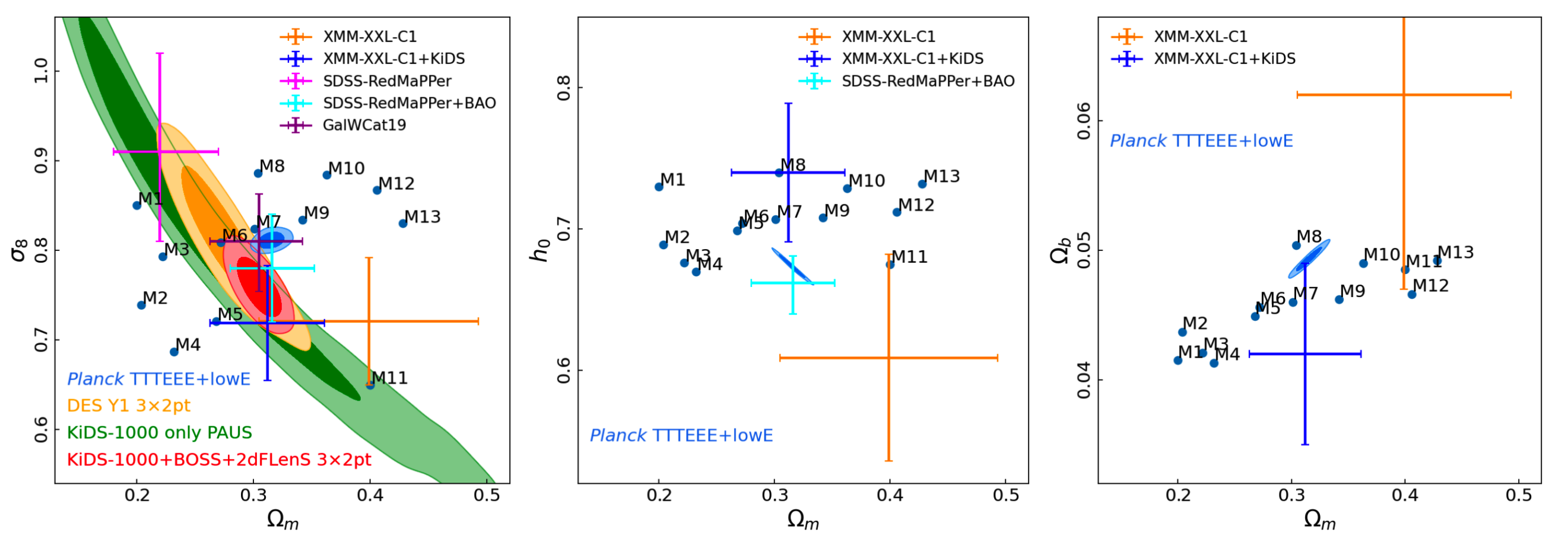}
\caption{Cosmological parameters map for the 13 cosmological models. Blue points show the flat $\Lambda$CDM models in the multi-cosmology runs. For comparison, the error bars show the constraints from the XMM-XXL C1 cluster abundance alone \citep{2018A&A...620A..10P} and plus KiDS tomographic weak lensing \citep{2017MNRAS.465.1454H} joint analysis, the SDSS RedMaPPer cluster abundance alone and plus BAO joint analysis \citep{2019MNRAS.488.4779C}, the GalWCal19 cluster abundance \citep{2020ApJ...901...90A}. {Contours show the marginalized posterior distributions of CMB constraints \citep{2020A&A...641A...6P}, $3\times2$pt analysis from DES Y1\citep{2018PhRvD..98d3526A}, $3\times2$pt analyses from KiDS-1000 with BOSS and 2dFLenS \citep{2021A&A...646A.140H}, and KiDS-1000 spec-$z$ fiducial constraints \citep{2022A&A...664A.170V} -- see legend bottom left, in each panel.}}
\label{fig:para_map}
\end{figure*}

\section{Data}
\label{sec:data}
In the previous section, we have anticipated that the main aim of this work is to demonstrate the ability of a machine learning method to predict cosmological parameters, starting from the observables of a set of galaxy clusters. In this section, we introduce the set of multi-cosmology simulations adopted to train such a tool. The galaxy cluster catalogs derived from these simulations represent the ``observational-like'' data (the {\it features}) to start from, to first train the machine learning method and then 
test the predictions of the cosmological parameters (the {\it targets}). In particular, we explain how we define the training and the test samples used to train and evaluate the performances of the proposed ML tool. We also briefly discuss the limitations of the current simulation set and the need to expand the coverage of the cosmological parameter space for real applications.

\subsection{Multi-cosmology simulations}
\label{sec:magneticum}
Magneticum simulations are based on the $N$-body code \textit{P-GADGET3}, which is the successor of the code \textit{P-GADGET2} \citep{2005Natur.435..629S,2005MNRAS.364.1105S,2009MNRAS.398.1150B}, from which it differs for a space-filling curve aware neighbor search \citep{2016pcre.conf..411R} and an improved Smoothed Particle Hydrodynamics (SPH) solver \citep{2016MNRAS.455.2110B}. The physics of these simulations are presented in a series of separate method papers: e.g., \cite{2005MNRAS.361..776S} discusses the treatment of radiative cooling, heating, ultraviolet (UV) background, star formation, and stellar feedback processes; \cite{2007MNRAS.382.1050T} describes in details the chemical evolution and enrichment model, while \cite{2010MNRAS.401.1670F};
and \cite{2014MNRAS.442.2304H} present the prescriptions for the black hole growth and active galactic nuclei (AGNs) feedback.

Halos are identified using the friends-of-friends (FOF) algorithm with linking length $b=0.16$. The spherical overdensity (SO) virial masses \citep{1998ApJ...495...80B} are computed using the SUBFIND algorithm \citep{2001MNRAS.328..726S,2009MNRAS.399..497D}.

In this paper, we focus on the multi-cosmology simulations of the Magneticum project \cite[][S+20 hereafter]{2016MNRAS.463.1797Dolag16,2020MNRAS.494.3728S}. The original simulation set includes 15 flat $\Lambda$CDM cosmological models (C1, C2, ..., C15) that run with the same initial conditions, and same feedback circumstances, but different configurations of four cosmological parameters, namely, the mass density parameter $\Omega_m$, the power spectrum normalization $\sigma_8$, the ``reduced'' Hubble constant $h_0$, defined as $H_0/100$ km/s/Mpc, and the baryon density parameter $\Omega_b$ (see S+20, Table 1). Each simulation uses a large size ($\sim896\ h_{0}^{-1}$Mpc) box, containing $1512^3$ dark matter particles and an equal number of gas particles. The mass of the dark matter particles is $1.3\times 10^{10}\ h^{-1}\rm{M_{\odot}}$ and the initial mass of gas particles is $2.6\times 10^{9}\  h^{-1}\rm{M_{\odot}}$.

For each simulation, only halos with $M_{vir}>2\times10^{14}\ \rm{M_{\odot}}$ are selected to avoid spurious detections due to resolution and other numerical effects. 
The catalogs of the selected clusters are obtained for different redshift snapshots, i.e. $z=0.00, 0.14, 0.29, 0.47, 0.67, 0.90$. Taken as a whole, the numbers of identified haloes vary significantly among these 15 cosmological models due to different configurations of cosmological parameters (see S+20, Table 2). Considering that the identified haloes generated by C1 and C2 are too few (i.e., 1245 and 4810, respectively) to construct an informative sample for the machine learning training process, we decide to use only the other 13 cosmological models, C3, C4, ..., C15, and denote them as M1, M2, ..., M13 in this paper and consider M6, the one with the WMAP7 best-fitting configuration \citep{2011ApJS..192...18K}, as the fiducial cosmology consistently with the Magneticum project.

The cosmological parameters of M1$\sim$M13 are specified in Table \ref{tab:para_map} and shown in Fig. \ref{fig:para_map}, together with cosmological constraints obtained by different surveys and methods: 
{CMB power spectra constraints \citep{2020A&A...641A...6P}, $3\times2$pt analysis from DES Y1\citep{2018PhRvD..98d3526A}, $3\times2$pt analyses from KiDS-1000 with BOSS and 2dFLenS \citep{2021A&A...646A.140H}, KiDS-1000 spec-$z$ fiducial constraints \citep{2022A&A...664A.170V}, XMM-XXL C1 cluster abundance alone \citep{2018A&A...620A..10P} and adding KiDS tomographic weak lensing joint analysis \citep{2017MNRAS.465.1454H}, SDSS RedMaPPer cluster abundance alone and adding BAO joint analysis \citep{2019MNRAS.488.4779C}, GalWCal19 cluster abundance \citep{2020ApJ...901...90A}. 
From Fig. \ref{fig:para_map}, we can see that the cosmological parameter ranges covered by the M1 $\sim$ M13 simulations, i.e.  $0.200 < \Omega_m < 0.428$, $0.650 < \sigma_8 < 0.886$, $0.670 < h_{0} < 0.740$ and $0.0413 < \Omega_b < 0.0504$, embrace the core of the confidence contours of most of the constraints of the above-mentioned experiments, especially in the $\Omega_m-\sigma_8$ space, while the constraints on $h_0$ and $\Omega_b$ are sometimes more scattered. This means that, in principle, the current Magneticum set of simulations is not fully representative of the overall variation of the cosmological parameters compatible with all observations. This limitation, together with the sparse coverage of the parameter space allowed by the current simulation set, does not make it optimal for applications to real data. However, with this paper, we want to make a first step toward the application to real data and test the suitability of the method for the kind of catalogs we expect to collect from current and future observations (see e.g. 
eFEDS, \citealt{2022arXiv220712429C}). On the other hand, if we demonstrate that with such a limited sample of simulations, ML is able to make predictions on the cosmological parameters underlying some cluster observations, then we can expect that the method will be even more effective when the simulation sample will be expanded to a wider range of parameters and a more fine coarse coverage of the parameter space. Hence, besides testing the suitability of this novel approach to infer cosmology from cluster observations, another outcome of the proof-of-concept test, discussed in this work, is to concretely motivate the investment in more extended simulation set-ups to offer flexible and accurate inferences.

\subsection{Features and labels}
\label{sec:features and labels}
Each of the selected clusters has corresponding features and a label.
The labels are the cosmological models they come from, i.e., M1 $\sim$ M13. The features are the physical properties of the identified clusters in each simulation, namely:
\begin{enumerate}
\setlength{\itemsep}{2pt}
\item $R$: the radius of the cluster, i.e., the comoving radius of a sphere centered at the minimum of the potential encompassing a given mean overdensity, in $h^{-1} (1+z)^{-1}$kpc.
\item $M_{*}$: the stellar mass of the cluster, i.e., the sum of the mass of all star particles within the mean overdensity radius, $R$, defined above, in $h^{-1}\rm{M_{\odot}}$.
\item $M_{g}$: the gas mass of the cluster, i.e., the sum of the mass of all gas particles within $R$, in $h^{-1}\rm{M_{\odot}}$.
\item $M_{t}$: the total mass of the cluster, i.e., the sum of the mass of all star, gas, and dark matter particles within $R$, in $h^{-1}\rm{M_{\odot}}$.
\item $L_{g}$: the gas luminosity of the cluster, i.e., the X-ray bolometric gas luminosity within $R$, in $10^{44}$erg/s.
\item $T_{g}$: the gas temperature of the cluster, i.e., the mass-weighted gas temperature within $R$, in keV.
\item $\sigma_{v}$: the velocity dispersion of the cluster, i.e., the mass-weighted velocity dispersion of all particles belonging to a FOF halo, in km/s.
\item $z$: the redshift of the cluster.
\end{enumerate}

All these features are continuous variables except for $z$, which only has 6 discrete values (0, 0.14, 0.29, 0.47, 0.67, 0.9). 
From the definitions above, we see that $M_{*}$, $M_{g}$, $M_{t}$, $L_{g}$ and $T_{g}$ are $R$-dependent quantities, i.e., they are integrated within a given overdensity radius, while $\sigma_{v}$ is independent of $R$ and has one value per halo (S+20). Magneticum simulations provide 6 typical definitions for radius. 
In addition to the standard virial radius, $R_{vir}$, at which the mean density crosses the one of a theoretical virialized homogeneous top-hat overdensity \citep{1998ApJ...495...80B}, 
there are radii corresponding to cluster densities which are 200 times ($R_{200M}$) and 500 times ($R_{500M}$) the mean matter density of the Universe at the cluster's redshift.
Furthermore, there are the $R_{200C}$ and $R_{500C}$ radii that are similar to $R_{200M}$ and $R_{500M}$, but based on the critical density of the Universe.
In principle, we could use any of these radius definitions, as we can find a mapping of the values of cluster features between different definitions of characteristic radii by assuming a theoretical halo density profile (e.g., NFW profile, \citealt{1996ApJ...462..563N}). 
However, to be consistent with the usual choices in previous literature (e.g. \citealt{2022A&A...661A...2L}), we adopt $R_{500C}$ as the reference radius, and all quantities related to this radius in the rest of this analysis.

\begin{figure*}
\centering
\includegraphics[scale=0.8,trim=0 0.5cm 0 0]{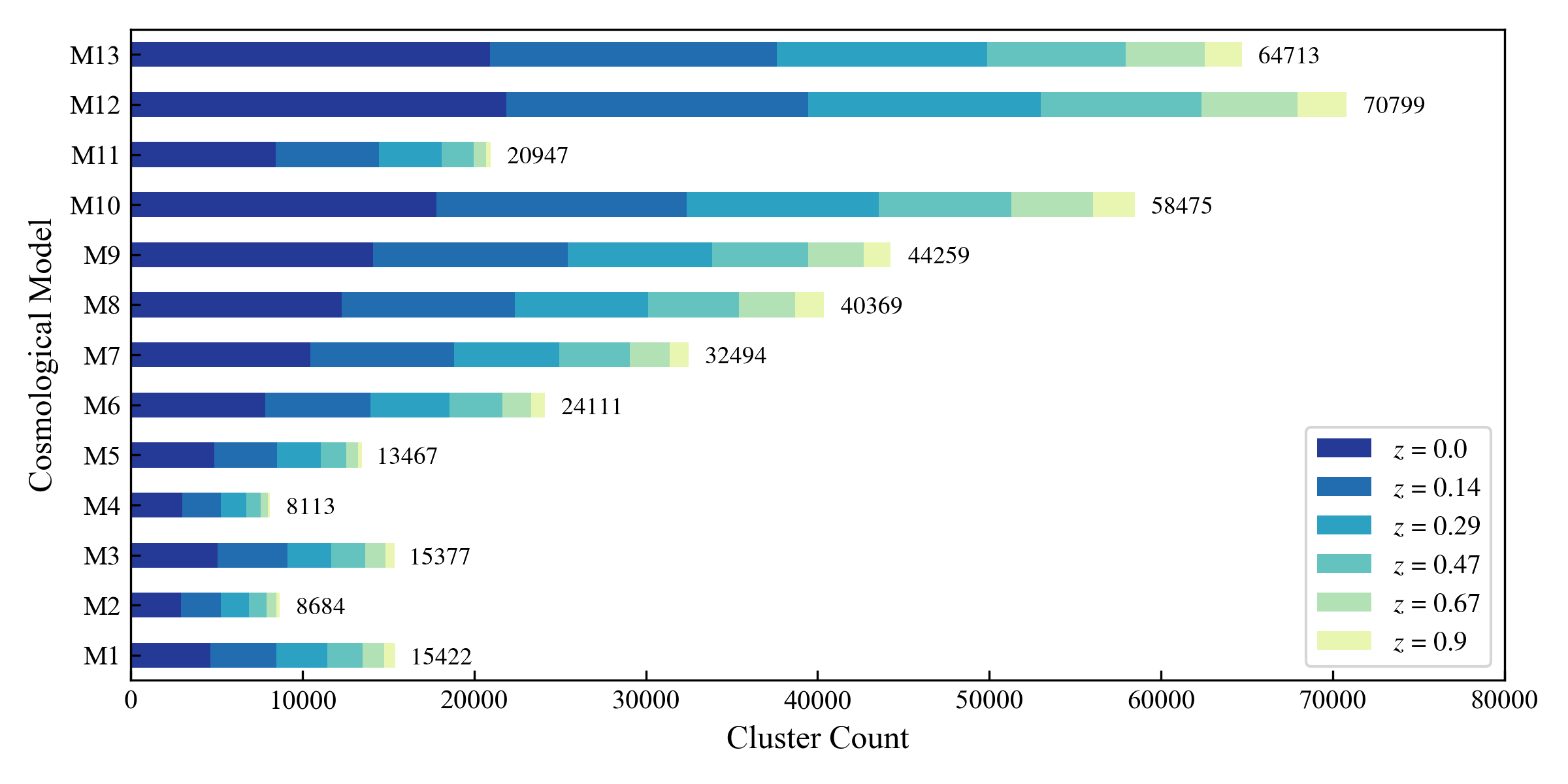}
\caption{The cluster count in each cosmological model and at each redshift. All of these clusters have already undergone the data pre-processing described in Sect. \ref{sec:pre-process}. The y-axis lists the different cosmological models, while the cluster counts are displayed in horizontal bars. The different redshifts are represented by different colors, as shown in the legend. As can be seen, the number of galaxy clusters in these 13 models varies significantly, with M12 (70799) having almost 9 times as many clusters as M4 (8113). To balance the training sample among different cosmologies, we adopt an undersampling, as described in Sect. \ref{sec:algorithm}.}
\label{fig:number_map}
\end{figure*}

\subsection{Pre-processed data}
\label{sec:pre-process}
Data pre-processing refers to cleaning, transformation, integration, normalization, and other operations on the raw data before using machine learning algorithms to make the data more suitable for the training and testing of machine learning models. 
Through the inspection of the original data, we found that there are problems such as outliers and heavy-tailed distribution that, if not cleaned, can affect the training and prediction of the model to a certain extent. Hence, we decided to select all quantities defined within $R_{500C}$ and deleted clusters with obviously non-physical properties, like negative $M_*$ or $\sigma_v$, likely coming from artifacts of the FOF algorithm (0.2\% over all simulations).

After this first cleaning step, being the features in simulations quite idealistic as, for instance, they do not have measurement errors, we decided to implement some ``rough'' observational realism. We artificially add Gaussian errors to mock a measurement process and make the quantities extracted from the simulation more similar to real cluster observations. 
As for the measurement errors, we have checked typical cluster observables from the literature and used eFEDS for reference. E.g., in \cite{2022A&A...661A...7B},  $M_{g}$, $L_{g}$, $T_{g}$, have typical relative errors of the order of 1$\%$, 2$\%$, 4$\%$ or less, respectively. For $M_{t}$, \cite{2022A&A...661A...2L} provides errors of the order of 1$\%$, which might be a little optimistic if compared to typical mass errors from weak lensing. To be conservative, we decided to adopt 5$\%$ relative errors as a reference experiment over all cataloged features discussed in Sect. \ref{sec:data} except for $z$ which are assumed here to be spectroscopic redshift with negligible errors. However, we will also consider more conservative errors of the order of 10\% for all features and up to 30\% for the total mass. This latter takes into account the largest errors obtained in weak lensing analyses of mid-low mass clusters  (see e.g. \citealt{2018ApJ...860L...4Sereno2018}).
After adding Gaussian noise to features other than $z$, we further performed logarithmic processing to solve the heavy-tailed distribution problem and make the predictive performance of subsequent machine learning models more stable.

The ``mock'' observations have been implemented by re-assigning, to each cluster, the ``observed'' physical quantities ($R, M_*, M_g, M_t, L_g, T_g, \sigma_V$), assuming Gaussian errors. This is done by randomly drawing the observed quantities from a Normal distribution centered in their original (true) value and with standard deviation corresponding to the adopted relative errors (in turn, 5\%, for the reference experiment, or smaller/larger, as discussed above).
This produces catalogs of observable-like features we will use for training and testing the ML tool (see Sect. \ref{sec:training and test}).
To give an overview of the final catalogs provided by the Magneticum multi-cosmology sample, we first visualize the cluster count as the function of both the cosmological model and redshift, in Fig. \ref{fig:number_map}.
The different cosmological models are listed on the y-axis, and for each horizontal bar showing the cluster counts, different colors represent different redshifts as in the legend.
As expected, we see that the total number of galaxy clusters in different universes varies greatly due to cosmological parameters. For example, M12 and M13 reach more than 60,000 clusters up to $z=0.9$, while M2 and M4 have fewer than 9,000 galaxy clusters in a volume of the same size. As the M1$\sim$M13 models are listed with increasing $\Omega_m$values, this is mainly the impact of the mass density of the Universe making the cluster collapse more effective.

\begin{figure*}
\hspace{-0.15cm}
\includegraphics[scale=0.33]
{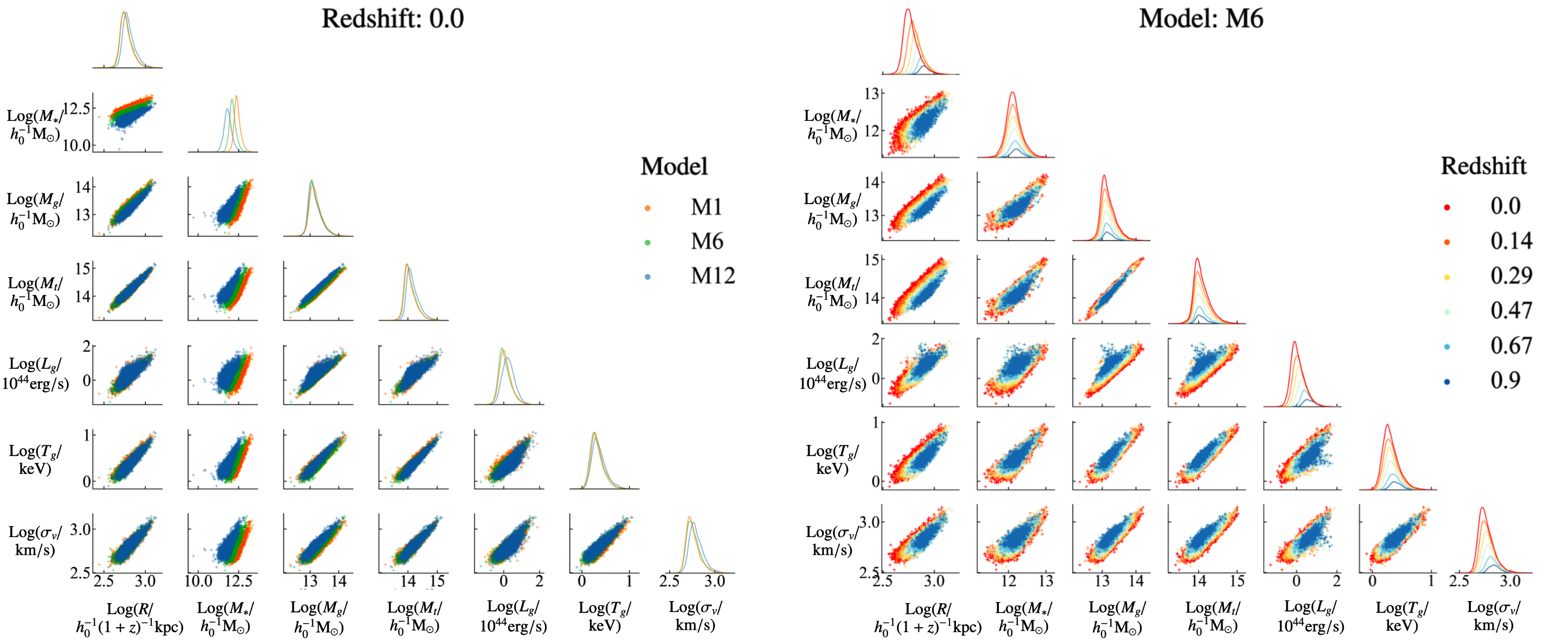}
\caption{The cluster features with 5\% measurement errors. These panels show all possible correlations (i.e., scaling relations) among the 7 features for 3 cosmologies (M1, M6, M12) at redshift $z=0$ (left), and as a function of the redshift ($z=0,0.14,0.29,0.47,0.67,0.9$) for M6 (right). Left panel: for a fixed redshift, we can see how the slope of the scaling relations is affected by cosmology, in particular for the scaling relation related to the stellar mass, $M_*$, and gas mass $M_g$, while all other scaling relations are more mixed. The corner histogram also shows how the number of clusters changes in a given cosmological volume.
Right panel: for a given cosmological model, apart from the differences in number counts, the galaxy clusters at different redshifts show similar power-law structures but with offsets driven by redshifts. 
This "cluster feature map" gives an overall impression of scatters and variations of cluster features among different cosmologies, which is the cornerstone of the method that uses galaxy cluster features to predict cosmology based on machine learning. For more details on definitions and accessibility of these cluster features, see Sect. \ref{sec:features and labels} and Sect. \ref{sec:pre-process}, respectively.}
\label{fig:feature_corner}
\end{figure*}

In Fig. \ref{fig:feature_corner} we also show all possible correlations (scaling relations) among the 7 features for 3 cosmologies at redshift $z=0$ (left), and as a function of the redshift for M6 (right), with M6 being the reference cosmology for Magneticum (see Sect. \ref{sec:magneticum}). This ``cluster feature map'' gives an impression of the scatter and the variation the ML method needs to be sensitive to, to distinguish different cosmologies and make correct predictions. Overall, from the figure we can see that some of the correlations are clearly distinguishable as a function of the cosmology at a fixed redshift (e.g. the correlations with $M_*$ or the $M_g$-$M_t$ in the left panel), while other correlations are rather mixed (e.g. the correlations involving the size, $R$). However, besides correlations, we can see that the expected distributions are different (see corner histograms), meaning that also the cluster densities in the parameter space can be used to distinguish cosmologies. We can also see that, for a given cosmology, there is a clear evolution of almost correlations with redshift (right panel). We expect the ML tools we intend to develop here can efficiently capture these features in the cluster catalogs.

It is worth noting that most of these features are standard products of cluster surveys, e.g. $M_{*}$, $M_{g}$, $L_{g}$, $T_{g}$ \citep{2009A&A...498..361P,2009ApJ...692.1033V,2013A&A...555A..30B,2019ApJ...871...50B}, while some other quantities are harder to get in real observations. For example, with respect to imaging and X-ray observations, only the most massive clusters can be used to derive precise total mass $M_{t}$ (e.g. with weak lensing measurements). Similarly, $\sigma_{v}$ needs time-consuming spectroscopical campaigns, and generally, these are also limited to a few tens of cluster members, although upcoming large all-sky redshift surveys (DESI: \citealt{2016arXiv161100036D}, WEAVE: \citealt{2012SPIE.8446E..0PD}, 4MOST: \citealt{2019Msngr.175....3D}) will soon produce rather large catalogs of clusters internal kinematics.

Hence, in this work, we have the chance to optimize the number of observables that are needed to constrain the cosmology.
By performing a ``feature importance'' analysis, we can check if ML can fully exploit the cosmological information encoded in some features, and their scaling relations, with respect to others, 
for example, checking the impact of the quantities that are observationally more difficult to obtain, e.g. $M_t$ and $\sigma_v$.

\section{The Machine Learning Cluster Cosmology Algorithm}
\label{sec:method}
In Sect. \ref{sec:data} we have introduced the multi-cosmological simulation data and related cosmology labels and described the 8 observational-like cluster features.
In this section, we present the full Machine Learning Cluster Cosmology Algorithm (MLCCA, hereafter), which we train to predict the best cosmology given a set of cluster observations ({\it mock catalog}, hereafter).
As anticipated, for this proof-of-concept we want to first demonstrate if an ML tool can recognize what cosmological simulation a given dataset has been extracted from. The basic idea is to produce random {\it mock catalogs} extracted from one of the M1$\sim$M13 simulations (including clusters from different redshifts) and let the MLCCA decide from which simulation this has been picked, on the basis of the correlations among the features (scaling relations as in Fig. \ref{fig:feature_corner}). 
This can be treated as a typical classification problem, where a machine learning classifier can predict the probability that a dataset belongs to different cosmological models.
This is the most obvious choice, given the limited number of cosmologies, although we will test also regression algorithms in the near future.

Classification-wise, due to the similarity of cosmological scaling relations in adjacent parameter spaces, the classification itself will have an error. This essentially produces uncertainties in the inference of cosmological parameters.
Also, by sparsely sampling the cosmological parameter space (see Fig. \ref{fig:para_map}), we can check whether the MLCCA can learn a pattern among the scaling relations in the cosmological parameter space and interpolate data coming from a ``cosmology'' (meaning a simulation) that is not included in the training.
We quantify each of these steps by proper evaluation metrics defined in Sect. \ref{sec:metric}. The final goal is to build an algorithm that, starting from cluster catalogs, can return confidence contours of the four cosmological parameters ($\Omega_m, \sigma_8, h_0, \Omega_b$) used as labels in the ML training.

\subsection{Machine learning classifiers}
\label{sec:algorithm}
Broadly speaking, the task of the classifier will be to issue the probability for a given cluster $i$ to belong to a given cosmological model $j$.
Machine learning classifiers are mainly divided into two types: tree models and neural networks. In this work, we want to use {\it tree models} which are generally more robust and possess better interpretability than neural networks \citep{2001MachL..45....5B}.
In particular, we are interested in ensemble learning on tree models, which is a way to optimize the accuracy of single-tree models. The improvement of the performance, here, is obtained by constructing a set of tree models and then classifying new data points by taking a (weighted) vote on their predictions \citep{2000Ensemble}, hence overcoming the non-optimal performance (underfitting, overfitting, etc.) of each individual tree model. 

To perform the classification on 13 cosmological models based on available features, we consider four typical ensemble tree models, i.e., Random Forest (RF, \citealt{2001MachL..45....5B}), Extra Trees (ET, \citealt{ET}), Light Gradient Boosting machine (LGB, \citealt{LGB}) and eXtreme Gradient Boosting (XGB, \citealt{XGB}).
To select the most appropriate model for this project, we evaluated the above 4 machine learning models and selected the best option using appropriate evaluation criteria such as {Accuracy} and {Logloss} as described in Sect. \ref{sec:classifier metrics}. 
We anticipate here that LGB is the best solution, as it is discussed in detail in Sect. \ref{sec:class_selection}.

\subsection{Training and test samples}
\label{sec:training and test}
For the training phase, we use the cluster features as the input to obtain the label of the predicted cosmological model as the output. In particular, we adopt a multi-class classification, which directly gives the probability that a cluster may belong to any of 13 available models, and take the model with the highest probability as the predicted model.

Regarding the construction of the training sample, the number of galaxy clusters in different cosmologies varies greatly due to the influence of cosmology itself on large-scale formation, as shown in Fig. \ref{fig:number_map}. This uneven distribution can likely force the model prediction to skew toward categories with a higher number of samples \citep{2004Data}. To correct this effect, we apply an under-sampling method, i.e., we reduce the size of the samples in the majority classes to balance the datasets of the smaller classes. 
Since all selected cosmologies have more than 8000 galaxy clusters, we randomly draw 7000 galaxy clusters for each cosmology as training samples. 
We stress here that this is a rather brute-force approach driven by low $\Omega_m$ cosmologies, producing a low number of clusters, that strongly penalizes the predictive power for more populated cosmologies in Fig. \ref{fig:number_map}.
We have decided to accept this drawback in order to keep the largest number of cosmologies for this first test based on the current Magneticum sample.
For the testing phase, in order to make full use of the left behind non-training objects for each cosmology, we randomly selected 20 times 700 clusters, to obtain 20 different test samples with no overlap with the corresponding 7000 clusters which make up the
training sample. 
Each test sample (i.e., {\it mock catalog}) can be regarded as a sample representative of typical observational catalogs currently available for cosmological tests (see e.g.\ \citealt{2018A&A...620A...5A,2020MNRAS.492.4528S}).

\subsection{Evaluation metrics}
\label{sec:metric}
Here, we introduce the metrics to assess the three main tasks of this paper: 1) selecting the best classifier capable of performing the multi-class analysis of the {\it mock catalogs}; 2) classifying  {\it mock catalogs} belonging to different cosmological models; 3) predicting cosmological parameters for a certain galaxy cluster {\it mock catalog}. 
In all cases, we first train the ML tool using an ensemble of clusters with the same size from each of $m$ cosmological models distinguished by their labels (the 4 cosmological parameters).
Then, we use a test set that contains $n$ clusters coming from the same cosmology to finally measure the performance of the results.
All the corresponding quantities of model $j$ ($j\in\{1,2,...,m\}$) and cluster $i$ ($i\in\{1,2,...,n\}$) are defined as follows:
\begin{enumerate}
\setlength{\itemsep}{3pt}
\item $\{ \theta_{j}\}$: cosmological parameters of model $j$;
\item $\{ X_{i}\}$: features of cluster $i$; 
\item $y_i$: true cosmological model of cluster $i$;
\item $\hat{y}_i$: predicted cosmological model of cluster $i$;
\item $\{ \theta_{i}\}$: true cosmological parameters of cluster $i$;
\item $\{{\mu}_{i}\}$: mean values of predicted cosmological parameters of cluster $i$;
\item $\{{\sigma}_{i}\}$: standard deviations of predicted cosmological parameters of cluster $i$;
\item $P({\theta_j}|X_{i})$: probability that cluster $i$ belongs to model $j$, which is the outcome of the classifier,
\end{enumerate}
where cosmological parameters $\theta, \mu \in\{\Omega_m, \sigma_8, \Omega_b, h_0\}$, cluster features $X \in\{R, M_{t}, M_{*}, M_{g}, L_{g}, T_{g}, \sigma_{v}, z\}$, model labels $y, \hat{y} \in\{1,2,...,m\}$ and the sum of predicted probabilities for each cluster $\sum_{j = 1}^{m} P({\theta_j}|X_{i}) =1$.

\subsubsection{Classifier metrics}
\label{sec:classifier metrics}
For the classifiers' performances, we include the following evaluators: 1) {Accuracy} and 2) {Logloss}. By {Accuracy} we indicate the proportion of all correctly classified samples ($N(\hat{y}_i=y_i)$) in all samples ($n$). To estimate that we use the following equation:
\begin{equation}
{\rm Accuracy} = \frac{N(\hat{y}_i=y_i)}{n}
\end{equation}
ranging from 0 to 1. The closer to 1, the better the classifier performance on the whole. The {Logloss} represents the average probability (in logarithm) of a cluster being correctly classified. The equation defining this is:
\begin{equation}
{\rm Logloss} = -\frac{1}{n}\displaystyle\sum_{i = 1}^{n}\displaystyle\sum_{j = 1}^{m} \delta_{j,y_i} \log (P({\theta_j}|X_{i})),
\end{equation}
where $\delta_{j,y_i}$ equals 1 if $j=y_i$ and 0 otherwise. The lower and upper limits of probability are set as $10^{-15}$ and $1$, respectively, to avoid infinity in the logarithm. The Logloss also ranges from 0 to 1, and the closer to 0, the better the classifier performance.

\begin{figure*}[h]
\centering
\includegraphics[scale=0.3]{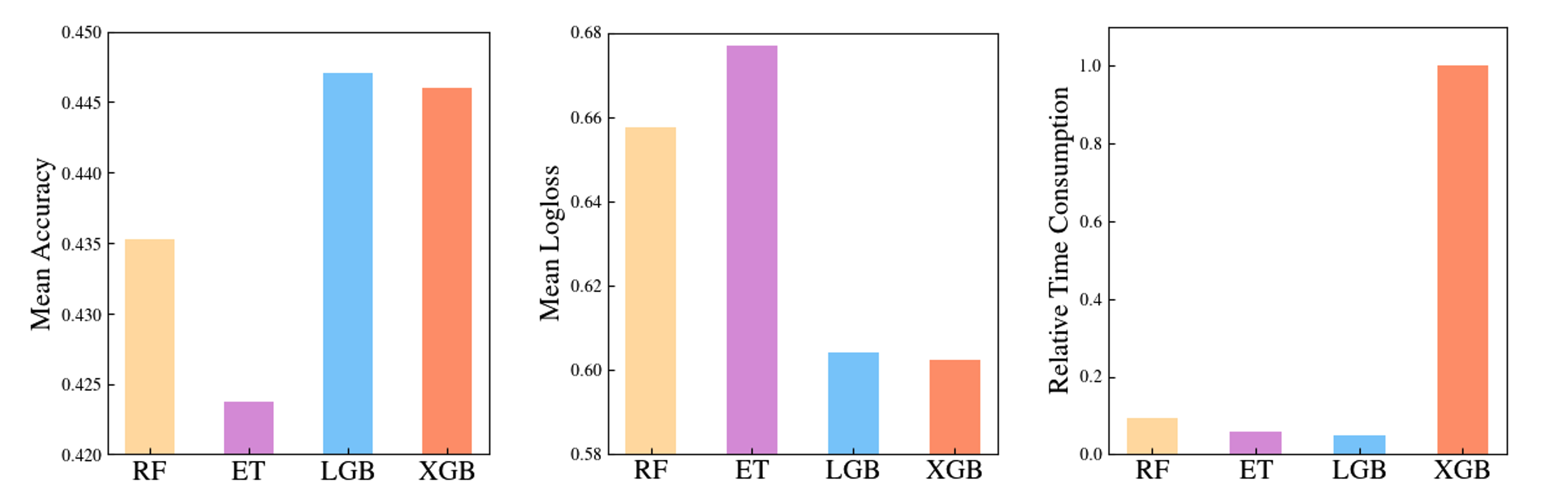}
\caption{Performance comparison of four classifiers (RF, ET, LGB, XGB) for baseline configurations in terms of mean accuracy, mean logloss, and relative time consumption during the cross-validation process.}
\label{fig:pipeline-classifier}
\end{figure*}

\subsubsection{Classification metrics}
\label{sec:class_met}
Once we have defined the best classifier, we can proceed with assessing the performance of the classification. 
This will be based on the \emph{Recall}, which represents the ratio of correctly predicted samples with respect to the total sample.

For each model, the classifier returns a true/false binary outcome and will produce four different results, in terms of correct (positive) or incorrect (negative) prediction: (1) TP: Truly predict positive to be Positive; (2) FP: Falsely predict negative to be Positive; (3) TN: Truly predict negative to be Negative; (4) FN: Falsely predict positive to be Negative. The \emph{Recall} of model $j$ ($j\in\{1,2,...,m\}$) is defined as the fraction of the correctly classified $j$ samples in all real $j$ samples, as follows:
\begin{equation}
{\rm Recall_j} = \frac{N(\hat{y}_i=y_i=j)}{N(y_i=j)}=\rm{\frac{TP_j}{TP_j+FN_j}}
\label{eq:recall}
\end{equation}
This ranges from 0 to 1, and the closer it is to 1, the better the classifier performance on model $j$ is. 
As we are dealing with a multi-classification problem, the TP, FP, TN, and FN are defined in Eq. \ref{eq:recall} with respect to the maximum probability received by each cluster $i$ among the 13 $j$ cosmologies. In principle, we could use a lower threshold to account for a reasonably significant probability for the ML tool to ``recognize'' a cluster to belong to a given cosmology, but this would alter the final distribution of the recall and arbitrarily reduce the ``errors'' on the classification\footnote{We have tested a series of lower threshold like 0.1, 0.2, 0.3 and checked that this would increase the TPs and reduce the FNs, overall improving the Recall.}. On the other hand, assuming no lower threshold we can stress test the overall method by minimizing its accuracy and checking if it can really produce correct classifications and cosmological parameter estimates.

\subsubsection{Cosmological parameter metrics}
\label{sec:para metric}
After classification, for each cluster $i$, we use the probability that cluster $i$ belongs to model $j$, $P({\theta_j}|X_{i})$, to infer its cosmological parameters. For each individual cluster, in principle, we can define the mean and standard deviation of a certain parameter as
\begin{equation}
\mu_{i}  = \displaystyle\sum_{j = 1}^{m} P({\theta_j}|X_{i}) \cdot  \theta_j
\end{equation}
\begin{equation}
\sigma_{i}^2  = \displaystyle\sum_{j = 1}^{m} P({\theta_j}|X_{i}) \cdot  (\theta_j-\mu_i)^2,
\end{equation}
where $P({\theta_j}|X_{i})$ is considered as a probability distribution.
Using the same $P({\theta_j}|X_{i})$, in order to account for asymmetric errors, we decide to compute the lower 16\% percentile, the median, and the upper 84\% percentile, roughly corresponding to 1-$\sigma$ lower bound, $\sigma_l$, median $\hat{\theta}_m$, and 1-$\sigma$ upper bound, $\sigma_u$, respectively.
Then, we use 1) {Bias} and 2) {Score} to evaluate the parameter predictions.
The {Bias} represents the deviation between the predicted median and the true value, i.e.,
\begin{equation}
{\rm Bias} = \hat{\theta}_m-\theta.
\end{equation}
The {Score} is short for {Standard Score}, which represents the magnitude of {Bias} relative to a confidence interval.
\begin{equation}
{\rm Score}=
\begin{cases}
\displaystyle{\frac{\hat{\theta}_m-\theta}{\sigma_l}}& \text{ when $\hat{\theta}_m > \theta$ } \\
\displaystyle{\frac{\hat{\theta}_m-\theta}{\sigma_u}}& \text{ when $\hat{\theta}_m < \theta$ }.
\end{cases}
\end{equation}

We can finally obtain the marginalized 2D 1-$\sigma$ and 2-$\sigma$ confidence contours of all combinations of the 4 parameters, as the 68\% and 95\% enclosed probability of the probability distribution function (PDF) of the cluster catalog (see also Appendix \ref{sec:appA} for more details). This latter can be defined as PDF$=\sum_{i = 1}^{n}G(\mu_i, \sigma_i)=1$, assuming a Gaussian distribution, $G(\mu, \sigma)$, for the cluster individual parameter estimates. We stress here that this returns a conservative estimate of the uncertainties of the parameter, fully capturing the uncertainties in the classification encoded in the $\sigma_i$.

\begin{figure}
\hspace{-0.3cm}
\includegraphics[scale=0.53]{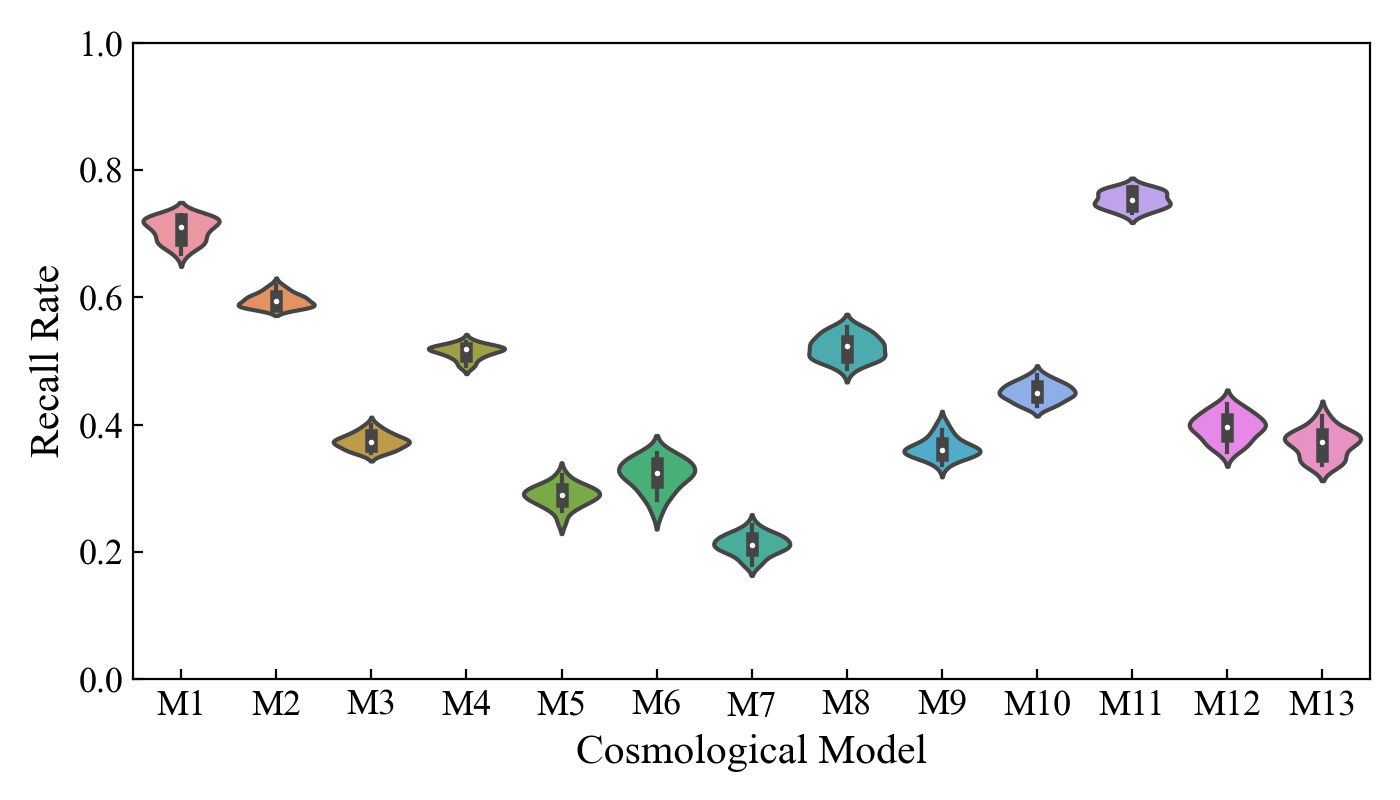}
\caption{Recall rate over the 20 test samples (i.e. {\it mock catalogs}) used in each cosmological model. 
The recall rate represents the proportion of galaxy clusters that are correctly classified into all clusters. 
The shape of a ``violin'', i.e., the width as a function of the Recall rate, represents the probability distribution of recall of the 20 test samples (see text for details). The white dot in the center of the violin represents the median recall. As can be seen, the median recall rate displays distinct variations among different cosmological models.}
\label{fig:recall}
\end{figure}

\section{Results}
\label{sec:result}
In this section, we show the results of 1) selecting the best classifier, 2) mock catalog classification, and 3) cosmological parameter estimates.
We first choose the best classifier for the MLCCA, according to the performance evaluation discussed in Sect. \ref{sec:metric}. Then, we apply the MLCCA to the test sample described in Sect. \ref{sec:training and test} and assess its performance, including the accuracy and precision of the cosmological parameter estimates, in the perspective of future applications over real datasets.

\begin{figure}
\hspace{-0.5cm}
\includegraphics[scale=0.47,trim=1cm 0 0 0]{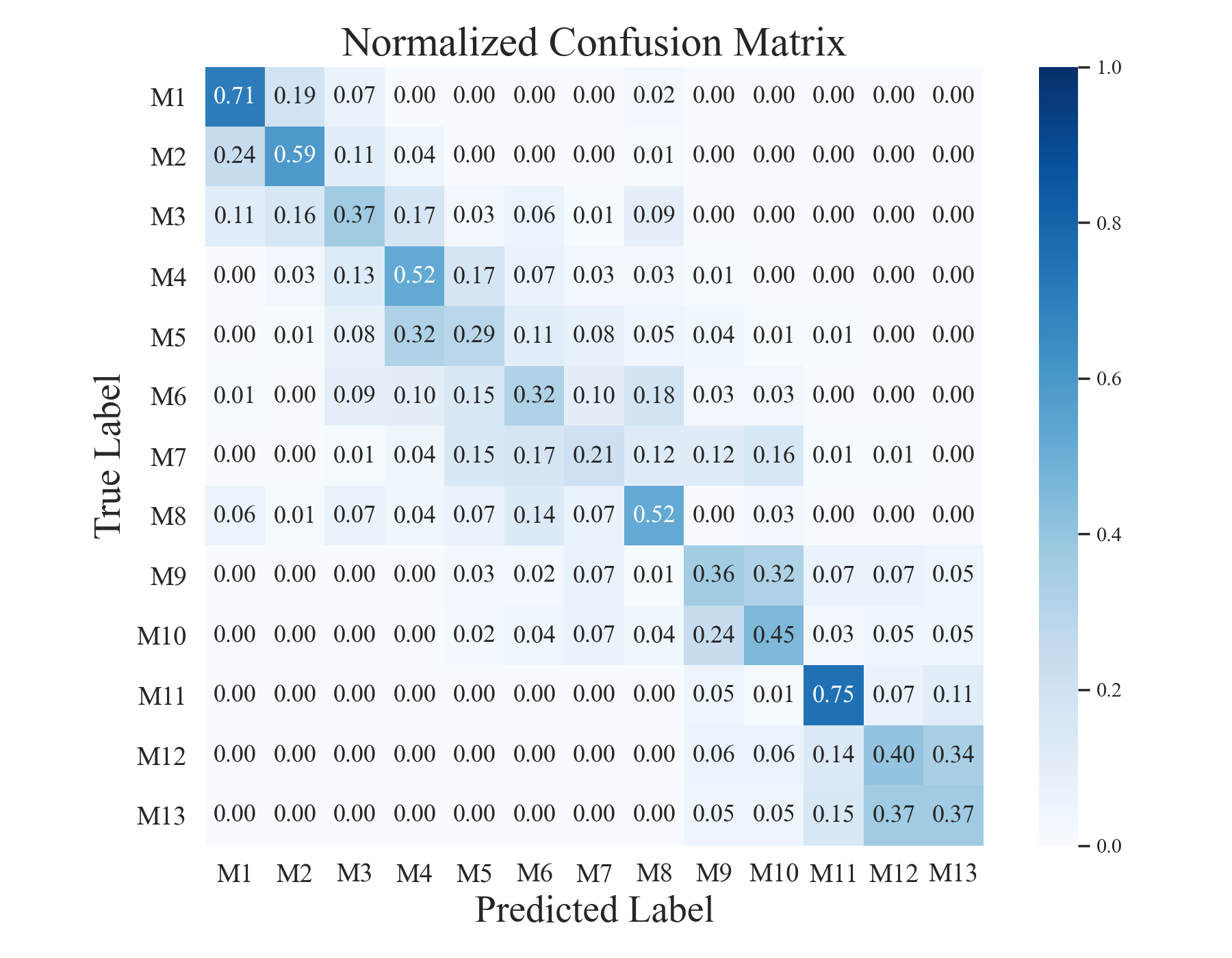}
\caption{Normalized confusion matrix for test samples (i.e. {\it mock catalogs}). Each row of this matrix represents a test sample taken from a certain cosmology (containing 700 galaxy clusters), where each cell represents the fraction of galaxy clusters classified as belonging to the x-label cosmology. The diagonal of the matrix represents the recall rate (i.e., the fraction of clusters correctly classified) coinciding with the median recall of each violin in Fig. \ref{fig:recall}. The non-diagonal elements of the matrix represent the fraction of clusters that have been misclassified to other universes. As can be seen from Fig. \ref{fig:recall} and this figure, machine learning has a low recall rate and large misclassified fractions for the central models (such as M5, M6, M7, and M9), indicating that these cosmologies have more overlap with neighboring cosmologies.}
\label{fig:confusion}
\end{figure}

\subsection{Selecting a proper classifier}
\label{sec:class_selection}
We start by using the four classifiers (RF, ET, LGB, XGB) to perform a first-round test on the training sample with 5-fold cross-validation. That is, in 5 subsequent experiments, we rotate 4/5 of the sample as a training sample, and the other 1/5 as a test sample to calculate the results, and then take the mean of the 5 test experiments as the final result.
In Fig. \ref{fig:pipeline-classifier} we show the three indicators discussed in Sect. \ref{sec:classifier metrics}, i.e. the mean {Accuracy} and mean {Logloss}. We also show the computing time needed during the cross-validation process as a further indicator of the efficiency of the method. 
We find that the LGB has the highest mean {Accuracy} and the 2nd lowest mean {Logloss} with minimal time consumption. Therefore, we identify LGB as the best machine learning classifier among the four considered in our analysis, as it possesses clear advantages due to the fast training, high accuracy, and low memory footprint.
These performances come from its ability to discretize continuous features through a histogram-based decision tree algorithm and to use distributed gradient boosting decision trees (GBDT), which are specifically efficient to improve training efficiency.
To further optimize the LGB and reach a higher {Accuracy}, we use Optuna \citep{2019Optuna}, which is an automated hyperparameter tuning framework, to mainly adjust {\tt learning rate} and {\tt n$\_$estimators}, that are strictly related to {Accuracy}. We finally find that the combination of {\tt learning rate = 0.07} and {\tt n$\_$estimators = 150} can improve the {Accuracy} and also reduce the {Logloss} with respect to the default configuration with {\tt learning rate = 0.1} and {\tt n$\_$estimators = 100}. However, the mean {Accuracy} of 5-fold cross-validation for the latter is 0.447 while for the optimized version is 0.449. Also, the mean {Logloss} for the default configuration is 0.604 while for the optimized version is 0.602. Hence, from default to optimized LGB, the {Accuracy} has increased by 0.002 and the {Logloss} has decreased by 0.002. These are small changes, which prove that there is not much freedom in the set-up of the network and the final performances are fully dominated by the intrinsic complexity of the data and how these reflect the cosmological information encoded in them.

\begin{figure}
\hspace{-0.5cm}
\includegraphics[scale=0.28,trim=0 0 0 0]{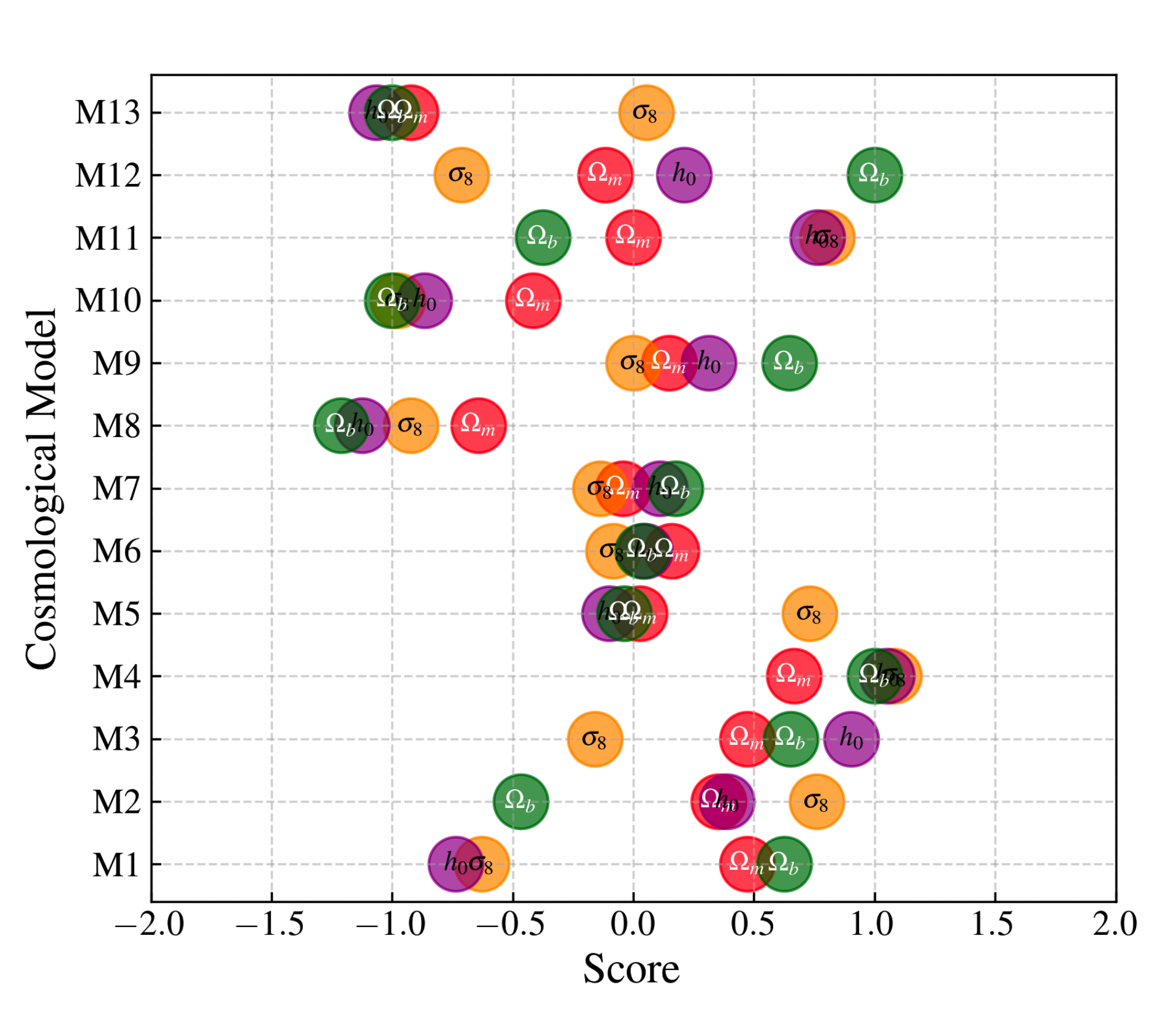}
\caption{\emph{Score} values (x-axis) for the different cosmological models (y-axis) showing the distribution of the estimated cosmological parameters represented with different colors. Negative and positive \emph{Score} values indicate underprediction and overprediction, respectively. As can be seen, almost all parameters are predicted by the MLCCA within $1\sigma$ from their true value. Notably, for cosmologies at the center of the parameter space, such as M5, M6, M7, and M9, the MLCCA method can accurately recover the four cosmological parameters well within the 1$\sigma$ level.}
\label{fig:score}
\end{figure}

\subsection{Classifying cosmological models}
\label{sec:class_models}
We now apply the MLCCA based on the optimized LGB to the test samples of 13 cosmological models respectively. 
In Fig. \ref{fig:recall}, we show the statistics of the overall recall rate over the 20 test samples used in each cosmology. For each cosmological model, due to the variance among 20 test sets, the recall distribution has a certain fluctuation, which we quantify with a ``violin'' plot, where the width of each violin represents the probability at a certain recall level. The ``median'' recall rate varies from different cosmological models, with lower recall rates found for cosmologies that have more overlap with neighbor cosmological models, given a larger chance that the classifier assigns a cluster to some close cosmology.

For each mock test sample from a given cosmology from the violin diagram above, in Fig. \ref{fig:confusion} we show the median confusion matrix, showing the ``median'' fraction of a given cluster sample that has been classified on each cosmology, color-coded by the density of the allocated cluster in a given sample. A perfect classifier would return a series of 1 along the diagonal, while in Fig. \ref{fig:confusion} we see this is not the case, as the confusion matrix mirrors the situation seen in Fig. \ref{fig:recall}. In particular, we can see that for simulations with larger overlaps with close cosmologies, there is a larger spread or recall cluster from each sample.
However, in all cases (except M5\footnote{Note that the close off-diagonal bin has a recall rate which is larger but consistent with the diagonal one within Poissonian noise. As we will show in Appendix \ref{sec:appB}, this does not impact an unbiased cosmological parameter prediction.}), the classifier assigns the majority of the cluster of the sample to the correct cosmology (along the diagonal), while the misclassified clusters still carry on their cosmological
information. As we will see in the next sections, this cosmological information remains encoded in the classification probabilities
among all these cosmological models and effectively impacts the 
recovery of the true cosmological parameters, as well as their uncertainties.

\subsection{Inferring cosmological parameters}
\label{sec:infer para}
We can now check the performance of the MLCCA in the prediction of the cosmological parameters from the test sample, using the metrics described in Sect. \ref{sec:metric}. 

In Fig. \ref{fig:score}, we start by showing the {\it Score} of the predicted cosmological parameters (reported on the x-axis) for all cosmological models (y-axis). This plot gives in one glance the accuracy and precision for each cosmological parameter as a function of the ``true'' cosmology the mock catalog is originally extracted from. For instance, for the catalog extracted from M13 (on the top row), only $\sigma_8$ is constrained at less than 1$\sigma$ level, while the other parameters are off the scale, i.e. are ``biased'' by $\sim 1\sigma$. Similarly for M1 (bottom row) none of the parameters is constrained with accuracy better than 0.5$\sigma$. On the other hand, for models like M5, M6, M7 and M9, the MLCCA  correctly recovers $\Omega_m$, $\sigma_8$, $h_0$ and $\Omega_b$ with the true values all well within $1\sigma$ confidence intervals
of the prediction ranges. Overall, the models lying in the bulk of the parameter space covered by the Magneticum multi-cosmology simulations obtain a $|{Score}|<0.5$ for most of the cosmological parameters, especially $\Omega_m$ and $\sigma_8$.
Besides, there is a mild trend that the farther a parameter is from the parameter space bulk in Fig. \ref{fig:para_map}, the larger the probability of being under/overestimated.

We can have a better perception of the remarkable accuracy and precision of the recovered parameters from the corner plot in Fig. \ref{fig:M6}, where we draw the confidence contours for M6. 
As mentioned before, the {\it mock catalog} from M6 cosmology, used to derive these constraints, contains $R$, $M_{*}$, $M_{g}$, $M_{t}$, $L_{g}$, $T_{g}$, $\sigma_{v}$ and $z$ values for 700 galaxy clusters, having all, except the redshift, relative error of 5\%. 
The predicted values are 
$(0.279_{-0.039}^{+0.041}, 0.806_{-0.066}^{+0.060}, 0.705_{-0.021}^{+0.021}, 0.0457_{-0.0028}^{+0.0027})$,
respectively. 
They are all consistent with the true values of $(\Omega_m, \sigma_8, h_0, \Omega_b)$ of M6, which are $(0.272, 0.809, 0.704, 0.0456)$, within the estimated errors.
The corresponding $1\sigma$ relative precisions are 14\% for $\Omega_m$, 8\% for $\sigma_8$, 3\% for $h_0$, 6\% for $\Omega_b$. 
These constraints are somehow tighter for $\Omega_m$ but similar to the ones on $\sigma_8$ of the ones obtained by using joint analyses of the cluster abundance and the weak-lensing mass calibration (22\% for $\Omega_m$ and 8\% for $\sigma_8$ in, e.g., \citealt{2022arXiv220712429C}). This can be due to the error size adopted here, which might be optimistic for some parameters, although they are still more conservative than the ones from \citet{2022arXiv220712429C}.
In Sect. \ref{sec:errors}, we will check the impact of even more conservative errors and see that the parameter precisions are little affected, except for $\Omega_m$.

\begin{figure}
\hspace{-0.5cm}
\includegraphics[scale=0.39]{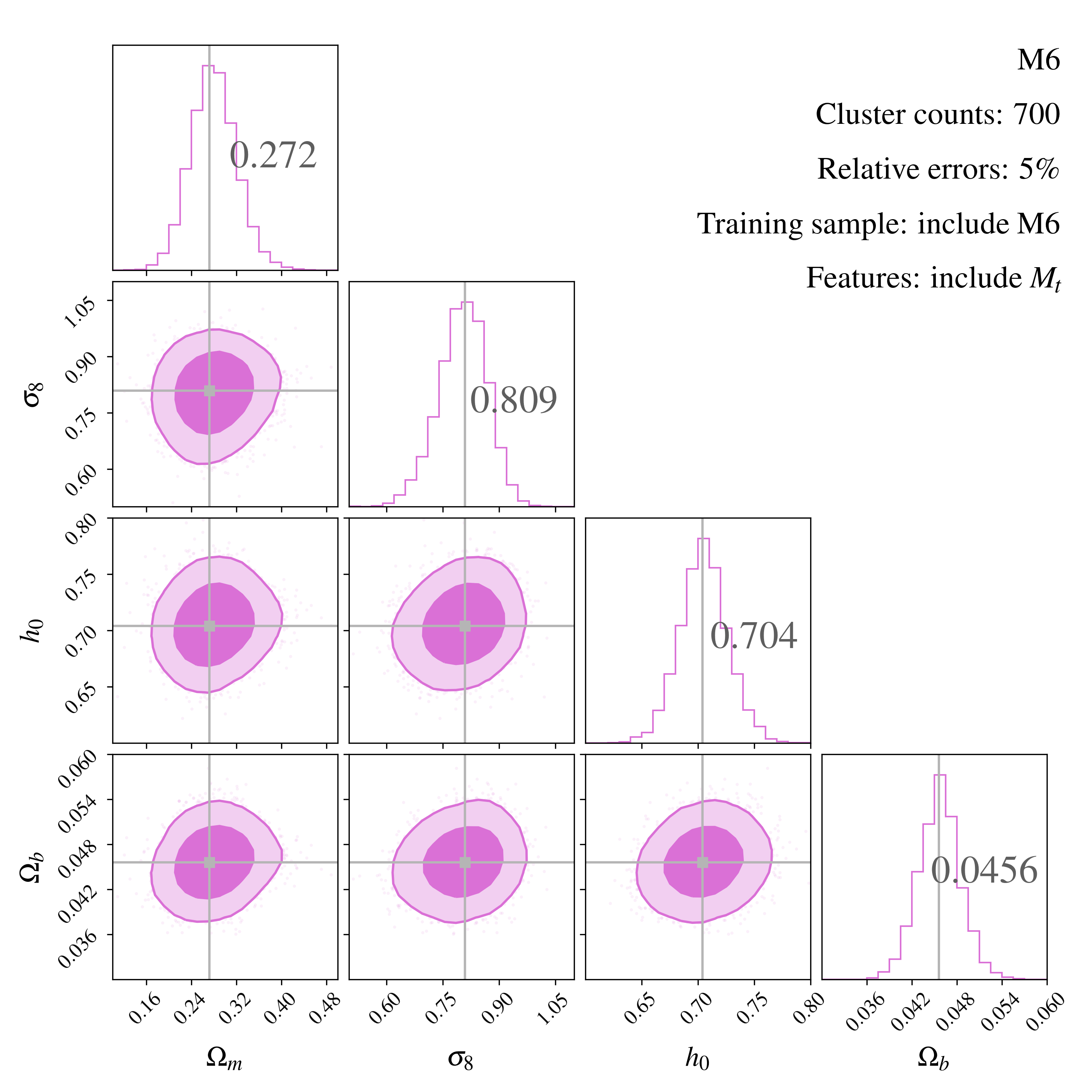}
\caption{Cosmological parameters of M6 inferred by the MLCCA. The contours enclose 1- and 2-$\sigma$ confidence intervals for the cosmological parameters of each 2D projection. The histograms on the diagonal represent the posterior probability distribution of the four cosmological parameters. The gray lines in the figure represent the true values of the various parameters of M6 cosmology, with the true value of each parameter shown in the posterior probability diagrams. It can be seen that all cosmological parameters are within the 1$\sigma$ confidence interval.}
\label{fig:M6}
\end{figure}

We finally remark that, for a certain cosmological model, both the accuracy of the classification and the estimated parameters are related to its position in the parameter space (i.e., the parameter distribution: Fig. \ref{fig:para_map}).
Some of the more extreme cosmologies, such as M1 and M11, are at the edge of the sampled parameter space, so they are easier to recognize by classifiers and therefore have higher classification accuracy (see confusion matrix: Fig. \ref{fig:confusion}). At the same time, though, due to their position on the edge of the parameter space, the misclassified clusters are oddly distributed, as they are mixed with cosmology located more likely on the same side of the parameter space (at least in some projections), resulting in an overall overestimation or underestimation of some parameters with a larger overlap. For instance, M1 and M11 lie in the opposite edges of the $\Omega_m-\sigma_8$ and  $\Omega_m-h_0$ projections in Fig. \ref{fig:para_map}, which makes them easy to classify (recall rate larger than 0.7 in Fig. \ref{fig:confusion}); however, from Tab. \ref{tab:para_map}, M1 seats on the minimum of the $\Omega_m$ range and close to the maximum of $\sigma_8$ and these parameters are overestimated and underestimated\footnote{M1 is also close to the minimum of the $\Omega_b$ and has a large $h_0$, so these parameters are also biased.}, respectively (see Fig. \ref{fig:score}), while M11 has a minimum in both $\sigma_8$ and $h_0$, which are overestimated and is the second ranked in $\Omega_b$ (see Tab. \ref{tab:para_map}), which is underestimated (Fig. \ref{fig:score}). For cosmologies in the bulk of the parameter space, such as M5, M6, M7, and M9, despite a lower classification accuracy, the misclassified clusters are more evenly distributed on both sides of the parameter space, hence producing a more balanced parameter prediction, with a smaller bias. This can be seen in Fig. \ref{fig:score}, where the accuracy of the prediction of the four parameters of M5, M6, M7, and M9 is obviously better than that of other models (see also contour plots in Appendix \ref{sec:appB}).

We, therefore, conclude that the MLCCA algorithm works better for cosmological predictions in the center of the sampled parameter space.
More precisely, for a specific cosmological model, {\it the MLCCA can efficiently recover the true cosmological parameters, provided that the training set, made by a series of multi-cosmology hydro-simulations, evenly covers the cosmological parameter space around the true cosmology}. 
This represents the main results of this paper as it strongly suggests increasing the number of cosmologies covered by large-volume, mid-resolution hydro-simulations, to fully apply this method to real data in the future.

\section{Robustness and Systematics}
\label{sec:discussion}
In the previous sections, we demonstrated the ability of the MLCCA to recover the cosmological parameters by giving a mock catalog of 700 clusters randomly distributed in redshift, for which seven specific observational quantities are given. 
In this section, we want to check the robustness of this result and discuss the impact of some assumptions made in our analysis and by the properties of the simulations adopted.
To be more specific we will consider: 1) the ability of the MLCCA to predict the cosmology of the test sample in the case this is not covered in the training sample, in fact by testing the capability to interpolate between different cosmologies in a grid of parameters; 2) the accuracy of the MLCCA predictions excluding some relevant features, in particular, the total mass; 3) the impact of the size of the measurement errors; 4) the impact of the simulation resolution.  

\begin{figure}
\hspace{-0.5cm}
\includegraphics[scale=0.39]{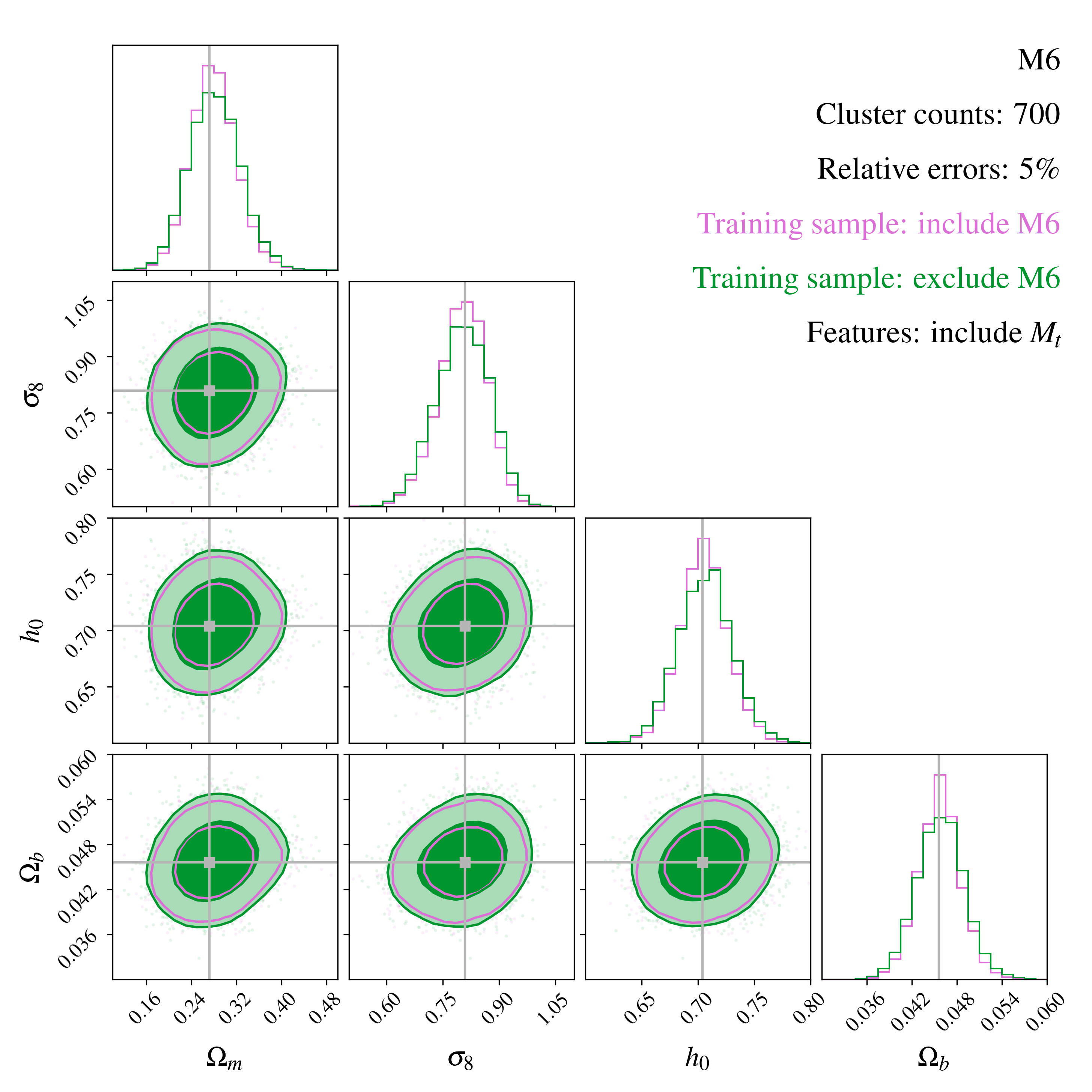}
\caption{Parameter constraints for an M6 {\it mock catalog} obtained by training a model using the training sample that includes (purple) or excludes (green) the M6 cosmology. This graph is the same type as Fig. \ref{fig:M6}. In both cases, all cosmological parameters are in the 1$\sigma$ region, indicating that our method has the potential to be applied to the cosmology where each cosmological parameter is roughly located in the center of the parameter space of the training sample, but the specific configuration is unknown.}
\label{fig:M6 include exclude}
\end{figure}

\begin{figure}
\centering
\includegraphics[scale=0.45,trim=0.7cm 0 0.3cm 0]{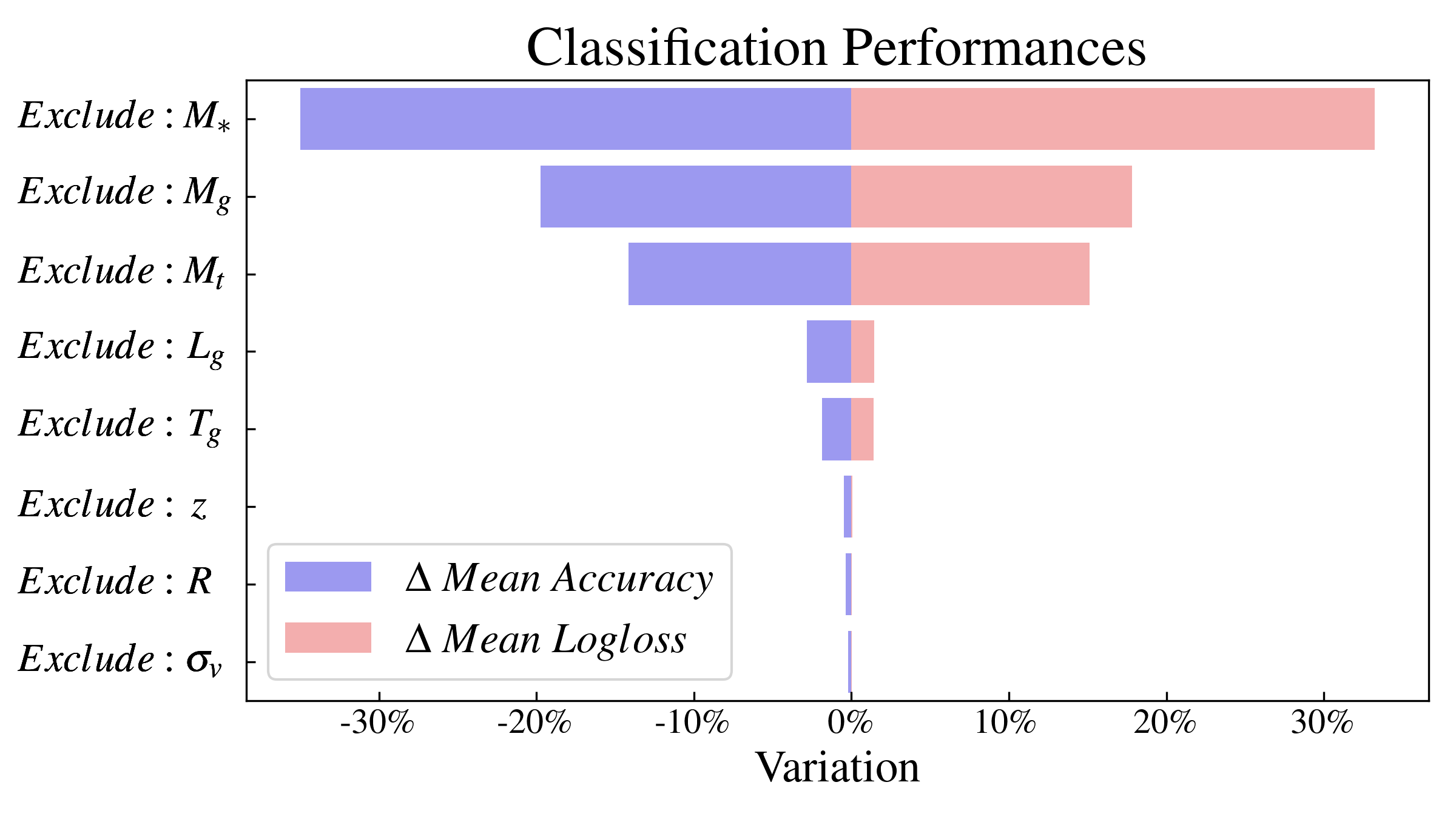}
\caption{Comparisons between the performance of the model retrained after excluding a certain feature and the performance of the model before exclusion. The red bars and purple bars represent the percentage change in mean {Accuracy} and mean {Logloss} during the 5-fold cross-validation process respectively. As can be seen, stellar mass, $M_*$, and gas mass, $M_g$, have the most substantial impact on the overall performance of the classifier, indicating their crucial importance in MLCCA inference.}
\label{fig:variance percentage}
\end{figure}

\subsection{Excluding a certain cosmology}
The cosmological parameters of the real Universe may not be the same as any of the cosmological models in a given simulation set. In this case, we need to check if the machine learning trained with various existing cosmological models can still accurately predict a model that has not been directly learned before.
In Fig. \ref{fig:M6 include exclude}, we show the distribution of the predicted cosmological parameters from a mock catalog from M6 using an MLCCA trained on all the cosmologies in Table \ref{tab:para_map}, except M6 itself.
The predicted values are 
$(0.281_{-0.044}^{+0.046}, 0.805_{-0.075}^{+0.070}, 0.707_{-0.024}^{+0.023}, 0.0458_{-0.0032}^{+0.0030})$, respectively. 
They are all consistent with the true values of $(\Omega_m, \sigma_8, h_0, \Omega_b)$ of M6, which are $(0.272, 0.809, 0.704, 0.0456)$, within the estimated errors. This is a remarkable result, showing the ability of the MLCCA to interpolate even over a sparse grid of simulations around the true cosmology the test sample belongs.

\subsection{Excluding a certain feature}
\label{sec:exclude}
Ensemble algorithms based on tree models are commonly used to measure the ``feature importance''. This evaluates the influence of features on the final model accuracy and loss. However, this does not give any information on how the features are related to the final prediction results. To measure the impact of the individual features in the final predictions, we adopt a more direct experiment-based approach, by comparing the performance of the model retrained after excluding a certain feature with the performance of the model including the full set of features.

\begin{figure}
\hspace{-0.4cm}
\includegraphics[scale=0.39]{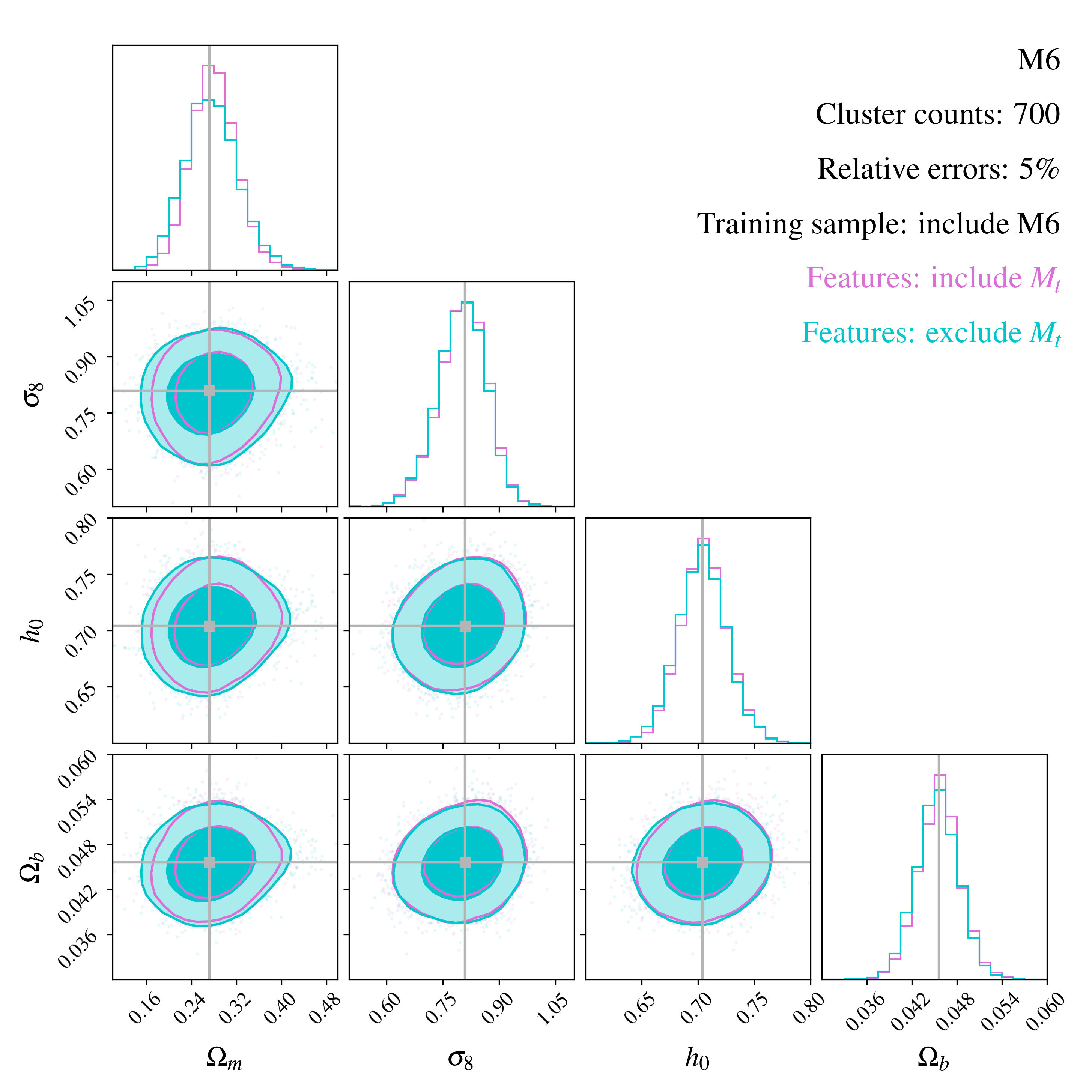}
\caption{Parameter constraints for an M6 {\it mock catalog} obtained by training a model using the training sample that includes (purple) or excludes (cyan) the total mass feature. This graph is the same as Fig. \ref{fig:M6}. In both cases, all cosmological parameters are in the 1$\sigma$ region, indicating that our method could achieve high limiting accuracy for cosmological parameters without using total mass.}
\label{fig:M6 include exclude Mt}
\end{figure}

In Fig. \ref{fig:variance percentage}, we report the variation in the percentage of the mean Accuracy and Logloss over the 5-fold cross-validation process, by excluding each of the features in turn. It is evident, that the ``mass features'' (i.e. $M_*$, $M_g$, and $M_t$) are the ones most affecting the results. 
For example, excluding stellar mass $M_*$ will cause a $35\%$ reduction in mean {Accuracy} and a $33\%$ increase in mean {Logloss}.
Excluding the gas mass, $M_g$, the {Accuracy} is reduced by $20\%$ and the {Logloss} increased by $18\%$, while without the total mass $M_t$, the {Accuracy} is reduced by $14\%$ and the {Logloss} increased $15\%$. On the other hand, excluding the gas luminosity $L_g$ or the gas temperature $T_g$ would not affect the {Accuracy} or {Logloss} by more than 3$\%$. The redshift $z$, the radius $R$, and the velocity dispersion $\sigma_v$, surprisingly rank the lowest with the combined influence on the overall results amounting only to $\sim 1\%$. This is likely because most of the information encoded in these features is also contained in the other features above (e.g. $\sigma_v$ is a proxy of the total mass). However, we need to remark on two facts here. First, this ``feature importance'' analysis is related to the simultaneous constraints of all the cosmological parameters together, while possibly the individual parameters can be more sensitive to a certain feature (e.g. $h_0$ being more sensitive to $M_*$\footnote{Despite stellar ages are not included in the simulation features, it is possible that the assembly of stellar masses in clusters is tightly correlated with the age of the universe, with stars being cosmological clocks (see e.g. \citealt{Jimenez02}).} and $z$). This is a test that is beyond the purposes of the current paper and we will address it in forthcoming analyses.
Second, this ``feature importance'' is related to the classification, which is not related to the ability to constrain the cosmology, as stressed above. Hence, we need to check if the absence of an important feature in classification can yet allow us to recover true cosmology.

\begin{figure}
\hspace{-0.5cm}
\includegraphics[scale=0.39]{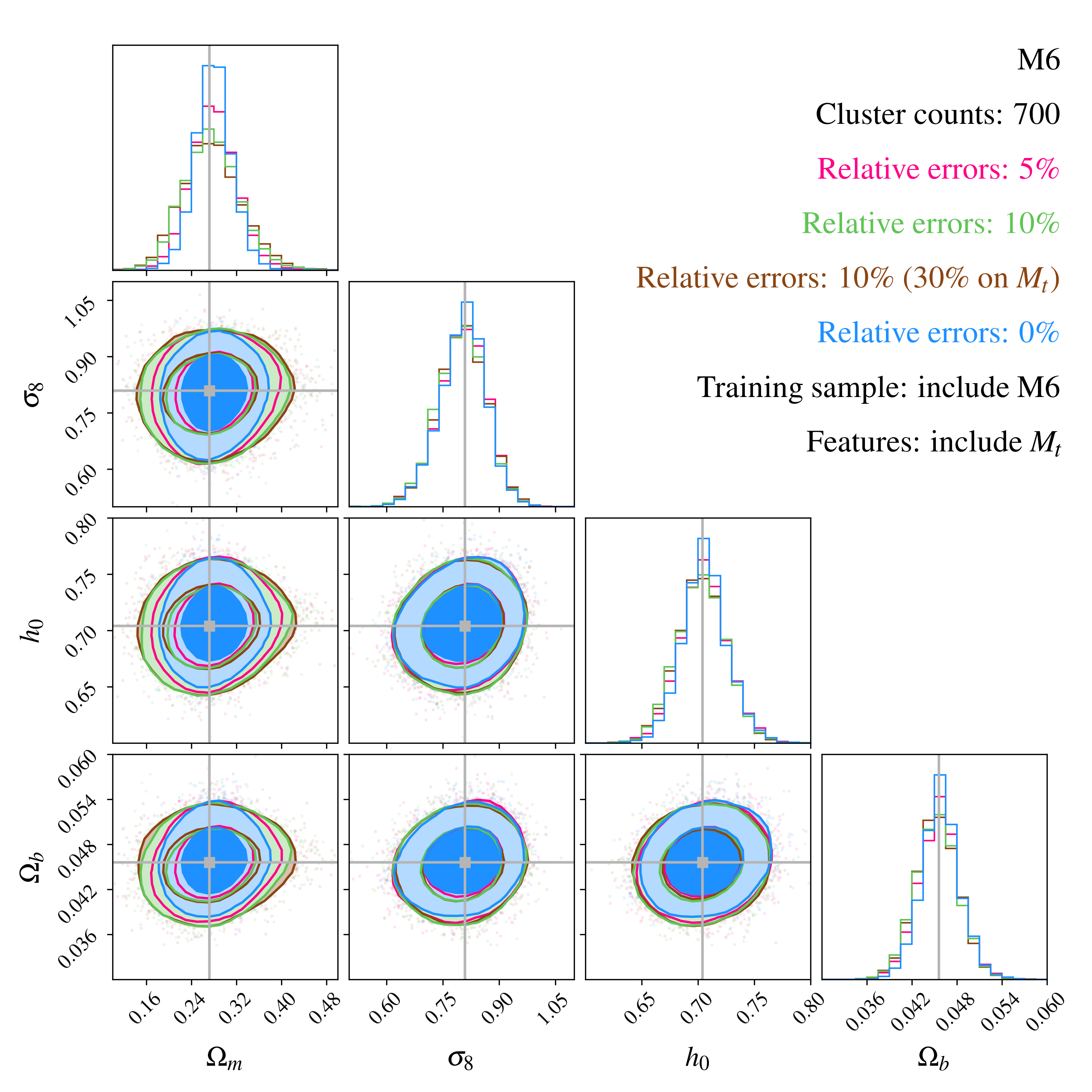}
\caption{The parameter constraining results for an M6 {\it mock catalog} obtained by training a model using the training sample adding 5\% (pink) or 10\% (green) or 0\% (blue) errors on  $M_*$, $M_g$, $M_t$, $L_g$, $T_g$, $R$, $\sigma_v$, and adding 10\% errors on  $M_*$, $M_g$, $L_g$, $T_g$, $R$, $\sigma_v$ while adding 30\% errors on $M_t$ (brown). This graph is the same type as Fig. \ref{fig:M6}. In these 4 cases, all cosmological parameters are in the 1$\sigma$ region, indicating that our method has relatively good robustness to the error degree of features.}
\label{fig:M6 5err 2err}
\end{figure}

In Fig. \ref{fig:M6 include exclude Mt}, as an example, we show the results of excluding total mass $M_t$ from the list of the features used to train and predict the cosmological parameters for M6 (our reference cosmology). The reason to check the impact of the absence of the total mass among the catalog features is that the mass is among the more uncertain quantities to estimate from observations (see Sect. \ref{sec:intro}). In this case, the confidence contours are still quite similar to the case of including $M_t$, except for the $\Omega_m$ contours and posterior probability, which look more broadened.
For M6, again, the predicted values for $(\Omega_m, \sigma_8, h_0, \Omega_b)$ are 
$(0.274_{-0.045}^{+0.048}, 0.802_{-0.065}^{+0.061}, 0.704_{-0.021}^{+0.021}, 0.0454_{-0.0028}^{+0.0028})$,
against the true values of M6, that are 
$(0.272, 0.809, 0.704, 0.0456)$.
This indicates that excluding $M_t$ would somehow affect the accuracy of classification, but produce a limited impact on the parameter constraints, except for the $\Omega_m$ precision. 
This means that the cosmological information about all parameters is still encoded in some other features that are directly accessible in observation (like stellar mass $M_*$ and gas mass $M_g$). Therefore, this experiment shows, specifically, that artificial intelligence can help extract information from multi-wavelength features to infer cosmological parameters even without the total mass.

\subsection{The impact of the measurement errors}
\label{sec:errors}
To take into account the measurement errors of cluster features in the real observation, we added 5$\%$ Gaussian errors to the simulation data. As discussed in Sect. \ref{sec:pre-process}, this was a conservative choice for most of the features, or even optimistic for others (see e.g. the total mass from weak lensing). Hence, we are interested to consider a wider range of statistical errors and check whether, by improving the precision of observations (smaller errors), one obtains tighter constraints on classification and cosmological parameter inferences and vice versa for larger observational errors. The uncertainties on the observed quantities equally impact traditional methods, e.g. the mass function of galaxy clusters, where higher/lower accuracy of cluster features produces more/less accurate cosmological results. 
In Fig. \ref{fig:M6 5err 2err}, we show the confidence intervals for the prediction of the four cosmological parameters where we consider the extreme case of 0\% errors for all features, which provides information on the uncertainties inherent to the ML model. We also consider the pessimistic cases where we assume 10\% errors for all features or 30\% errors for the $M_t$ and 10\% on other observables (see Sect. \ref{sec:pre-process}). These are shown against the reference case with 5$\%$ errors in overall quantities. 
The predictions for the peaks are almost identical in all these cases, implying a rather resilient accuracy, while the confidence contours are slightly shrunken in the 0\% case and expanded in the 10\% case, as expected, for all parameters. However, the 0\% errors allow an improvement in terms of accuracy of $\Omega_m$, by $\sim21\%$, which is reasonably good, but not significant improvements for the other parameters. On the other hand, for the case of 10\% errors, we observe a significant degradation of the $\Omega_m$ precision ($\sim23\%$ larger than the 5\% error case), but, again, no sensible changes for the other parameters, which are recovered with similar precision. Finally, the extreme case of 30\% on the $M_t$ does not show a catastrophic impact on the size of the contours of $\Omega_m$, that increases by $\sim38\%$ with respect to the 5\% error case and by $\sim12\%$ with respect to the 10\% error case. This is possibly due to the fact that the scaling relations, to which $\Omega_m$ is sensitive, are more tightly distributed with respect to the ones the other parameters are sensitive to. Hence larger measurement errors increase the overlap among scaling relations sensitive to $\Omega_m$ more than the ones of the other parameters. 
Finally, we can also argue that the measurement errors of $M_t$ have little effect on the cosmological parameter predictions, because this is a ``less important feature'' than $M_*$ and $M_g$ and the model performs well when $M_*$ and $M_g$ have 10\% errors and $M_t$ is much noisier than other features. 
Interestingly, we find that either including noisy $M_t$ estimates (as just discussed) or excluding $M_t$ from the catalogs (as discussed in Sect. \ref{sec:exclude}), leads to similar results.

\subsection{The impact of the simulation resolution}
In the previous section, we discussed measurement errors as a basic implementation of ``observational realism''. This latter element has larger ramifications than simple measurement errors and it tracks back to the definition of the observational quantities in simulations and how the observational conditions can affect the inferred physical measurements in synthetic datasets (see e.g. \citealt{2019MNRAS.490.5390Bottrell}, \citealt{2021MNRAS.508.3321Tang}). However, there are other profound implications related to the technical aspects of simulations and the way these are calibrated to observations, that might affect the proper training of machine learning tools and impact their application to real data. For instance, one problem is the ``resolution convergence''. It is known that any given property of a simulated halo may not be fully converged at any given mass/spatial resolution \citep{2017MNRAS.465.3291W,2018MNRAS.473.4077P}. 
Due to the different impact of the sub-grid physics (e.g. \citealt{2010ApJ...713..535C-subgrid}), both stellar masses and star formation rates can increase with better resolution for dark matter haloes of a fixed mass. This has been proven, e.g., in TNG simulations$\footnote{\url{https://www.tng-project.org/}}$ \citep{2018MNRAS.475..648P}.

\begin{figure}
\centering
\includegraphics[scale=0.45, trim=0 0.5cm 0 0.2cm]{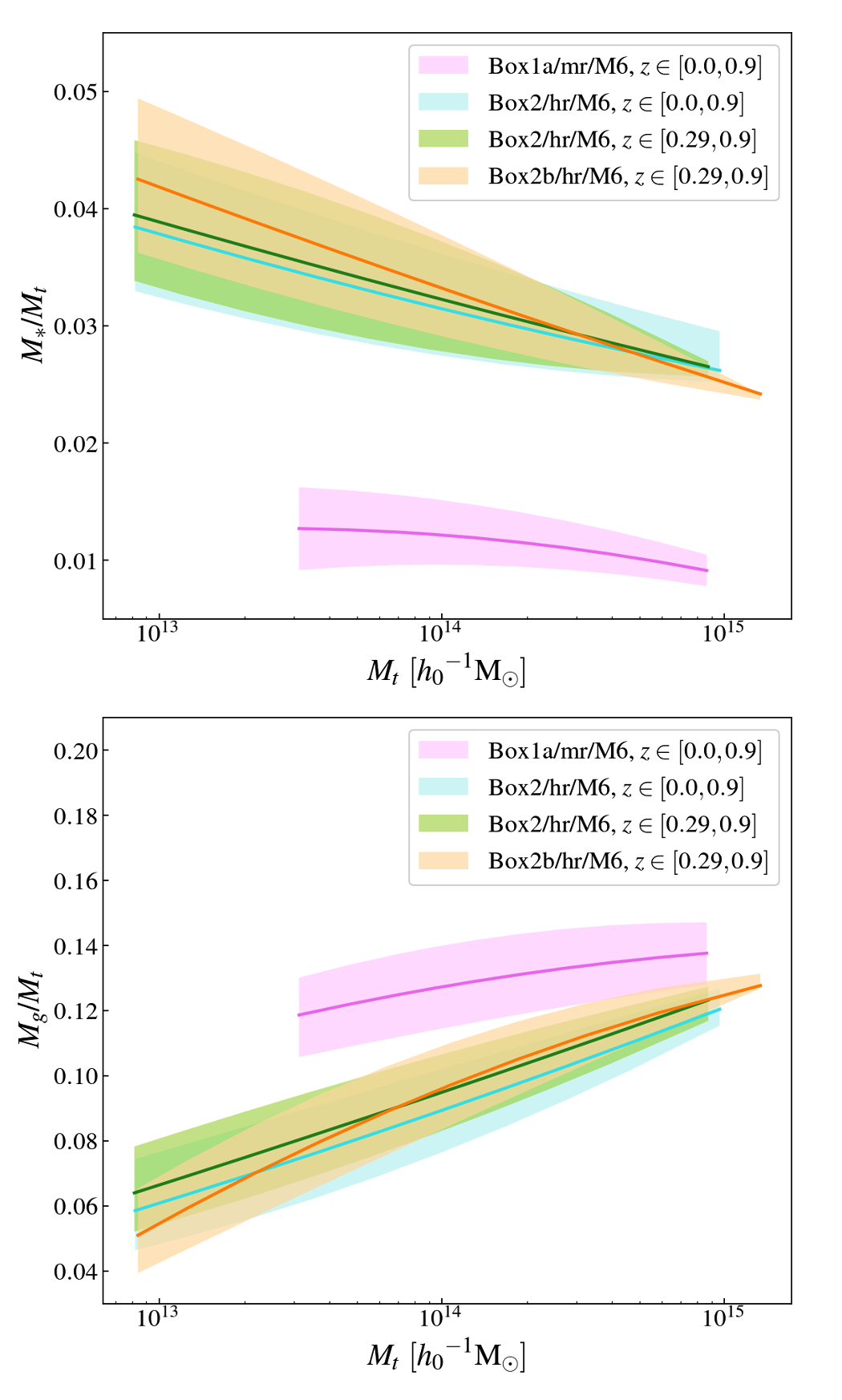}
\caption{The stellar mass ratio $M_*/M_t$ (the upper panel) and gas mass ratio $M_g/M_t$ (the lower panel) as a function of the total mass $M_t$. Different colors represent different simulation boxes and different redshift intervals. Box1a is of medium resolution while Box2 and Box2b are of high resolution.}
\label{fig:stellar and gas ratio}
\end{figure}

To check this effect in Magneticum simulations, we have derived the distribution of high-resolution ({\it hr}) cluster features with respect to the mid-resolution ({\it mr}) simulations.
M1$-$M13 all have {\it mr} simulations in the {\it mr} box, Box1a.
For M6 (the fiducial cosmology considered in Magneticum), additional simulations are available in {\it hr} boxes, for instance, Box2 and Box2b. The sizes of Box1a/mr, Box2/hr and Box2b/hr are $\sim896\ h_{0}^{-1}$Mpc, $\sim352\ h_{0}^{-1}$Mpc and $\sim640\ h_{0}^{-1}$Mpc, respectively. More details of these 3 boxes can be found at the Magneticum website$\footnote{\url{http://magneticum.org/simulations.html}}$.

In Fig. \ref{fig:stellar and gas ratio}, we calculate the stellar mass ratio $M_*/M_t$ (top) and gas mass ratio $M_g/M_t$ (bottom) as the function of total mass $M_t$ both for the M6 medium resolution simulations (Box1a/mr) and two high-resolution simulation boxes (Box2/hr, Box2b/hr). We stress here, in particular, that the Box2/hr/M6 simulation not only shares the same cosmology and feedback but also covers the same redshift interval of Box1a/mr/M6, while the Box2b/hr/M6 covers a higher redshift range ($z\geq0.29$). As expected, the stellar mass ratios and gas mass ratios are quite sensitive to the resolution levels. 
The higher the resolution, the smaller the stellar mass and gas mass at a fixed total mass.
Statistically, for clusters with masses between $2\times 10^{13}$ $h^{-1}\mathrm{M_{\odot}}$
and $10^{15}$ $h^{-1}\mathrm{M_{\odot}}$ in M6 cosmology, stellar mass averages 3$\%$ of the total mass at high resolution, while at a medium resolution, this percentage decreases to 1.2$\%$. On the other hand, the gas mass ratio seems to rise with decreasing resolution with about 10$\%$ at high resolution and 13$\%$ at medium resolution. This is consistent with what has been found in TNG simulations for haloes with total mass $\log M_t/M_\odot>14$ (\citealt{2018MNRAS.475..648P}). 
We also observe that different volumes (Box2/hr and Box2b/hr), show a sensitive tilt. To check if this is due to the lack of low-redshift data for Box2b (which is limited to $z\geq0.29$) or to cosmic variance, in Fig. \ref{fig:stellar and gas ratio} we also add the stellar and gas mass fractions for Box2/hr for redshifts $z\geq0.29$ only, consistently with Box2b/hr. As we can see this latter is slightly offset with respect to the case including clusters down to $z=0$, hence we conclude that the tilt possibly comes from cosmic variance. We notice though that the larger variance comes from $M_t<10^{14} h^{-1}M_\odot$ and is of the order of $1\%$.

\begin{figure}
\centering
\includegraphics[scale=0.28, trim=0.5cm 1cm 0 0.5cm]{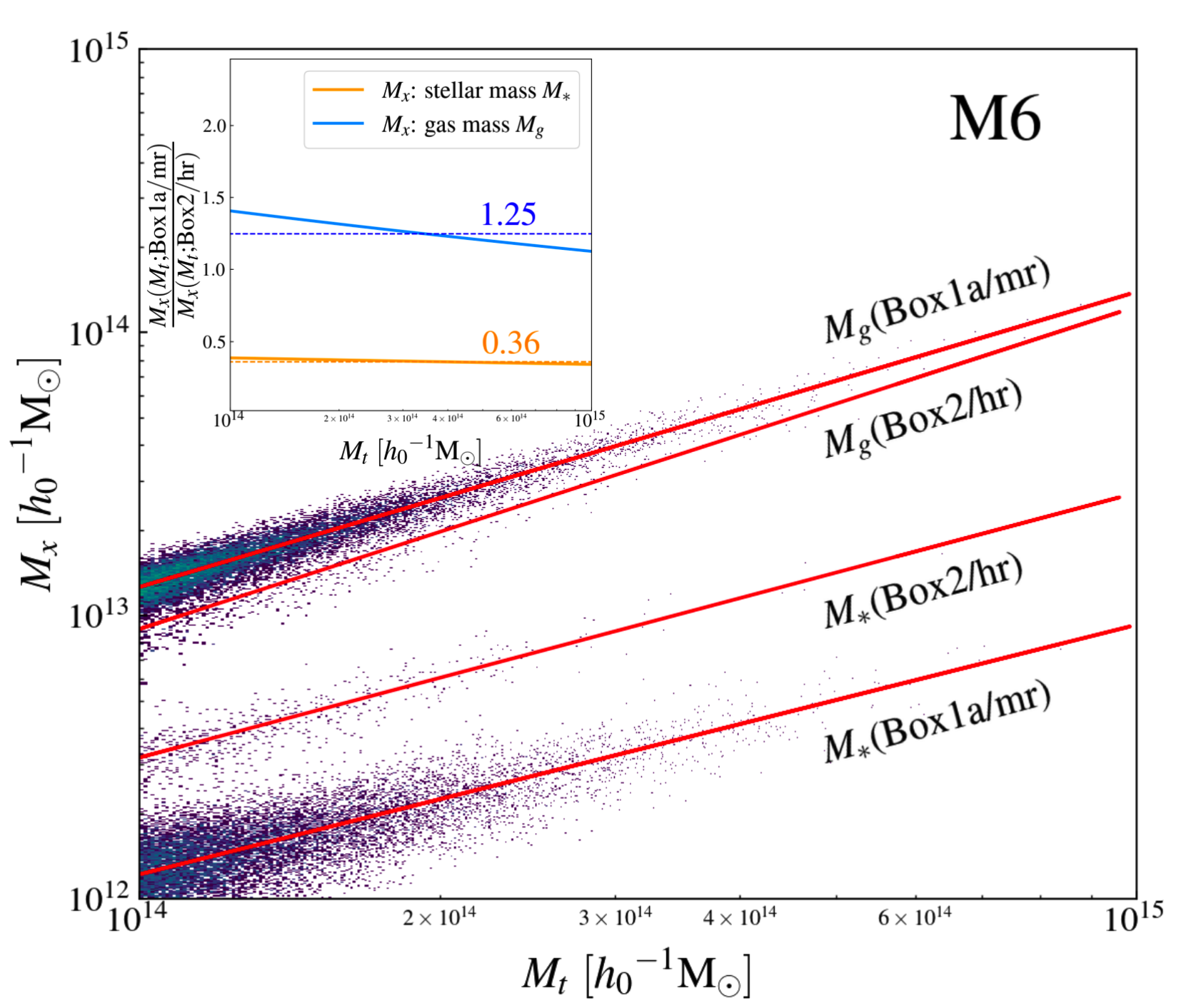}
\caption{$M_g \sim M_t$ and $M_g \sim M_t$ relationships in medium-resolution simulation (Box1a/mr) and high-resolution simulation (Box2/hr). The subplot in the upper left corner represents the conversion coefficients of stellar mass (blue) and gas mass (orange) between medium-resolution and high-resolution for a fixed total mass, corresponding to Eq. \ref{equ:conversion}.}
\label{fig:M6 mr hr mapping}
\end{figure}

In general, other features in {\it hr}, such as gas luminosity and temperature, also show deviations from those in {\it mr}. This raises the question of which resolution should be taken as the best representation of reality. 
This is certainly a question we will need to address when applying the MLCCA to real data, as we will need to ensure that the algorithm is trained over simulations for which the calibration of the relevant scaling relations and resolution do conspire to match observations. 
We anticipate here that this is not a simple task as observations do not provide an obvious indication about the ``ground truth'', having clusters a stellar mass fraction varying from 0.5\% to 3\% (see e.g. \citealt{2018MNRAS.478.3072C}), i.e. a scatter well beyond either the {\it mr} or {\it hr} relations in Fig. \ref{fig:stellar and gas ratio}. 
The obvious warning emerging from the question above is that we need to keep the subgrid-physics under control in simulations to produce predictions, given a baryon physics recipe, resolution-independent (see e.g. \citealt{2015MNRAS.447..178Murante}).
However, in the perspective of our proof-of-concept experiment, this yet important ``realism'' aspect is irrelevant as long as the training and the test sample are extracted from the same knowledge base provided by the same simulations with the same stellar mass or gas mass fraction. While it becomes relevant if one needs to train on a simulation with a resolution different from the one from which the test sample is extracted.
In this case, one can use a ``rescaling procedure'' (see e.g. \citealt{2018MNRAS.475..648P}) by applying a resolution correction factor to align the physical quantities from different resolution boxes. Of course, this is a workaround needed in order to compensate resolution effect and make the simulation predictions consistent at all resolution levels. From the point of view of this work, there is no particular reason why one wants to mix simulations of different resolutions, however, it can still be useful to check if the naif ``rescaling procedure'' makes the MLCCA predictions insensitive to the resolution correction.  

Indeed, according to the ``independent identically distribution'' hypothesis in machine learning inferences, any model can have reliable predictions only when the feature distributions of the test sample are comparable with those of the training sample. Hence, if we use a test sample from {\it hr} simulations, we expect the MLCCA trained on {\it mr} to fail, because the net effect of the resolution is to scale up/down the $M_*-M_t$ and the $M_g-M_t$ relations, similarly to what the different cosmology do at a fixed resolution (see Fig. \ref{fig:feature_corner}). This is a general problem that we would also face using real data where, in the case of the nonuniform definition of the observed quantities, the real features and the training feature can have deviations even if they come from the same cosmology, as the {\it hr} and {\it mr} mock catalogs do as shown in Fig. \ref{fig:stellar and gas ratio}.

\begin{figure}
\hspace{-0.5cm}
\includegraphics[scale=0.39,trim=0 0 0 0.5cm]{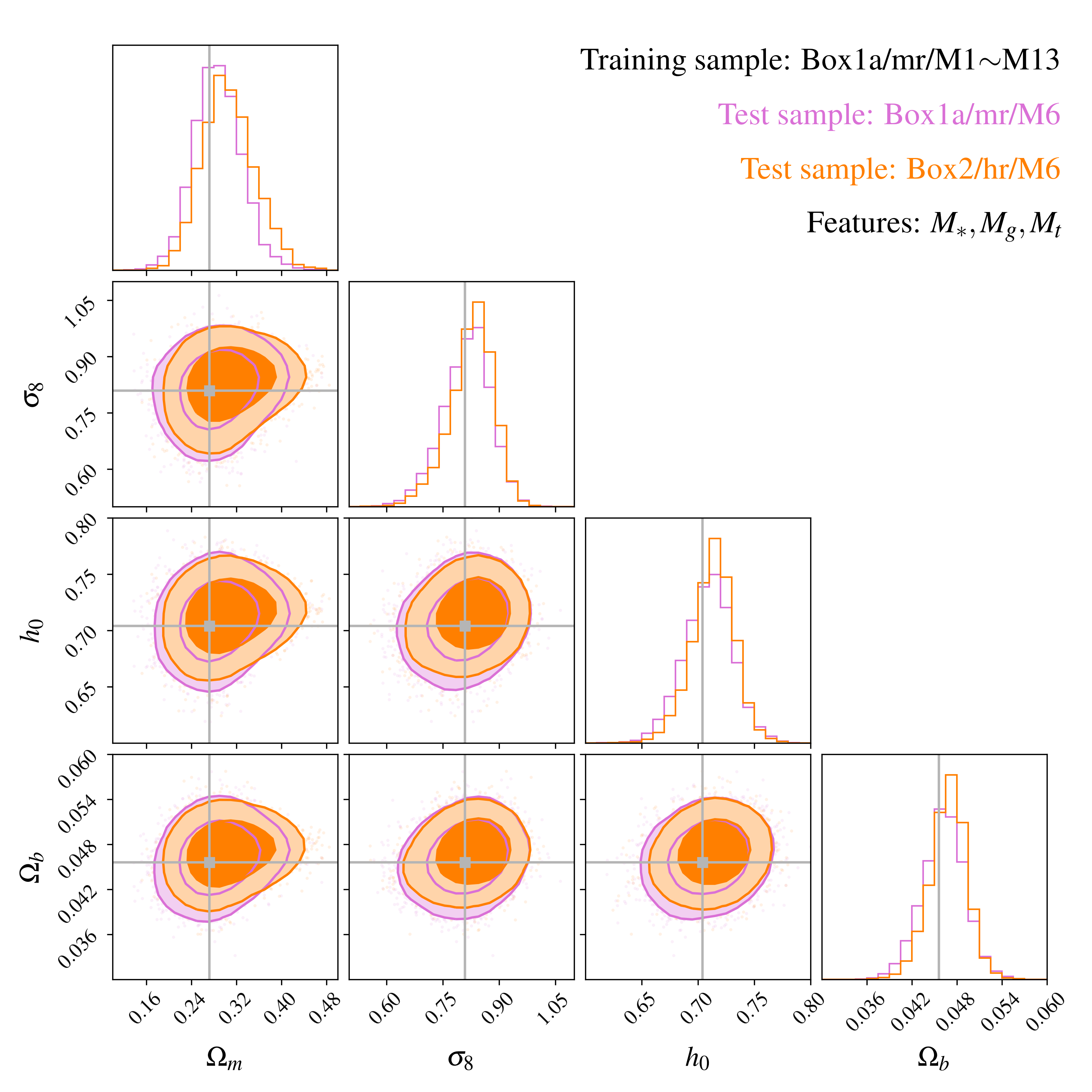}
\caption{Parameter constraints for a medium-resolution M6 {\it mock catalog} (Box1a/mr/M6, purple) and a high-resolution M6 {\it mock catalog} (Box2/hr/M6, orange) obtained by training a model using the medium-resolution training sample. This graph is the same type as Fig. \ref{fig:M6}. In both cases, we guarantee that the total mass of the cluster ranges from $10^{14}$ to $10^{15}$ $h^{-1}\mathrm{M_{\odot}}$. It can be seen that the high-resolution prediction values are higher than the real values, but all cosmological parameters are still in the 1$\sigma$ region, indicating that our method has certain application potential for different resolution cosmology.}
\label{fig:M6 mr hr}
\end{figure}

Following \citet{2018MNRAS.475..648P}, we adopt a heuristic correction to convert {\it hr} cluster features into their {\it mr} versions that approximately reproduce the {\it mr} training sample. 
First, from both Box1a/mr/M6 and Box2/hr/M6, we select clusters with the total mass within $10^{14}\sim10^{15}$ $h^{-1}\mathrm{M_{\odot}}$ to mitigate the effects of resolution on a too wide mass range and assume a constant correction. Second, despite different Box2/hr/M6 features explicitly varying from those of Box1a/mr/M6, we only adjust two of the most important features ($M_*$ and $M_g$ as from Fig. \ref{fig:variance percentage}), conservatively.
In Fig. \ref{fig:M6 mr hr mapping} we can see how the correlation of these two quantities changes as a function of $M_t$ in the different boxes/resolutions. We can adopt a mass-modulated conversion strategy to derive a conversion coefficient that can reflect the resolution-induced feature drift. For brevity, we assume that the conversion coefficient is the average of multiples of $M_x$/mr and $M_x$/hr obtained at each fixed $M_t$. Accordingly, for clusters whose $M_t$ within $10^{14}\sim10^{15}$ $h^{-1}\mathrm{M_{\odot}}$, their hypothetical {\it mr} versions of $M_x$ ($M_*$ or $M_g$) can be approximately obtained by multiplying the conversion coefficient $\alpha$ and their original {\it hr} versions as follows.
\begin{align}
\label{equ:conversion}
M_x(M_t;\mathrm{Box2/mr}) &\approx \left\langle \frac{M_x(\mathrm{M_t;Box1a/mr})}{M_x(\mathrm{M_t;Box2/hr})} \right\rangle \cdot M_x(M_t;\mathrm{Box2/hr}) \notag \\
&\approx\hspace{0.1cm}\alpha \cdot M_x(M_t;\mathrm{Box2/hr}).
\end{align}
From Fig. \ref{fig:M6 mr hr mapping}, we find that the best fit $\alpha$ is 0.36 and 1.25 for $M_*$ and $M_g$, respectively. We further apply these two coefficients to derive hypothetical Box2/mr clusters and have checked that their 3 features ($M_*$, $M_g$ $\&$ $M_t$) finally use these features to make the cosmological parameter predictions.

Fig. \ref{fig:M6 mr hr} shows parameter constraints from $M_*$, $M_g$, and $M_t$ for Box1a/mr and Box2/hr (converted to Box2/mr version). Compared to the true cosmological configuration of M6 ($\Omega_m:0.272,\sigma_8:0.809,h_0:0.704,\Omega_b:0.0456$), the predictions for Box1a/mr and Box2/hr are:
\begin{itemize}
\item[]$(0.286_{-0.040}^{+0.043}, 0.818_{-0.068}^{+0.056}, 0.710_{-0.023}^{+0.020}, 0.0463_{-0.0029}^{+0.0027})$,
\item[]$(0.303_{-0.043}^{+0.050}, 0.832_{-0.063}^{+0.050}, 0.714_{-0.019}^{+0.017}, 0.0470_{-0.0025}^{+0.0022})$,
\end{itemize}
\noindent respectively, i.e. yet consistent within the errors. 

We find the overall predictions made over Box2/hr are similar to that of Box1a/mr, especially for $h_0$ and $\Omega_b$, which demonstrates that our conversion strategy maintains most of the inner-correlations among the three mass quantities ($M_*$, $M_g$, $M_t$). However, MLCCA overestimates all parameters of both boxes, especially the $\Omega_m$ and $\sigma_8$. This residual discrepancy might come from the fact that changing $M_*$ and $M_g$, without changing the total mass, substantially alters the baryon fraction of the sample and, intrinsically, the underlying cosmology of the cluster catalog. 
This test shows that we cannot straightforwardly generalize the results, obtained from mid-resolution to high-resolution, as this would imply corrections on the features that might introduce biases in the predicted cosmology. This suggests that to avoid systematics, one should train the algorithm using features from numerically converged simulations.

\section{Summary and Conclusions}
\label{sec:conclusion}
In this paper, we have introduced and tested a first proof-of-concept machine learning pipeline which is able to predict the cosmological parameters starting from mock catalogs of galaxy clusters' physical parameters, namely the stellar mass, $M_*$, gas mass, $M_g$, total mass, $M_t$, gas luminosity, $L_g$, and temperature, $T_g$, sizes, $R_{500c}$, velocity dispersion, $\sigma_v$, and redshift, $z$. These are typical observables (or features) we expect to collect from current and future imaging surveys in optical and NIR (e.g. Rubin/LSST, CSST, and {\it Euclid}), spectroscopical surveys (e.g. DESI and 4MOST), and X-ray surveys (e.g. eROSITA). We have used the mock catalogs of galaxy clusters extracted from the multi-cosmology set of Magneticum hydrodynamical simulations, which spans a limited volume in the ($\Omega_m,\sigma_8,h_0,\Omega_b$) parameter space, centered around the WMAP7 cosmology. There are 15 different simulations available, which also include some variations of the feedback recipe from AGN and supernovae. We used only 13 of them, excluding 2 cosmologies with too few clusters to use as training samples, and also skipped the inclusion of multi-feedback for this first test, as there were only 4 simulations with 2 feedback recipes available. Again, these are too few to be used for a meaningful test.
The mock catalogs, including measurement error, are used to train an optimized Light Gradient Boosting machine (LGB) network to classify the cluster catalogs and predict the cosmological parameters. Based on this optimized LGB network, we have built a Machine Learning Cluster Cosmology Pipeline (MLCCA). 
The MLCCA has proven to be very effective in predicting the right set of cosmological parameters although the classification of the individual clusters to belong to the right cosmology suffers from the similarity of the scaling relations of close cosmologies. 
Here below, we summarize the main results of the application of the MLCCA to mock catalogs of 700 clusters from different cosmologies:
\begin{enumerate}
    \item The MLCCA can accurately predict the true cosmological parameters corresponding to the cosmological simulation the catalogs are drawn from. Despite the limited coverage in the parameter space, for cosmological models in the center of the parameter space, the classification recall rate is between $\sim0.2$ and $0.4$, but the predictions of the cosmological parameters are tighter. Typical 1-$\sigma$ level are 14\% for $\Omega_m$, 8\% for $\sigma_8$, 3\% for $h_0$, 6\% for $\Omega_b$. For cosmological models at the edge of parametric space, the classification accuracy increases because there is not any confusion with cluster properties from close models, but the cosmological parameters are slightly biased. This is clearly a ``border effect'' due to the training sample, rather than the true under-performance of the MLCCA. This leads us to conclude that more mid-resolution hydro-simulation Magneticum-like are needed to make the MLCCA effectively applicable to real data. 
    
    \item In order to fully check the performance of the MLCCA and, in particular, the ability to extrapolate to cosmologies that are not included in the training sample (this is a situation that might happen also if one uses a regular grid of cosmologies), we have tested the ability to recover the cosmology over a mock catalog taken from a cosmology (specifically we tried M6 and M7) that was not included in the training set and found that the MLCCA can recover the cosmological parameter with comparable accuracy and precision as the case where the training contains the mock catalog cosmology.
    
    \item We have tested the impact of the measurement errors, particularly how the recall rate of the classifier and the uncertainties on the cosmological parameters would be affected. We have found that for errors of the order of 2\%, the 1-$\sigma$ contours are shrunk by $\sim18\%$, while for larger errors, i.e. 10\%, only $\Omega_m$ show large degradation of the precision with typical 1-$\sigma$ contours widened by up to $\sim20-40$\%. Note that the current accuracy can be strongly affected by two main factors: 1) the limited size of the training sample, and 2) the limited number of the {it mock catalog} sizes, which we need to check with larger volumes of multi-cosmology simulations.
    
    \item We have tested the resilience of the MLCCA for missing features, i.e. in case cluster catalogs do not contain one or more of the observations used for the main experiment as at point 1) above. Also in this case, the MLCCA can correctly recover the cosmological parameter even if the ``mass features'' are missing, despite the fact that these are the most important features for the classification. We have understood that by the ability of the ML tool to still extract relevant cosmological information from the scaling relations involving all other features.  Among all features, stellar mass and gas mass have the greatest weight on accuracy for the classification.

    \item Finally, we have checked the effect of simulation resolution, as this latter produces a sensitive impact on the stellar and gas mass of clusters, due to the different effects of the sub-grid physics \citep{2018MNRAS.475..648P}. In particular, we have tested whether simple ``rescaling'' of the major cluster features can leave the predictions of the MLCCA unaltered and found that if one limits to only the major baryonic mass features (stars and gas) without also re-correcting the total mass, one ends with systematic effects. This calls for effective strategies to improve the sub-grid physics treatment in hydro-dynamical simulations to make their predictions more stable toward the change of resolution.
\end{enumerate}

This first application of cosmological inference from machine learning based on galaxy clusters shows that these tools have a rather strong predictive power, by efficiently cross-correlating features among different cosmological predictions. This is very promising for future applications making use of finer sampling of the cosmological and galaxy formation parameter space in future multi-cosmology hydro-dynamical simulation runs. And in the long term, this could help to fully exploit multi-wavelength observations from current and future surveys, to gain a more profound understanding of the true universe model.

This work follows a line of experiments trying to extract cosmological information from observational data using machine learning tools applied to multi-cosmology simulations. \cite{2022ApJ...929..132V} use the internal properties of a single galaxy simulated by the CAMELS project $\footnote{\url{https://www.camel-simulations.org/}}$ to predict cosmological parameters, especially $\Omega_m$ and $\sigma_8$. Their machine learning model could infer the value of $\Omega_m$ with a precision of $\delta\Omega_m / \Omega_m \simeq 10\% - 15\%$ (with a possible explanation that $\Omega_m$ could affect the dark matter content of galaxies and then further result in a unique change in the observables' manifold). However, they could not infer $\sigma_8$ due to the small non-linear scale of galaxies. 
Further works by CAMELS include quantifying the robustness of the ML model by testing on galaxies from different codes \citep{2023arXiv230406084E}, improving the inference on cosmological parameters by enlarging the simulation sets \citep{2023arXiv230402096N}, etc.

In our work, we show that galaxy clusters are very powerful in inferring cosmological parameters, mainly because of the stronger connection with large-scale structure formation, which is more sensitive to cosmology. Among the cluster features that we use, the underlying halo mass function has been widely proved to constrain $\Omega_m$ and $\sigma_8$, the gas mass (and baryonic mass in general) has been proved to sensitively depend on $\Omega_b$, while the stellar mass, velocity dispersion, and gas temperature have been proved to sensitively depend on $h_0$. In the next analyses, we expect to apply the MLCCA to upcoming sets of mid-resolution, large-volume hydrodynamical simulations, considering a wider range of cosmologies and, for each of them, different feedback recipes, to finally test the predictions of the cosmological parameters and baryonic physics at the same time. This will eventually allow us to move toward the first application to real data.

\begin{acknowledgements}
      NRN acknowledges financial support from the Research Fund for International Scholars of the National Science Foundation of China (NSFC), grant n. 12150710511 and from China Manned Space project n. CMS-CSST-2021-A01. SB acknowledges support from Fondazione ICSC - Spoke 3 Astrophysics and Cosmos Observations - National Recovery and Resilience Plan (PNRR) Project ID CN\_00000013 "Italian Research Center on High-Performance Computing, Big Data and Quantum Computing"  funded by MUR Missione 4 Componente 2 Investimento 1.4: "Potenziamento strutture di ricerca e creazione di "campioni nazionali di R\&S (M4C2-19 )" - Next Generation EU (NGEU) and INFN InDark Grant. XL acknowledges the science research grants from the China Manned Space Project with No. CMS-CSST-2021-A03, No. CMS-CSST-2021-B01. WL acknowledges financial support from NSFC, grant n. 12073089. KD acknowledges support by the COMPLEX project from the European Research Council (ERC) under the European Union’s Horizon 2020 research and innovation program grant agreement ERC-2019-AdG 882679 as well as support by the Deutsche Forschungsgemeinschaft (DFG, German Research Foundation) under Germany’s Excellence Strategy - EXC-2094 - 390783311. The calculations for the hydrodynamical simulations were carried out at the Leibniz Supercomputer Center (LRZ) under project pr83li (Magneticum).
\end{acknowledgements}

%
%

\bibliography{ref}

\begin{thebibliography}{108}
\expandafter\ifx\csname natexlab\endcsname\relax\def\natexlab#1{#1}\fi

\bibitem[{{Abbott} {et~al.}(2018){Abbott}, {Abdalla}, {Alarcon}, {Aleksi{\'c}}, {Allam}, {Allen}, {Amara}, {Annis}, {Asorey}, {Avila}, {Bacon}, {Balbinot}, {Banerji}, {Banik}, {Barkhouse}, {Baumer}, {Baxter}, {Bechtol}, {Becker}, {Benoit-L{\'e}vy}, {Benson}, {Bernstein}, {Bertin}, {Blazek}, {Bridle}, {Brooks}, {Brout}, {Buckley-Geer}, {Burke}, {Busha}, {Campos}, {Capozzi}, {Carnero Rosell}, {Carrasco Kind}, {Carretero}, {Castander}, {Cawthon}, {Chang}, {Chen}, {Childress}, {Choi}, {Conselice}, {Crittenden}, {Crocce}, {Cunha}, {D'Andrea}, {da Costa}, {Das}, {Davis}, {Davis}, {De Vicente}, {DePoy}, {DeRose}, {Desai}, {Diehl}, {Dietrich}, {Dodelson}, {Doel}, {Drlica-Wagner}, {Eifler}, {Elliott}, {Elsner}, {Elvin-Poole}, {Estrada}, {Evrard}, {Fang}, {Fernandez}, {Fert{\'e}}, {Finley}, {Flaugher}, {Fosalba}, {Friedrich}, {Frieman}, {Garc{\'\i}a-Bellido}, {Garcia-Fernandez}, {Gatti}, {Gaztanaga}, {Gerdes}, {Giannantonio}, {Gill}, {Glazebrook}, {Goldstein}, {Gruen}, {Gruendl}, {Gschwend}, {Gutierrez}, {Hamilton},
  {Hartley}, {Hinton}, {Honscheid}, {Hoyle}, {Huterer}, {Jain}, {James}, {Jarvis}, {Jeltema}, {Johnson}, {Johnson}, {Kacprzak}, {Kent}, {Kim}, {King}, {Kirk}, {Kokron}, {Kovacs}, {Krause}, {Krawiec}, {Kremin}, {Kuehn}, {Kuhlmann}, {Kuropatkin}, {Lacasa}, {Lahav}, {Li}, {Liddle}, {Lidman}, {Lima}, {Lin}, {MacCrann}, {Maia}, {Makler}, {Manera}, {March}, {Marshall}, {Martini}, {McMahon}, {Melchior}, {Menanteau}, {Miquel}, {Miranda}, {Mudd}, {Muir}, {M{\"o}ller}, {Neilsen}, {Nichol}, {Nord}, {Nugent}, {Ogando}, {Palmese}, {Peacock}, {Peiris}, {Peoples}, {Percival}, {Petravick}, {Plazas}, {Porredon}, {Prat}, {Pujol}, {Rau}, {Refregier}, {Ricker}, {Roe}, {Rollins}, {Romer}, {Roodman}, {Rosenfeld}, {Ross}, {Rozo}, {Rykoff}, {Sako}, {Salvador}, {Samuroff}, {S{\'a}nchez}, {Sanchez}, {Santiago}, {Scarpine}, {Schindler}, {Scolnic}, {Secco}, {Serrano}, {Sevilla-Noarbe}, {Sheldon}, {Smith}, {Smith}, {Smith}, {Soares-Santos}, {Sobreira}, {Suchyta}, {Tarle}, {Thomas}, {Troxel}, {Tucker}, {Tucker}, {Uddin}, {Varga},
  {Vielzeuf}, {Vikram}, {Vivas}, {Walker}, {Wang}, {Wechsler}, {Weller}, {Wester}, {Wolf}, {Yanny}, {Yuan}, {Zenteno}, {Zhang}, {Zhang}, {Zuntz}, \& {Dark Energy Survey Collaboration}}]{2018PhRvD..98d3526A}
{Abbott}, T.~M.~C., {Abdalla}, F.~B., {Alarcon}, A., {et~al.} 2018, \prd, 98, 043526

\bibitem[{{Abbott} {et~al.}(2020){Abbott}, {Aguena}, {Alarcon}, {Allam}, {Allen}, {Annis}, {Avila}, {Bacon}, {Bechtol}, {Bermeo}, {Bernstein}, {Bertin}, {Bhargava}, {Bocquet}, {Brooks}, {Brout}, {Buckley-Geer}, {Burke}, {Carnero Rosell}, {Carrasco Kind}, {Carretero}, {Castander}, {Cawthon}, {Chang}, {Chen}, {Choi}, {Costanzi}, {Crocce}, {da Costa}, {Davis}, {De Vicente}, {DeRose}, {Desai}, {Diehl}, {Dietrich}, {Dodelson}, {Doel}, {Drlica-Wagner}, {Eckert}, {Eifler}, {Elvin-Poole}, {Estrada}, {Everett}, {Evrard}, {Farahi}, {Ferrero}, {Flaugher}, {Fosalba}, {Frieman}, {Garc{\'\i}a-Bellido}, {Gatti}, {Gaztanaga}, {Gerdes}, {Giannantonio}, {Giles}, {Grandis}, {Gruen}, {Gruendl}, {Gschwend}, {Gutierrez}, {Hartley}, {Hinton}, {Hollowood}, {Honscheid}, {Hoyle}, {Huterer}, {James}, {Jarvis}, {Jeltema}, {Johnson}, {Johnson}, {Kent}, {Krause}, {Kron}, {Kuehn}, {Kuropatkin}, {Lahav}, {Li}, {Lidman}, {Lima}, {Lin}, {MacCrann}, {Maia}, {Mantz}, {Marshall}, {Martini}, {Mayers}, {Melchior}, {Mena-Fern{\'a}ndez},
  {Menanteau}, {Miquel}, {Mohr}, {Nichol}, {Nord}, {Ogando}, {Palmese}, {Paz-Chinch{\'o}n}, {Plazas}, {Prat}, {Rau}, {Romer}, {Roodman}, {Rooney}, {Rozo}, {Rykoff}, {Sako}, {Samuroff}, {S{\'a}nchez}, {Sanchez}, {Saro}, {Scarpine}, {Schubnell}, {Scolnic}, {Serrano}, {Sevilla-Noarbe}, {Sheldon}, {Smith}, {Smith}, {Suchyta}, {Swanson}, {Tarle}, {Thomas}, {To}, {Troxel}, {Tucker}, {Varga}, {von der Linden}, {Walker}, {Wechsler}, {Weller}, {Wilkinson}, {Wu}, {Yanny}, {Zhang}, {Zhang}, {Zuntz}, \& {DES Collaboration}}]{2020PhRvD.102b3509A}
{Abbott}, T.~M.~C., {Aguena}, M., {Alarcon}, A., {et~al.} 2020, \prd, 102, 023509

\bibitem[{{Abdullah} {et~al.}(2020){Abdullah}, {Klypin}, \& {Wilson}}]{2020ApJ...901...90A}
{Abdullah}, M.~H., {Klypin}, A., \& {Wilson}, G. 2020, \apj, 901, 90

\bibitem[{{Adami} {et~al.}(2018){Adami}, {Giles}, {Koulouridis}, {Pacaud}, {Caretta}, {Pierre}, {Eckert}, {Ramos-Ceja}, {Gastaldello}, {Fotopoulou}, {Guglielmo}, {Lidman}, {Sadibekova}, {Iovino}, {Maughan}, {Chiappetti}, {Alis}, {Altieri}, {Baldry}, {Bottini}, {Birkinshaw}, {Bremer}, {Brown}, {Cucciati}, {Driver}, {Elmer}, {Ettori}, {Evrard}, {Faccioli}, {Granett}, {Grootes}, {Guzzo}, {Hopkins}, {Horellou}, {Lef{\`e}vre}, {Liske}, {Malek}, {Marulli}, {Maurogordato}, {Owers}, {Paltani}, {Poggianti}, {Polletta}, {Plionis}, {Pollo}, {Pompei}, {Ponman}, {Rapetti}, {Ricci}, {Robotham}, {Tuffs}, {Tasca}, {Valtchanov}, {Vergani}, {Wagner}, {Willis}, \& {XXL Consortium}}]{2018A&A...620A...5A}
{Adami}, C., {Giles}, P., {Koulouridis}, E., {et~al.} 2018, \aap, 620, A5

\bibitem[{Akiba {et~al.}(2019)Akiba, Sano, Yanase, Ohta, \& Koyama}]{2019Optuna}
Akiba, T., Sano, S., Yanase, T., Ohta, T., \& Koyama, M. 2019, ACM

\bibitem[{{Allen} {et~al.}(2011){Allen}, {Evrard}, \& {Mantz}}]{2011ARA&A..49..409A}
{Allen}, S.~W., {Evrard}, A.~E., \& {Mantz}, A.~B. 2011, \araa, 49, 409

\bibitem[{{Armitage} {et~al.}(2019){Armitage}, {Kay}, \& {Barnes}}]{2019MNRAS.484.1526A}
{Armitage}, T.~J., {Kay}, S.~T., \& {Barnes}, D.~J. 2019, \mnras, 484, 1526

\bibitem[{{Bahar} {et~al.}(2022){Bahar}, {Bulbul}, {Clerc}, {Ghirardini}, {Liu}, {Nandra}, {Pacaud}, {Chiu}, {Comparat}, {Ider-Chitham}, {Klein}, {Liu}, {Merloni}, {Migkas}, {Okabe}, {Ramos-Ceja}, {Reiprich}, {Sanders}, \& {Schrabback}}]{2022A&A...661A...7B}
{Bahar}, Y.~E., {Bulbul}, E., {Clerc}, N., {et~al.} 2022, \aap, 661, A7

\bibitem[{{Beck} {et~al.}(2016){Beck}, {Murante}, {Arth}, {Remus}, {Teklu}, {Donnert}, {Planelles}, {Beck}, {F{\"o}rster}, {Imgrund}, {Dolag}, \& {Borgani}}]{2016MNRAS.455.2110B}
{Beck}, A.~M., {Murante}, G., {Arth}, A., {et~al.} 2016, \mnras, 455, 2110

\bibitem[{{Biviano} {et~al.}(2013){Biviano}, {Rosati}, {Balestra}, {Mercurio}, {Girardi}, {Nonino}, {Grillo}, {Scodeggio}, {Lemze}, {Kelson}, {Umetsu}, {Postman}, {Zitrin}, {Czoske}, {Ettori}, {Fritz}, {Lombardi}, {Maier}, {Medezinski}, {Mei}, {Presotto}, {Strazzullo}, {Tozzi}, {Ziegler}, {Annunziatella}, {Bartelmann}, {Benitez}, {Bradley}, {Brescia}, {Broadhurst}, {Coe}, {Demarco}, {Donahue}, {Ford}, {Gobat}, {Graves}, {Koekemoer}, {Kuchner}, {Melchior}, {Meneghetti}, {Merten}, {Moustakas}, {Munari}, {Reg{\H{o}}s}, {Sartoris}, {Seitz}, \& {Zheng}}]{2013A&A...558A...1Biviano13}
{Biviano}, A., {Rosati}, P., {Balestra}, I., {et~al.} 2013, \aap, 558, A1

\bibitem[{{Bleem} {et~al.}(2015){Bleem}, {Stalder}, {de Haan}, {Aird}, {Allen}, {Applegate}, {Ashby}, {Bautz}, {Bayliss}, {Benson}, {Bocquet}, {Brodwin}, {Carlstrom}, {Chang}, {Chiu}, {Cho}, {Clocchiatti}, {Crawford}, {Crites}, {Desai}, {Dietrich}, {Dobbs}, {Foley}, {Forman}, {George}, {Gladders}, {Gonzalez}, {Halverson}, {Hennig}, {Hoekstra}, {Holder}, {Holzapfel}, {Hrubes}, {Jones}, {Keisler}, {Knox}, {Lee}, {Leitch}, {Liu}, {Lueker}, {Luong-Van}, {Mantz}, {Marrone}, {McDonald}, {McMahon}, {Meyer}, {Mocanu}, {Mohr}, {Murray}, {Padin}, {Pryke}, {Reichardt}, {Rest}, {Ruel}, {Ruhl}, {Saliwanchik}, {Saro}, {Sayre}, {Schaffer}, {Schrabback}, {Shirokoff}, {Song}, {Spieler}, {Stanford}, {Staniszewski}, {Stark}, {Story}, {Stubbs}, {Vanderlinde}, {Vieira}, {Vikhlinin}, {Williamson}, {Zahn}, \& {Zenteno}}]{2015ApJS..216...27B}
{Bleem}, L.~E., {Stalder}, B., {de Haan}, T., {et~al.} 2015, \apjs, 216, 27

\bibitem[{{Bocquet} {et~al.}(2019){Bocquet}, {Dietrich}, {Schrabback}, {Bleem}, {Klein}, {Allen}, {Applegate}, {Ashby}, {Bautz}, {Bayliss}, {Benson}, {Brodwin}, {Bulbul}, {Canning}, {Capasso}, {Carlstrom}, {Chang}, {Chiu}, {Cho}, {Clocchiatti}, {Crawford}, {Crites}, {de Haan}, {Desai}, {Dobbs}, {Foley}, {Forman}, {Garmire}, {George}, {Gladders}, {Gonzalez}, {Grandis}, {Gupta}, {Halverson}, {Hlavacek-Larrondo}, {Hoekstra}, {Holder}, {Holzapfel}, {Hou}, {Hrubes}, {Huang}, {Jones}, {Khullar}, {Knox}, {Kraft}, {Lee}, {von der Linden}, {Luong-Van}, {Mantz}, {Marrone}, {McDonald}, {McMahon}, {Meyer}, {Mocanu}, {Mohr}, {Morris}, {Padin}, {Patil}, {Pryke}, {Rapetti}, {Reichardt}, {Rest}, {Ruhl}, {Saliwanchik}, {Saro}, {Sayre}, {Schaffer}, {Shirokoff}, {Stalder}, {Stanford}, {Staniszewski}, {Stark}, {Story}, {Strazzullo}, {Stubbs}, {Vanderlinde}, {Vieira}, {Vikhlinin}, {Williamson}, \& {Zenteno}}]{2019ApJ...878...55B}
{Bocquet}, S., {Dietrich}, J.~P., {Schrabback}, T., {et~al.} 2019, \apj, 878, 55

\bibitem[{{Bocquet} {et~al.}(2020){Bocquet}, {Heitmann}, {Habib}, {Lawrence}, {Uram}, {Frontiere}, {Pope}, \& {Finkel}}]{2020ApJ...901....5B}
{Bocquet}, S., {Heitmann}, K., {Habib}, S., {et~al.} 2020, \apj, 901, 5

\bibitem[{{B{\"o}hringer} {et~al.}(2013){B{\"o}hringer}, {Chon}, {Collins}, {Guzzo}, {Nowak}, \& {Bobrovskyi}}]{2013A&A...555A..30B}
{B{\"o}hringer}, H., {Chon}, G., {Collins}, C.~A., {et~al.} 2013, \aap, 555, A30

\bibitem[{{Borgani} {et~al.}(1999){Borgani}, {Girardi}, {Carlberg}, {Yee}, \& {Ellingson}}]{1999ApJ...527..561B}
{Borgani}, S., {Girardi}, M., {Carlberg}, R.~G., {Yee}, H.~K.~C., \& {Ellingson}, E. 1999, \apj, 527, 561

\bibitem[{{Borgani} \& {Guzzo}(2001)}]{2001Natur.409...39B}
{Borgani}, S. \& {Guzzo}, L. 2001, \nat, 409, 39

\bibitem[{{Borgani} {et~al.}(2004){Borgani}, {Murante}, {Springel}, {Diaferio}, {Dolag}, {Moscardini}, {Tormen}, {Tornatore}, \& {Tozzi}}]{2004MNRAS.348.1078B}
{Borgani}, S., {Murante}, G., {Springel}, V., {et~al.} 2004, \mnras, 348, 1078

\bibitem[{{Bottrell} {et~al.}(2019){Bottrell}, {Hani}, {Teimoorinia}, {Ellison}, {Moreno}, {Torrey}, {Hayward}, {Thorp}, {Simard}, \& {Hernquist}}]{2019MNRAS.490.5390Bottrell}
{Bottrell}, C., {Hani}, M.~H., {Teimoorinia}, H., {et~al.} 2019, \mnras, 490, 5390

\bibitem[{{Boylan-Kolchin} {et~al.}(2009){Boylan-Kolchin}, {Springel}, {White}, {Jenkins}, \& {Lemson}}]{2009MNRAS.398.1150B}
{Boylan-Kolchin}, M., {Springel}, V., {White}, S. D.~M., {Jenkins}, A., \& {Lemson}, G. 2009, \mnras, 398, 1150

\bibitem[{{Breiman}(2001)}]{2001MachL..45....5B}
{Breiman}, L. 2001, Machine Learning, 45, 5

\bibitem[{{Bryan} \& {Norman}(1998)}]{1998ApJ...495...80B}
{Bryan}, G.~L. \& {Norman}, M.~L. 1998, \apj, 495, 80

\bibitem[{{Bulbul} {et~al.}(2019){Bulbul}, {Chiu}, {Mohr}, {McDonald}, {Benson}, {Bautz}, {Bayliss}, {Bleem}, {Brodwin}, {Bocquet}, {Capasso}, {Dietrich}, {Forman}, {Hlavacek-Larrondo}, {Holzapfel}, {Khullar}, {Klein}, {Kraft}, {Miller}, {Reichardt}, {Saro}, {Sharon}, {Stalder}, {Schrabback}, \& {Stanford}}]{2019ApJ...871...50B}
{Bulbul}, E., {Chiu}, I.~N., {Mohr}, J.~J., {et~al.} 2019, \apj, 871, 50

\bibitem[{{Chawak} {et~al.}(2023){Chawak}, {Villaescusa-Navarro}, {Echeverri Rojas}, {Ni}, {Hahn}, \& {Angles-Alcazar}}]{2023arXiv230912048C}
{Chawak}, C., {Villaescusa-Navarro}, F., {Echeverri Rojas}, N., {et~al.} 2023, arXiv e-prints, arXiv:2309.12048

\bibitem[{Chen \& Guestrin(2016)}]{XGB}
Chen, T. \& Guestrin, C. 2016, ACM

\bibitem[{{Chiu} {et~al.}(2018){Chiu}, {Mohr}, {McDonald}, {Bocquet}, {Desai}, {Klein}, {Israel}, {Ashby}, {Stanford}, {Benson}, {Brodwin}, {Abbott}, {Abdalla}, {Allam}, {Annis}, {Bayliss}, {Benoit-L{\'e}vy}, {Bertin}, {Bleem}, {Brooks}, {Buckley-Geer}, {Bulbul}, {Capasso}, {Carlstrom}, {Rosell}, {Carretero}, {Castander}, {Cunha}, {D'Andrea}, {da Costa}, {Davis}, {Diehl}, {Dietrich}, {Doel}, {Drlica-Wagner}, {Eifler}, {Evrard}, {Flaugher}, {Garc{\'\i}a-Bellido}, {Garmire}, {Gaztanaga}, {Gerdes}, {Gonzalez}, {Gruen}, {Gruendl}, {Gschwend}, {Gupta}, {Gutierrez}, {Hlavacek-L}, {Honscheid}, {James}, {Jeltema}, {Kraft}, {Krause}, {Kuehn}, {Kuhlmann}, {Kuropatkin}, {Lahav}, {Lima}, {Maia}, {Marshall}, {Melchior}, {Menanteau}, {Miquel}, {Murray}, {Nord}, {Ogando}, {Plazas}, {Rapetti}, {Reichardt}, {Romer}, {Roodman}, {Sanchez}, {Saro}, {Scarpine}, {Schindler}, {Schubnell}, {Sharon}, {Smith}, {Smith}, {Soares-Santos}, {Sobreira}, {Stalder}, {Stern}, {Strazzullo}, {Suchyta}, {Swanson}, {Tarle}, {Vikram}, {Walker},
  {Weller}, \& {Zhang}}]{2018MNRAS.478.3072C}
{Chiu}, I., {Mohr}, J.~J., {McDonald}, M., {et~al.} 2018, \mnras, 478, 3072

\bibitem[{{Chiu} {et~al.}(2022){Chiu}, {Klein}, {Mohr}, \& {Bocquet}}]{2022arXiv220712429C}
{Chiu}, I.-N., {Klein}, M., {Mohr}, J., \& {Bocquet}, S. 2022, arXiv e-prints, arXiv:2207.12429

\bibitem[{{Cohn} \& {Battaglia}(2020)}]{2020MNRAS.491.1575C}
{Cohn}, J.~D. \& {Battaglia}, N. 2020, \mnras, 491, 1575

\bibitem[{{Col{\'\i}n} {et~al.}(2010){Col{\'\i}n}, {Avila-Reese}, {V{\'a}zquez-Semadeni}, {Valenzuela}, \& {Ceverino}}]{2010ApJ...713..535C-subgrid}
{Col{\'\i}n}, P., {Avila-Reese}, V., {V{\'a}zquez-Semadeni}, E., {Valenzuela}, O., \& {Ceverino}, D. 2010, \apj, 713, 535

\bibitem[{{Costanzi} {et~al.}(2019){Costanzi}, {Rozo}, {Simet}, {Zhang}, {Evrard}, {Mantz}, {Rykoff}, {Jeltema}, {Gruen}, {Allen}, {McClintock}, {Romer}, {von der Linden}, {Farahi}, {DeRose}, {Varga}, {Weller}, {Giles}, {Hollowood}, {Bhargava}, {Bermeo-Hernandez}, {Chen}, {Abbott}, {Abdalla}, {Avila}, {Bechtol}, {Brooks}, {Buckley-Geer}, {Burke}, {Rosell}, {Kind}, {Carretero}, {Crocce}, {Cunha}, {da Costa}, {Davis}, {De Vicente}, {Diehl}, {Dietrich}, {Doel}, {Eifler}, {Estrada}, {Flaugher}, {Fosalba}, {Frieman}, {Garc{\'\i}a-Bellido}, {Gaztanaga}, {Gerdes}, {Giannantonio}, {Gruendl}, {Gschwend}, {Gutierrez}, {Hartley}, {Honscheid}, {Hoyle}, {James}, {Krause}, {Kuehn}, {Kuropatkin}, {Lima}, {Lin}, {Maia}, {March}, {Marshall}, {Martini}, {Menanteau}, {Miller}, {Miquel}, {Mohr}, {Ogando}, {Plazas}, {Roodman}, {Sanchez}, {Scarpine}, {Schindler}, {Schubnell}, {Serrano}, {Sevilla-Noarbe}, {Sheldon}, {Smith}, {Soares-Santos}, {Sobreira}, {Suchyta}, {Swanson}, {Tarle}, {Thomas}, \& {Wechsler}}]{2019MNRAS.488.4779C}
{Costanzi}, M., {Rozo}, E., {Simet}, M., {et~al.} 2019, \mnras, 488, 4779

\bibitem[{{Cui} {et~al.}(2012){Cui}, {Borgani}, {Dolag}, {Murante}, \& {Tornatore}}]{2012MNRAS.423.2279C}
{Cui}, W., {Borgani}, S., {Dolag}, K., {Murante}, G., \& {Tornatore}, L. 2012, \mnras, 423, 2279

\bibitem[{{Dalton} {et~al.}(2012){Dalton}, {Trager}, {Abrams}, {Carter}, {Bonifacio}, {Aguerri}, {MacIntosh}, {Evans}, {Lewis}, {Navarro}, {Agocs}, {Dee}, {Rousset}, {Tosh}, {Middleton}, {Pragt}, {Terrett}, {Brock}, {Benn}, {Verheijen}, {Cano Infantes}, {Bevil}, {Steele}, {Mottram}, {Bates}, {Gribbin}, {Rey}, {Rodriguez}, {Delgado}, {Guinouard}, {Walton}, {Irwin}, {Jagourel}, {Stuik}, {Gerlofsma}, {Roelfsma}, {Skillen}, {Ridings}, {Balcells}, {Daban}, {Gouvret}, {Venema}, \& {Girard}}]{2012SPIE.8446E..0PD}
{Dalton}, G., {Trager}, S.~C., {Abrams}, D.~C., {et~al.} 2012, in Society of Photo-Optical Instrumentation Engineers (SPIE) Conference Series, Vol. 8446, Ground-based and Airborne Instrumentation for Astronomy IV, ed. I.~S. {McLean}, S.~K. {Ramsay}, \& H.~{Takami}, 84460P

\bibitem[{{de Andres} {et~al.}(2022){de Andres}, {Cui}, {Ruppin}, {De Petris}, {Yepes}, {Gianfagna}, {Lahouli}, {Aversano}, {Dupuis}, {Jarraya}, \& {Vega-Ferrero}}]{2022NatAs...6.1325D}
{de Andres}, D., {Cui}, W., {Ruppin}, F., {et~al.} 2022, Nature Astronomy, 6, 1325

\bibitem[{{de Jong} {et~al.}(2019){de Jong}, {Agertz}, {Berbel}, {Aird}, {Alexander}, {Amarsi}, {Anders}, {Andrae}, {Ansarinejad}, {Ansorge}, {Antilogus}, {Anwand-Heerwart}, {Arentsen}, {Arnadottir}, {Asplund}, {Auger}, {Azais}, {Baade}, {Baker}, {Baker}, {Balbinot}, {Baldry}, {Banerji}, {Barden}, {Barklem}, {Barth{\'e}l{\'e}my-Mazot}, {Battistini}, {Bauer}, {Bell}, {Bellido-Tirado}, {Bellstedt}, {Belokurov}, {Bensby}, {Bergemann}, {Bestenlehner}, {Bielby}, {Bilicki}, {Blake}, {Bland-Hawthorn}, {Boeche}, {Boland}, {Boller}, {Bongard}, {Bongiorno}, {Bonifacio}, {Boudon}, {Brooks}, {Brown}, {Brown}, {Br{\"u}ggen}, {Brynnel}, {Brzeski}, {Buchert}, {Buschkamp}, {Caffau}, {Caillier}, {Carrick}, {Casagrande}, {Case}, {Casey}, {Cesarini}, {Cescutti}, {Chapuis}, {Chiappini}, {Childress}, {Christlieb}, {Church}, {Cioni}, {Cluver}, {Colless}, {Collett}, {Comparat}, {Cooper}, {Couch}, {Courbin}, {Croom}, {Croton}, {Daguis{\'e}}, {Dalton}, {Davies}, {Davis}, {de Laverny}, {Deason}, {Dionies}, {Disseau}, {Doel},
  {D{\"o}scher}, {Driver}, {Dwelly}, {Eckert}, {Edge}, {Edvardsson}, {Youssoufi}, {Elhaddad}, {Enke}, {Erfanianfar}, {Farrell}, {Fechner}, {Feiz}, {Feltzing}, {Ferreras}, {Feuerstein}, {Feuillet}, {Finoguenov}, {Ford}, {Fotopoulou}, {Fouesneau}, {Frenk}, {Frey}, {Gaessler}, {Geier}, {Gentile Fusillo}, {Gerhard}, {Giannantonio}, {Giannone}, {Gibson}, {Gillingham}, {Gonz{\'a}lez-Fern{\'a}ndez}, {Gonzalez-Solares}, {Gottloeber}, {Gould}, {Grebel}, {Gueguen}, {Guiglion}, {Haehnelt}, {Hahn}, {Hansen}, {Hartman}, {Hauptner}, {Hawkins}, {Haynes}, {Haynes}, {Heiter}, {Helmi}, {Aguayo}, {Hewett}, {Hinton}, {Hobbs}, {Hoenig}, {Hofman}, {Hook}, {Hopgood}, {Hopkins}, {Hourihane}, {Howes}, {Howlett}, {Huet}, {Irwin}, {Iwert}, {Jablonka}, {Jahn}, {Jahnke}, {Jarno}, {Jin}, {Jofre}, {Johl}, {Jones}, {J{\"o}nsson}, {Jordan}, {Karovicova}, {Khalatyan}, {Kelz}, {Kennicutt}, {King}, {Kitaura}, {Klar}, {Klauser}, {Kneib}, {Koch}, {Koposov}, {Kordopatis}, {Korn}, {Kosmalski}, {Kotak}, {Kovalev}, {Kreckel}, {Kripak}, {Krumpe},
  {Kuijken}, {Kunder}, {Kushniruk}, {Lam}, {Lamer}, {Laurent}, {Lawrence}, {Lehmitz}, {Lemasle}, {Lewis}, {Li}, {Lidman}, {Lind}, {Liske}, {Lizon}, {Loveday}, {Ludwig}, {McDermid}, {Maguire}, {Mainieri}, {Mali}, {Mandel}, {Mandel}, {Mannering}, {Martell}, {Martinez Delgado}, {Matijevic}, {McGregor}, {McMahon}, {McMillan}, {Mena}, {Merloni}, {Meyer}, {Michel}, {Micheva}, {Migniau}, {Minchev}, {Monari}, {Muller}, {Murphy}, {Muthukrishna}, {Nandra}, {Navarro}, {Ness}, {Nichani}, {Nichol}, {Nicklas}, {Niederhofer}, {Norberg}, {Obreschkow}, {Oliver}, {Owers}, {Pai}, {Pankratow}, {Parkinson}, {Paschke}, {Paterson}, {Pecontal}, {Parry}, {Phillips}, {Pillepich}, {Pinard}, {Pirard}, {Piskunov}, {Plank}, {Pl{\"u}schke}, {Pons}, {Popesso}, {Power}, {Pragt}, {Pramskiy}, {Pryer}, {Quattri}, {Queiroz}, {Quirrenbach}, {Rahurkar}, {Raichoor}, {Ramstedt}, {Rau}, {Recio-Blanco}, {Reiss}, {Renaud}, {Revaz}, {Rhode}, {Richard}, {Richter}, {Rix}, {Robotham}, {Roelfsema}, {Romaniello}, {Rosario}, {Rothmaier}, {Roukema}, {Ruchti},
  {Rupprecht}, {Rybizki}, {Ryde}, {Saar}, {Sadler}, {Sahl{\'e}n}, {Salvato}, {Sassolas}, {Saunders}, {Saviauk}, {Sbordone}, {Schmidt}, {Schnurr}, {Scholz}, {Schwope}, {Seifert}, {Shanks}, {Sheinis}, {Sivov}, {Sk{\'u}lad{\'o}ttir}, {Smartt}, {Smedley}, {Smith}, {Smith}, {Sorce}, {Spitler}, {Starkenburg}, {Steinmetz}, {Stilz}, {Storm}, {Sullivan}, {Sutherland}, {Swann}, {Tamone}, {Taylor}, {Teillon}, {Tempel}, {ter Horst}, {Thi}, {Tolstoy}, {Trager}, {Traven}, {Tremblay}, {Tresse}, {Valentini}, {van de Weygaert}, {van den Ancker}, {Veljanoski}, {Venkatesan}, {Wagner}, {Wagner}, {Walcher}, {Waller}, {Walton}, {Wang}, {Winkler}, {Wisotzki}, {Worley}, {Worseck}, {Xiang}, {Xu}, {Yong}, {Zhao}, {Zheng}, {Zscheyge}, \& {Zucker}}]{2019Msngr.175....3D}
{de Jong}, R.~S., {Agertz}, O., {Berbel}, A.~A., {et~al.} 2019, The Messenger, 175, 3

\bibitem[{{DESI Collaboration} {et~al.}(2016){DESI Collaboration}, {Aghamousa}, {Aguilar}, {Ahlen}, {Alam}, {Allen}, {Allende Prieto}, {Annis}, {Bailey}, {Balland}, {Ballester}, {Baltay}, {Beaufore}, {Bebek}, {Beers}, {Bell}, {Bernal}, {Besuner}, {Beutler}, {Blake}, {Bleuler}, {Blomqvist}, {Blum}, {Bolton}, {Briceno}, {Brooks}, {Brownstein}, {Buckley-Geer}, {Burden}, {Burtin}, {Busca}, {Cahn}, {Cai}, {Cardiel-Sas}, {Carlberg}, {Carton}, {Casas}, {Castander}, {Cervantes-Cota}, {Claybaugh}, {Close}, {Coker}, {Cole}, {Comparat}, {Cooper}, {Cousinou}, {Crocce}, {Cuby}, {Cunningham}, {Davis}, {Dawson}, {de la Macorra}, {De Vicente}, {Delubac}, {Derwent}, {Dey}, {Dhungana}, {Ding}, {Doel}, {Duan}, {Ealet}, {Edelstein}, {Eftekharzadeh}, {Eisenstein}, {Elliott}, {Escoffier}, {Evatt}, {Fagrelius}, {Fan}, {Fanning}, {Farahi}, {Farihi}, {Favole}, {Feng}, {Fernandez}, {Findlay}, {Finkbeiner}, {Fitzpatrick}, {Flaugher}, {Flender}, {Font-Ribera}, {Forero-Romero}, {Fosalba}, {Frenk}, {Fumagalli}, {Gaensicke}, {Gallo},
  {Garcia-Bellido}, {Gaztanaga}, {Pietro Gentile Fusillo}, {Gerard}, {Gershkovich}, {Giannantonio}, {Gillet}, {Gonzalez-de-Rivera}, {Gonzalez-Perez}, {Gott}, {Graur}, {Gutierrez}, {Guy}, {Habib}, {Heetderks}, {Heetderks}, {Heitmann}, {Hellwing}, {Herrera}, {Ho}, {Holland}, {Honscheid}, {Huff}, {Hutchinson}, {Huterer}, {Hwang}, {Illa Laguna}, {Ishikawa}, {Jacobs}, {Jeffrey}, {Jelinsky}, {Jennings}, {Jiang}, {Jimenez}, {Johnson}, {Joyce}, {Jullo}, {Juneau}, {Kama}, {Karcher}, {Karkar}, {Kehoe}, {Kennamer}, {Kent}, {Kilbinger}, {Kim}, {Kirkby}, {Kisner}, {Kitanidis}, {Kneib}, {Koposov}, {Kovacs}, {Koyama}, {Kremin}, {Kron}, {Kronig}, {Kueter-Young}, {Lacey}, {Lafever}, {Lahav}, {Lambert}, {Lampton}, {Landriau}, {Lang}, {Lauer}, {Le Goff}, {Le Guillou}, {Le Van Suu}, {Lee}, {Lee}, {Leitner}, {Lesser}, {Levi}, {L'Huillier}, {Li}, {Liang}, {Lin}, {Linder}, {Loebman}, {Luki{\'c}}, {Ma}, {MacCrann}, {Magneville}, {Makarem}, {Manera}, {Manser}, {Marshall}, {Martini}, {Massey}, {Matheson}, {McCauley}, {McDonald},
  {McGreer}, {Meisner}, {Metcalfe}, {Miller}, {Miquel}, {Moustakas}, {Myers}, {Naik}, {Newman}, {Nichol}, {Nicola}, {Nicolati da Costa}, {Nie}, {Niz}, {Norberg}, {Nord}, {Norman}, {Nugent}, {O'Brien}, {Oh}, {Olsen}, {Padilla}, {Padmanabhan}, {Padmanabhan}, {Palanque-Delabrouille}, {Palmese}, {Pappalardo}, {P{\^a}ris}, {Park}, {Patej}, {Peacock}, {Peiris}, {Peng}, {Percival}, {Perruchot}, {Pieri}, {Pogge}, {Pollack}, {Poppett}, {Prada}, {Prakash}, {Probst}, {Rabinowitz}, {Raichoor}, {Ree}, {Refregier}, {Regal}, {Reid}, {Reil}, {Rezaie}, {Rockosi}, {Roe}, {Ronayette}, {Roodman}, {Ross}, {Ross}, {Rossi}, {Rozo}, {Ruhlmann-Kleider}, {Rykoff}, {Sabiu}, {Samushia}, {Sanchez}, {Sanchez}, {Schlegel}, {Schneider}, {Schubnell}, {Secroun}, {Seljak}, {Seo}, {Serrano}, {Shafieloo}, {Shan}, {Sharples}, {Sholl}, {Shourt}, {Silber}, {Silva}, {Sirk}, {Slosar}, {Smith}, {Smoot}, {Som}, {Song}, {Sprayberry}, {Staten}, {Stefanik}, {Tarle}, {Sien Tie}, {Tinker}, {Tojeiro}, {Valdes}, {Valenzuela}, {Valluri}, {Vargas-Magana},
  {Verde}, {Walker}, {Wang}, {Wang}, {Weaver}, {Weaverdyck}, {Wechsler}, {Weinberg}, {White}, {Yang}, {Yeche}, {Zhang}, {Zhao}, {Zheng}, {Zhou}, {Zhou}, {Zhu}, {Zou}, \& {Zu}}]{2016arXiv161100036D}
{DESI Collaboration}, {Aghamousa}, A., {Aguilar}, J., {et~al.} 2016, arXiv e-prints, arXiv:1611.00036

\bibitem[{{Dietrich} {et~al.}(2019){Dietrich}, {Bocquet}, {Schrabback}, {Applegate}, {Hoekstra}, {Grandis}, {Mohr}, {Allen}, {Bayliss}, {Benson}, {Bleem}, {Brodwin}, {Bulbul}, {Capasso}, {Chiu}, {Crawford}, {Gonzalez}, {de Haan}, {Klein}, {von der Linden}, {Mantz}, {Marrone}, {McDonald}, {Raghunathan}, {Rapetti}, {Reichardt}, {Saro}, {Stalder}, {Stark}, {Stern}, \& {Stubbs}}]{2019MNRAS.483.2871D}
{Dietrich}, J.~P., {Bocquet}, S., {Schrabback}, T., {et~al.} 2019, \mnras, 483, 2871

\bibitem[{Dietterich(2000)}]{2000Ensemble}
Dietterich, T.~G. 2000, in International Workshop on Multiple Classifier Systems

\bibitem[{{Dolag} {et~al.}(2009){Dolag}, {Borgani}, {Murante}, \& {Springel}}]{2009MNRAS.399..497D}
{Dolag}, K., {Borgani}, S., {Murante}, G., \& {Springel}, V. 2009, \mnras, 399, 497

\bibitem[{{Dolag} {et~al.}(2016){Dolag}, {Komatsu}, \& {Sunyaev}}]{2016MNRAS.463.1797Dolag16}
{Dolag}, K., {Komatsu}, E., \& {Sunyaev}, R. 2016, \mnras, 463, 1797

\bibitem[{{Echeverri} {et~al.}(2023){Echeverri}, {Villaescusa-Navarro}, {Chawak}, {Ni}, {Hahn}, {Hernandez-Martinez}, {Teyssier}, {Angles-Alcazar}, {Dolag}, \& {Castro}}]{2023arXiv230406084E}
{Echeverri}, N., {Villaescusa-Navarro}, F., {Chawak}, C., {et~al.} 2023, arXiv e-prints, arXiv:2304.06084

\bibitem[{{Echeverri-Rojas} {et~al.}(2023){Echeverri-Rojas}, {Villaescusa-Navarro}, {Chawak}, {Ni}, {Hahn}, {Hern{\'a}ndez-Mart{\'\i}nez}, {Teyssier}, {Angl{\'e}s-Alc{\'a}zar}, {Dolag}, \& {Castro}}]{2023ApJ...954..125E}
{Echeverri-Rojas}, N., {Villaescusa-Navarro}, F., {Chawak}, C., {et~al.} 2023, \apj, 954, 125

\bibitem[{{Euclid Collaboration} {et~al.}(2023){Euclid Collaboration}, {Giocoli}, {Meneghetti}, {Rasia}, {Borgani}, {Despali}, {Lesci}, {Marulli}, {Moscardini}, {Sereno}, {Cui}, {Knebe}, {Yepes}, {Castro}, {Corasaniti}, {Pires}, {Castignani}, {Ingoglia}, {Schrabback}, {Pratt}, {Le Brun}, {Aghanim}, {Amendola}, {Auricchio}, {Baldi}, {Bodendorf}, {Bonino}, {Branchini}, {Brescia}, {Brinchmann}, {Camera}, {Capobianco}, {Carbone}, {Carretero}, {Castander}, {Castellano}, {Cavuoti}, {Cledassou}, {Congedo}, {Conselice}, {Conversi}, {Copin}, {Corcione}, {Courbin}, {Cropper}, {Da Silva}, {Degaudenzi}, {Dinis}, {Dubath}, {Dupac}, {Dusini}, {Farrens}, {Ferriol}, {Fosalba}, {Frailis}, {Franceschi}, {Fumana}, {Galeotta}, {Garilli}, {Gillis}, {Grazian}, {Grupp}, {Haugan}, {Holmes}, {Hornstrup}, {Jahnke}, {K{\"u}mmel}, {Kermiche}, {Kilbinger}, {Kunz}, {Kurki-Suonio}, {Ligori}, {Lilje}, {Lloro}, {Maiorano}, {Mansutti}, {Marggraf}, {Markovic}, {Massey}, {Maurogordato}, {Mei}, {Merlin}, {Meylan}, {Moresco}, {Munari}, {Niemi},
  {Nightingale}, {Nutma}, {Padilla}, {Paltani}, {Pasian}, {Pedersen}, {Pettorino}, {Polenta}, {Poncet}, {Popa}, {Raison}, {Renzi}, {Rhodes}, {Riccio}, {Romelli}, {Roncarelli}, {Rossetti}, {Saglia}, {Sapone}, {Sartoris}, {Schneider}, {Secroun}, {Serrano}, {Sirignano}, {Sirri}, {Stanco}, {Starck}, {Tallada-Cresp{\'\i}}, {Taylor}, {Tereno}, {Toledo-Moreo}, {Torradeflot}, {Tutusaus}, {Valentijn}, {Valenziano}, {Vassallo}, {Wang}, {Weller}, {Zamorani}, {Zoubian}, {Andreon}, {Bardelli}, {Boucaud}, {Bozzo}, {Colodro-Conde}, {Di Ferdinando}, {Fabbian}, {Farina}, {Israel}, {Keih{\"a}nen}, {Lindholm}, {Mauri}, {Neissner}, {Schirmer}, {Scottez}, {Tenti}, {Zucca}, {Akrami}, {Baccigalupi}, {Ballardini}, {Bernardeau}, {Biviano}, {Borlaff}, {Burigana}, {Cabanac}, {Cappi}, {Carvalho}, {Casas}, {Chambers}, {Cooray}, {Courtois}, {Davini}, {de la Torre}, {De Lucia}, {Desprez}, {Dole}, {Escartin}, {Escoffier}, {Ferrero}, {Finelli}, {Gabarra}, {Ganga}, {Garcia-Bellido}, {George}, {Giacomini}, {Gozaliasl}, {Hildebrandt}, {Hook},
  {JIMENEZ MU\textbackslash\{N\}OZ}, {Joachimi}, {Kajava}, {Kansal}, {Kirkpatrick}, {Legrand}, {Loureiro}, {Macias-Perez}, {Magliocchetti}, {Mainetti}, {Maoli}, {Marcin}, {Martinelli}, {Martinet}, {Martins}, {Matthew}, {Maurin}, {Metcalf}, {Monaco}, {Morgante}, {Nadathur}, {Nucita}, {Patrizii}, {Peel}, {Pollack}, {Popa}, {Porciani}, {Potter}, {P{\"o}ntinen}, {Reimberg}, {S{\'a}nchez}, {Sakr}, {Schneider}, {Sefusatti}, {Shulevski}, {Spurio Mancini}, {Stadel}, {Steinwagner}, {Valiviita}, {Veropalumbo}, {Viel}, \& {Zinchenko}}]{2023arXiv230200687E}
{Euclid Collaboration}, {Giocoli}, C., {Meneghetti}, M., {et~al.} 2023, arXiv e-prints, arXiv:2302.00687

\bibitem[{{Fabjan} {et~al.}(2010){Fabjan}, {Borgani}, {Tornatore}, {Saro}, {Murante}, \& {Dolag}}]{2010MNRAS.401.1670F}
{Fabjan}, D., {Borgani}, S., {Tornatore}, L., {et~al.} 2010, \mnras, 401, 1670

\bibitem[{{Falco} {et~al.}(2013){Falco}, {Mamon}, {Wojtak}, {Hansen}, \& {Gottl{\"o}ber}}]{2013MNRAS.436.2639F}
{Falco}, M., {Mamon}, G.~A., {Wojtak}, R., {Hansen}, S.~H., \& {Gottl{\"o}ber}, S. 2013, \mnras, 436, 2639

\bibitem[{Geurts {et~al.}(2006)Geurts, Ernst, \& Wehenkel}]{ET}
Geurts, P., Ernst, D., \& Wehenkel, L. 2006, Machine Learning, 63, 3

\bibitem[{{Giocoli} {et~al.}(2021){Giocoli}, {Marulli}, {Moscardini}, {Sereno}, {Veropalumbo}, {Gigante}, {Maturi}, {Radovich}, {Bellagamba}, {Roncarelli}, {Bardelli}, {Contarini}, {Covone}, {Harnois-D{\'e}raps}, {Ingoglia}, {Lesci}, {Nanni}, \& {Puddu}}]{2021A&A...653A..19Giocoli}
{Giocoli}, C., {Marulli}, F., {Moscardini}, L., {et~al.} 2021, \aap, 653, A19

\bibitem[{{Henson} {et~al.}(2017){Henson}, {Barnes}, {Kay}, {McCarthy}, \& {Schaye}}]{2017MNRAS.465.3361H}
{Henson}, M.~A., {Barnes}, D.~J., {Kay}, S.~T., {McCarthy}, I.~G., \& {Schaye}, J. 2017, \mnras, 465, 3361

\bibitem[{{Heymans} {et~al.}(2021){Heymans}, {Tr{\"o}ster}, {Asgari}, {Blake}, {Hildebrandt}, {Joachimi}, {Kuijken}, {Lin}, {S{\'a}nchez}, {van den Busch}, {Wright}, {Amon}, {Bilicki}, {de Jong}, {Crocce}, {Dvornik}, {Erben}, {Fortuna}, {Getman}, {Giblin}, {Glazebrook}, {Hoekstra}, {Joudaki}, {Kannawadi}, {K{\"o}hlinger}, {Lidman}, {Miller}, {Napolitano}, {Parkinson}, {Schneider}, {Shan}, {Valentijn}, {Verdoes Kleijn}, \& {Wolf}}]{2021A&A...646A.140H}
{Heymans}, C., {Tr{\"o}ster}, T., {Asgari}, M., {et~al.} 2021, \aap, 646, A140

\bibitem[{{Hildebrandt} {et~al.}(2017){Hildebrandt}, {Viola}, {Heymans}, {Joudaki}, {Kuijken}, {Blake}, {Erben}, {Joachimi}, {Klaes}, {Miller}, {Morrison}, {Nakajima}, {Verdoes Kleijn}, {Amon}, {Choi}, {Covone}, {de Jong}, {Dvornik}, {Fenech Conti}, {Grado}, {Harnois-D{\'e}raps}, {Herbonnet}, {Hoekstra}, {K{\"o}hlinger}, {McFarland}, {Mead}, {Merten}, {Napolitano}, {Peacock}, {Radovich}, {Schneider}, {Simon}, {Valentijn}, {van den Busch}, {van Uitert}, \& {Van Waerbeke}}]{2017MNRAS.465.1454H}
{Hildebrandt}, H., {Viola}, M., {Heymans}, C., {et~al.} 2017, \mnras, 465, 1454

\bibitem[{{Hilton} {et~al.}(2021){Hilton}, {Sif{\'o}n}, {Naess}, {Madhavacheril}, {Oguri}, {Rozo}, {Rykoff}, {Abbott}, {Adhikari}, {Aguena}, {Aiola}, {Allam}, {Amodeo}, {Amon}, {Annis}, {Ansarinejad}, {Aros-Bunster}, {Austermann}, {Avila}, {Bacon}, {Battaglia}, {Beall}, {Becker}, {Bernstein}, {Bertin}, {Bhandarkar}, {Bhargava}, {Bond}, {Brooks}, {Burke}, {Calabrese}, {Carrasco Kind}, {Carretero}, {Choi}, {Choi}, {Conselice}, {da Costa}, {Costanzi}, {Crichton}, {Crowley}, {D{\"u}nner}, {Denison}, {Devlin}, {Dicker}, {Diehl}, {Dietrich}, {Doel}, {Duff}, {Duivenvoorden}, {Dunkley}, {Everett}, {Ferraro}, {Ferrero}, {Fert{\'e}}, {Flaugher}, {Frieman}, {Gallardo}, {Garc{\'\i}a-Bellido}, {Gaztanaga}, {Gerdes}, {Giles}, {Golec}, {Gralla}, {Grandis}, {Gruen}, {Gruendl}, {Gschwend}, {Gutierrez}, {Han}, {Hartley}, {Hasselfield}, {Hill}, {Hilton}, {Hincks}, {Hinton}, {Ho}, {Honscheid}, {Hoyle}, {Hubmayr}, {Huffenberger}, {Hughes}, {Jaelani}, {Jain}, {James}, {Jeltema}, {Kent}, {Knowles}, {Koopman}, {Kuehn}, {Lahav},
  {Lima}, {Lin}, {Lokken}, {Loubser}, {MacCrann}, {Maia}, {Marriage}, {Martin}, {McMahon}, {Melchior}, {Menanteau}, {Miquel}, {Miyatake}, {Moodley}, {Morgan}, {Mroczkowski}, {Nati}, {Newburgh}, {Niemack}, {Nishizawa}, {Ogando}, {Orlowski-Scherer}, {Page}, {Palmese}, {Partridge}, {Paz-Chinch{\'o}n}, {Phakathi}, {Plazas}, {Robertson}, {Romer}, {Carnero Rosell}, {Salatino}, {Sanchez}, {Schaan}, {Schillaci}, {Sehgal}, {Serrano}, {Shin}, {Simon}, {Smith}, {Soares-Santos}, {Spergel}, {Staggs}, {Storer}, {Suchyta}, {Swanson}, {Tarle}, {Thomas}, {To}, {Trac}, {Ullom}, {Vale}, {Van Lanen}, {Vavagiakis}, {De Vicente}, {Wilkinson}, {Wollack}, {Xu}, \& {Zhang}}]{2021ApJS..253....3H}
{Hilton}, M., {Sif{\'o}n}, C., {Naess}, S., {et~al.} 2021, \apjs, 253, 3

\bibitem[{{Hirschmann} {et~al.}(2014){Hirschmann}, {Dolag}, {Saro}, {Bachmann}, {Borgani}, \& {Burkert}}]{2014MNRAS.442.2304H}
{Hirschmann}, M., {Dolag}, K., {Saro}, A., {et~al.} 2014, \mnras, 442, 2304

\bibitem[{Ho {et~al.}(2019)Ho, Rau, Ntampaka, Farahi, Trac, \& Póczos}]{Ho_2019}
Ho, M., Rau, M.~M., Ntampaka, M., {et~al.} 2019, The Astrophysical Journal, 887, 25

\bibitem[{{Hoekstra} {et~al.}(2015){Hoekstra}, {Herbonnet}, {Muzzin}, {Babul}, {Mahdavi}, {Viola}, \& {Cacciato}}]{2015MNRAS.449..685H}
{Hoekstra}, H., {Herbonnet}, R., {Muzzin}, A., {et~al.} 2015, \mnras, 449, 685

\bibitem[{{Ingoglia} {et~al.}(2022){Ingoglia}, {Covone}, {Sereno}, {Giocoli}, {Bardelli}, {Bellagamba}, {Castignani}, {Farrens}, {Hildebrandt}, {Joudaki}, {Jullo}, {Lanzieri}, {Lesci}, {Marulli}, {Maturi}, {Moscardini}, {Nanni}, {Puddu}, {Radovich}, {Roncarelli}, {Sapio}, \& {Schimd}}]{2022MNRAS.511.1484I}
{Ingoglia}, L., {Covone}, G., {Sereno}, M., {et~al.} 2022, \mnras, 511, 1484

\bibitem[{{Jimenez} \& {Loeb}(2002)}]{Jimenez02}
{Jimenez}, R. \& {Loeb}, A. 2002, \apj, 573, 37

\bibitem[{{Kobayashi} {et~al.}(2022){Kobayashi}, {Nishimichi}, {Takada}, \& {Miyatake}}]{2022PhRvD.105h3517K}
{Kobayashi}, Y., {Nishimichi}, T., {Takada}, M., \& {Miyatake}, H. 2022, \prd, 105, 083517

\bibitem[{{Kodi Ramanah} {et~al.}(2020){Kodi Ramanah}, {Wojtak}, {Ansari}, {Gall}, \& {Hjorth}}]{2020MNRAS.499.1985K}
{Kodi Ramanah}, D., {Wojtak}, R., {Ansari}, Z., {Gall}, C., \& {Hjorth}, J. 2020, \mnras, 499, 1985

\bibitem[{{Kodi Ramanah} {et~al.}(2021){Kodi Ramanah}, {Wojtak}, \& {Arendse}}]{2021MNRAS.501.4080K}
{Kodi Ramanah}, D., {Wojtak}, R., \& {Arendse}, N. 2021, \mnras, 501, 4080

\bibitem[{{Komatsu} {et~al.}(2011){Komatsu}, {Smith}, {Dunkley}, {Bennett}, {Gold}, {Hinshaw}, {Jarosik}, {Larson}, {Nolta}, {Page}, {Spergel}, {Halpern}, {Hill}, {Kogut}, {Limon}, {Meyer}, {Odegard}, {Tucker}, {Weiland}, {Wollack}, \& {Wright}}]{2011ApJS..192...18K}
{Komatsu}, E., {Smith}, K.~M., {Dunkley}, J., {et~al.} 2011, \apjs, 192, 18

\bibitem[{{Kravtsov} \& {Borgani}(2012)}]{2012ARA&A..50..353K}
{Kravtsov}, A.~V. \& {Borgani}, S. 2012, \araa, 50, 353

\bibitem[{{Lesci} {et~al.}(2022{\natexlab{a}}){Lesci}, {Marulli}, {Moscardini}, {Sereno}, {Veropalumbo}, {Maturi}, {Giocoli}, {Radovich}, {Bellagamba}, {Roncarelli}, {Bardelli}, {Contarini}, {Covone}, {Ingoglia}, {Nanni}, \& {Puddu}}]{2022A&A...659A..88L}
{Lesci}, G.~F., {Marulli}, F., {Moscardini}, L., {et~al.} 2022{\natexlab{a}}, \aap, 659, A88

\bibitem[{{Lesci} {et~al.}(2022{\natexlab{b}}){Lesci}, {Nanni}, {Marulli}, {Moscardini}, {Veropalumbo}, {Maturi}, {Sereno}, {Radovich}, {Bellagamba}, {Roncarelli}, {Bardelli}, {Castignani}, {Covone}, {Giocoli}, {Ingoglia}, \& {Puddu}}]{2022A&A...665A.100L}
{Lesci}, G.~F., {Nanni}, L., {Marulli}, F., {et~al.} 2022{\natexlab{b}}, \aap, 665, A100

\bibitem[{{Liu} {et~al.}(2022){Liu}, {Bulbul}, {Ghirardini}, {Liu}, {Klein}, {Clerc}, {{\"O}zsoy}, {Ramos-Ceja}, {Pacaud}, {Comparat}, {Okabe}, {Bahar}, {Biffi}, {Brunner}, {Br{\"u}ggen}, {Buchner}, {Ider Chitham}, {Chiu}, {Dolag}, {Gatuzz}, {Gonzalez}, {Hoang}, {Lamer}, {Merloni}, {Nandra}, {Oguri}, {Ota}, {Predehl}, {Reiprich}, {Salvato}, {Schrabback}, {Sanders}, {Seppi}, \& {Thibaud}}]{2022A&A...661A...2L}
{Liu}, A., {Bulbul}, E., {Ghirardini}, V., {et~al.} 2022, \aap, 661, A2

\bibitem[{{{\L}okas} {et~al.}(2006){{\L}okas}, {Wojtak}, {Gottl{\"o}ber}, {Mamon}, \& {Prada}}]{2006MNRAS.367.1463L}
{{\L}okas}, E.~L., {Wojtak}, R., {Gottl{\"o}ber}, S., {Mamon}, G.~A., \& {Prada}, F. 2006, \mnras, 367, 1463

\bibitem[{{Mantz} {et~al.}(2010){Mantz}, {Allen}, {Rapetti}, \& {Ebeling}}]{2010MNRAS.406.1759M}
{Mantz}, A., {Allen}, S.~W., {Rapetti}, D., \& {Ebeling}, H. 2010, \mnras, 406, 1759

\bibitem[{{Mantz} {et~al.}(2016){Mantz}, {Allen}, {Morris}, {von der Linden}, {Applegate}, {Kelly}, {Burke}, {Donovan}, \& {Ebeling}}]{2016MNRAS.463.3582M}
{Mantz}, A.~B., {Allen}, S.~W., {Morris}, R.~G., {et~al.} 2016, \mnras, 463, 3582

\bibitem[{{Maturi} {et~al.}(2019){Maturi}, {Bellagamba}, {Radovich}, {Roncarelli}, {Sereno}, {Moscardini}, {Bardelli}, \& {Puddu}}]{2019MNRAS.485..498M}
{Maturi}, M., {Bellagamba}, F., {Radovich}, M., {et~al.} 2019, \mnras, 485, 498

\bibitem[{{Melchior} {et~al.}(2017){Melchior}, {Gruen}, {McClintock}, {Varga}, {Sheldon}, {Rozo}, {Amara}, {Becker}, {Benson}, {Bermeo}, {Bridle}, {Clampitt}, {Dietrich}, {Hartley}, {Hollowood}, {Jain}, {Jarvis}, {Jeltema}, {Kacprzak}, {MacCrann}, {Rykoff}, {Saro}, {Suchyta}, {Troxel}, {Zuntz}, {Bonnett}, {Plazas}, {Abbott}, {Abdalla}, {Annis}, {Benoit-L{\'e}vy}, {Bernstein}, {Bertin}, {Brooks}, {Buckley-Geer}, {Carnero Rosell}, {Carrasco Kind}, {Carretero}, {Cunha}, {D'Andrea}, {da Costa}, {Desai}, {Eifler}, {Flaugher}, {Fosalba}, {Garc{\'\i}a-Bellido}, {Gaztanaga}, {Gerdes}, {Gruendl}, {Gschwend}, {Gutierrez}, {Honscheid}, {James}, {Kirk}, {Krause}, {Kuehn}, {Kuropatkin}, {Lahav}, {Lima}, {Maia}, {March}, {Martini}, {Menanteau}, {Miller}, {Miquel}, {Mohr}, {Nichol}, {Ogando}, {Romer}, {Sanchez}, {Scarpine}, {Sevilla-Noarbe}, {Smith}, {Soares-Santos}, {Sobreira}, {Swanson}, {Tarle}, {Thomas}, {Walker}, {Weller}, \& {Zhang}}]{2017MNRAS.469.4899M}
{Melchior}, P., {Gruen}, D., {McClintock}, T., {et~al.} 2017, \mnras, 469, 4899

\bibitem[{{Munari} {et~al.}(2014){Munari}, {Biviano}, \& {Mamon}}]{2014A&A...566A..68M}
{Munari}, E., {Biviano}, A., \& {Mamon}, G.~A. 2014, \aap, 566, A68

\bibitem[{{Murante} {et~al.}(2015){Murante}, {Monaco}, {Borgani}, {Tornatore}, {Dolag}, \& {Goz}}]{2015MNRAS.447..178Murante}
{Murante}, G., {Monaco}, P., {Borgani}, S., {et~al.} 2015, \mnras, 447, 178

\bibitem[{{Navarro} {et~al.}(1996){Navarro}, {Frenk}, \& {White}}]{1996ApJ...462..563N}
{Navarro}, J.~F., {Frenk}, C.~S., \& {White}, S. D.~M. 1996, \apj, 462, 563

\bibitem[{{Ni} {et~al.}(2023){Ni}, {Genel}, {Angl{\'e}s-Alc{\'a}zar}, {Villaescusa-Navarro}, {Jo}, {Bird}, {Di Matteo}, {Croft}, {Chen}, {de Santi}, {Gebhardt}, {Shao}, {Pandey}, {Hernquist}, \& {Dave}}]{2023arXiv230402096N}
{Ni}, Y., {Genel}, S., {Angl{\'e}s-Alc{\'a}zar}, D., {et~al.} 2023, arXiv e-prints, arXiv:2304.02096

\bibitem[{{Ntampaka} {et~al.}(2015){Ntampaka}, {Trac}, {Sutherland}, {Battaglia}, {P{\'o}czos}, \& {Schneider}}]{2015ApJ...803...50N}
{Ntampaka}, M., {Trac}, H., {Sutherland}, D.~J., {et~al.} 2015, \apj, 803, 50

\bibitem[{{Pacaud} {et~al.}(2018){Pacaud}, {Pierre}, {Melin}, {Adami}, {Evrard}, {Galli}, {Gastaldello}, {Maughan}, {Sereno}, {Alis}, {Altieri}, {Birkinshaw}, {Chiappetti}, {Faccioli}, {Giles}, {Horellou}, {Iovino}, {Koulouridis}, {Le F{\`e}vre}, {Lidman}, {Lieu}, {Maurogordato}, {Moscardini}, {Plionis}, {Poggianti}, {Pompei}, {Sadibekova}, {Valtchanov}, \& {Willis}}]{2018A&A...620A..10P}
{Pacaud}, F., {Pierre}, M., {Melin}, J.~B., {et~al.} 2018, \aap, 620, A10

\bibitem[{{Pillepich} {et~al.}(2018{\natexlab{a}}){Pillepich}, {Nelson}, {Hernquist}, {Springel}, {Pakmor}, {Torrey}, {Weinberger}, {Genel}, {Naiman}, {Marinacci}, \& {Vogelsberger}}]{2018MNRAS.475..648P}
{Pillepich}, A., {Nelson}, D., {Hernquist}, L., {et~al.} 2018{\natexlab{a}}, \mnras, 475, 648

\bibitem[{{Pillepich} {et~al.}(2018{\natexlab{b}}){Pillepich}, {Springel}, {Nelson}, {Genel}, {Naiman}, {Pakmor}, {Hernquist}, {Torrey}, {Vogelsberger}, {Weinberger}, \& {Marinacci}}]{2018MNRAS.473.4077P}
{Pillepich}, A., {Springel}, V., {Nelson}, D., {et~al.} 2018{\natexlab{b}}, \mnras, 473, 4077

\bibitem[{{Planck Collaboration} {et~al.}(2016){Planck Collaboration}, {Ade}, {Aghanim}, {Arnaud}, {Ashdown}, {Aumont}, {Baccigalupi}, {Banday}, {Barreiro}, {Bartlett}, {Bartolo}, {Battaner}, {Battye}, {Benabed}, {Beno{\^\i}t}, {Benoit-L{\'e}vy}, {Bernard}, {Bersanelli}, {Bielewicz}, {Bock}, {Bonaldi}, {Bonavera}, {Bond}, {Borrill}, {Bouchet}, {Bucher}, {Burigana}, {Butler}, {Calabrese}, {Cardoso}, {Catalano}, {Challinor}, {Chamballu}, {Chary}, {Chiang}, {Christensen}, {Church}, {Clements}, {Colombi}, {Colombo}, {Combet}, {Comis}, {Couchot}, {Coulais}, {Crill}, {Curto}, {Cuttaia}, {Danese}, {Davies}, {Davis}, {de Bernardis}, {de Rosa}, {de Zotti}, {Delabrouille}, {D{\'e}sert}, {Diego}, {Dolag}, {Dole}, {Donzelli}, {Dor{\'e}}, {Douspis}, {Ducout}, {Dupac}, {Efstathiou}, {Elsner}, {En{\ss}lin}, {Eriksen}, {Falgarone}, {Fergusson}, {Finelli}, {Forni}, {Frailis}, {Fraisse}, {Franceschi}, {Frejsel}, {Galeotta}, {Galli}, {Ganga}, {Giard}, {Giraud-H{\'e}raud}, {Gjerl{\o}w}, {Gonz{\'a}lez-Nuevo}, {G{\'o}rski},
  {Gratton}, {Gregorio}, {Gruppuso}, {Gudmundsson}, {Hansen}, {Hanson}, {Harrison}, {Henrot-Versill{\'e}}, {Hern{\'a}ndez-Monteagudo}, {Herranz}, {Hildebrandt}, {Hivon}, {Hobson}, {Holmes}, {Hornstrup}, {Hovest}, {Huffenberger}, {Hurier}, {Jaffe}, {Jaffe}, {Jones}, {Juvela}, {Keih{\"a}nen}, {Keskitalo}, {Kisner}, {Kneissl}, {Knoche}, {Kunz}, {Kurki-Suonio}, {Lagache}, {L{\"a}hteenm{\"a}ki}, {Lamarre}, {Lasenby}, {Lattanzi}, {Lawrence}, {Leonardi}, {Lesgourgues}, {Levrier}, {Liguori}, {Lilje}, {Linden-V{\o}rnle}, {L{\'o}pez-Caniego}, {Lubin}, {Mac{\'\i}as-P{\'e}rez}, {Maggio}, {Maino}, {Mandolesi}, {Mangilli}, {Maris}, {Martin}, {Mart{\'\i}nez-Gonz{\'a}lez}, {Masi}, {Matarrese}, {McGehee}, {Meinhold}, {Melchiorri}, {Melin}, {Mendes}, {Mennella}, {Migliaccio}, {Mitra}, {Miville-Desch{\^e}nes}, {Moneti}, {Montier}, {Morgante}, {Mortlock}, {Moss}, {Munshi}, {Murphy}, {Naselsky}, {Nati}, {Natoli}, {Netterfield}, {N{\o}rgaard-Nielsen}, {Noviello}, {Novikov}, {Novikov}, {Oxborrow}, {Paci}, {Pagano}, {Pajot},
  {Paoletti}, {Partridge}, {Pasian}, {Patanchon}, {Pearson}, {Perdereau}, {Perotto}, {Perrotta}, {Pettorino}, {Piacentini}, {Piat}, {Pierpaoli}, {Pietrobon}, {Plaszczynski}, {Pointecouteau}, {Polenta}, {Popa}, {Pratt}, {Pr{\'e}zeau}, {Prunet}, {Puget}, {Rachen}, {Rebolo}, {Reinecke}, {Remazeilles}, {Renault}, {Renzi}, {Ristorcelli}, {Rocha}, {Roman}, {Rosset}, {Rossetti}, {Roudier}, {Rubi{\~n}o-Mart{\'\i}n}, {Rusholme}, {Sandri}, {Santos}, {Savelainen}, {Savini}, {Scott}, {Seiffert}, {Shellard}, {Spencer}, {Stolyarov}, {Stompor}, {Sudiwala}, {Sunyaev}, {Sutton}, {Suur-Uski}, {Sygnet}, {Tauber}, {Terenzi}, {Toffolatti}, {Tomasi}, {Tristram}, {Tucci}, {Tuovinen}, {T{\"u}rler}, {Umana}, {Valenziano}, {Valiviita}, {Van Tent}, {Vielva}, {Villa}, {Wade}, {Wandelt}, {Wehus}, {Weller}, {White}, {Yvon}, {Zacchei}, \& {Zonca}}]{2016A&A...594A..24P}
{Planck Collaboration}, {Ade}, P.~A.~R., {Aghanim}, N., {et~al.} 2016, \aap, 594, A24

\bibitem[{{Planck Collaboration} {et~al.}(2020){Planck Collaboration}, {Aghanim}, {Akrami}, {Ashdown}, {Aumont}, {Baccigalupi}, {Ballardini}, {Banday}, {Barreiro}, {Bartolo}, {Basak}, {Battye}, {Benabed}, {Bernard}, {Bersanelli}, {Bielewicz}, {Bock}, {Bond}, {Borrill}, {Bouchet}, {Boulanger}, {Bucher}, {Burigana}, {Butler}, {Calabrese}, {Cardoso}, {Carron}, {Challinor}, {Chiang}, {Chluba}, {Colombo}, {Combet}, {Contreras}, {Crill}, {Cuttaia}, {de Bernardis}, {de Zotti}, {Delabrouille}, {Delouis}, {Di Valentino}, {Diego}, {Dor{\'e}}, {Douspis}, {Ducout}, {Dupac}, {Dusini}, {Efstathiou}, {Elsner}, {En{\ss}lin}, {Eriksen}, {Fantaye}, {Farhang}, {Fergusson}, {Fernandez-Cobos}, {Finelli}, {Forastieri}, {Frailis}, {Fraisse}, {Franceschi}, {Frolov}, {Galeotta}, {Galli}, {Ganga}, {G{\'e}nova-Santos}, {Gerbino}, {Ghosh}, {Gonz{\'a}lez-Nuevo}, {G{\'o}rski}, {Gratton}, {Gruppuso}, {Gudmundsson}, {Hamann}, {Handley}, {Hansen}, {Herranz}, {Hildebrandt}, {Hivon}, {Huang}, {Jaffe}, {Jones}, {Karakci}, {Keih{\"a}nen},
  {Keskitalo}, {Kiiveri}, {Kim}, {Kisner}, {Knox}, {Krachmalnicoff}, {Kunz}, {Kurki-Suonio}, {Lagache}, {Lamarre}, {Lasenby}, {Lattanzi}, {Lawrence}, {Le Jeune}, {Lemos}, {Lesgourgues}, {Levrier}, {Lewis}, {Liguori}, {Lilje}, {Lilley}, {Lindholm}, {L{\'o}pez-Caniego}, {Lubin}, {Ma}, {Mac{\'\i}as-P{\'e}rez}, {Maggio}, {Maino}, {Mandolesi}, {Mangilli}, {Marcos-Caballero}, {Maris}, {Martin}, {Martinelli}, {Mart{\'\i}nez-Gonz{\'a}lez}, {Matarrese}, {Mauri}, {McEwen}, {Meinhold}, {Melchiorri}, {Mennella}, {Migliaccio}, {Millea}, {Mitra}, {Miville-Desch{\^e}nes}, {Molinari}, {Montier}, {Morgante}, {Moss}, {Natoli}, {N{\o}rgaard-Nielsen}, {Pagano}, {Paoletti}, {Partridge}, {Patanchon}, {Peiris}, {Perrotta}, {Pettorino}, {Piacentini}, {Polastri}, {Polenta}, {Puget}, {Rachen}, {Reinecke}, {Remazeilles}, {Renzi}, {Rocha}, {Rosset}, {Roudier}, {Rubi{\~n}o-Mart{\'\i}n}, {Ruiz-Granados}, {Salvati}, {Sandri}, {Savelainen}, {Scott}, {Shellard}, {Sirignano}, {Sirri}, {Spencer}, {Sunyaev}, {Suur-Uski}, {Tauber}, {Tavagnacco},
  {Tenti}, {Toffolatti}, {Tomasi}, {Trombetti}, {Valenziano}, {Valiviita}, {Van Tent}, {Vibert}, {Vielva}, {Villa}, {Vittorio}, {Wandelt}, {Wehus}, {White}, {White}, {Zacchei}, \& {Zonca}}]{2020A&A...641A...6P}
{Planck Collaboration}, {Aghanim}, N., {Akrami}, Y., {et~al.} 2020, \aap, 641, A6

\bibitem[{Prati {et~al.}(2004)Prati, Batista, \& Monard}]{2004Data}
Prati, R.~C., Batista, G., \& Monard, M.~C. 2004, in Proceedings of the 4th Indian International Conference on Artificial Intelligence, IICAI 2009, Tumkur, Karnataka, India, December 16-18, 2009

\bibitem[{{Pratt} {et~al.}(2019){Pratt}, {Arnaud}, {Biviano}, {Eckert}, {Ettori}, {Nagai}, {Okabe}, \& {Reiprich}}]{2019SSRv..215...25P}
{Pratt}, G.~W., {Arnaud}, M., {Biviano}, A., {et~al.} 2019, \ssr, 215, 25

\bibitem[{{Pratt} {et~al.}(2009){Pratt}, {Croston}, {Arnaud}, \& {B{\"o}hringer}}]{2009A&A...498..361P}
{Pratt}, G.~W., {Croston}, J.~H., {Arnaud}, M., \& {B{\"o}hringer}, H. 2009, \aap, 498, 361

\bibitem[{Qi(2017)}]{LGB}
Qi, M. 2017, in Neural Information Processing Systems

\bibitem[{{Ragagnin} {et~al.}(2023){Ragagnin}, {Fumagalli}, {Castro}, {Dolag}, {Saro}, {Costanzi}, \& {Bocquet}}]{2023A&A...675A..77R}
{Ragagnin}, A., {Fumagalli}, A., {Castro}, T., {et~al.} 2023, \aap, 675, A77

\bibitem[{{Ragagnin} {et~al.}(2016){Ragagnin}, {Tchipev}, {Bader}, {Dolag}, \& {Hammer}}]{2016pcre.conf..411R}
{Ragagnin}, A., {Tchipev}, N., {Bader}, M., {Dolag}, K., \& {Hammer}, N.~J. 2016, in Advances in Parallel Computing, 411--420

\bibitem[{{Remus} {et~al.}(2017){Remus}, {Dolag}, {Naab}, {Burkert}, {Hirschmann}, {Hoffmann}, \& {Johansson}}]{2017MNRAS.464.3742Remus17}
{Remus}, R.-S., {Dolag}, K., {Naab}, T., {et~al.} 2017, \mnras, 464, 3742

\bibitem[{{Rykoff} {et~al.}(2016){Rykoff}, {Rozo}, {Hollowood}, {Bermeo-Hernandez}, {Jeltema}, {Mayers}, {Romer}, {Rooney}, {Saro}, {Vergara Cervantes}, {Wechsler}, {Wilcox}, {Abbott}, {Abdalla}, {Allam}, {Annis}, {Benoit-L{\'e}vy}, {Bernstein}, {Bertin}, {Brooks}, {Burke}, {Capozzi}, {Carnero Rosell}, {Carrasco Kind}, {Castander}, {Childress}, {Collins}, {Cunha}, {D'Andrea}, {da Costa}, {Davis}, {Desai}, {Diehl}, {Dietrich}, {Doel}, {Evrard}, {Finley}, {Flaugher}, {Fosalba}, {Frieman}, {Glazebrook}, {Goldstein}, {Gruen}, {Gruendl}, {Gutierrez}, {Hilton}, {Honscheid}, {Hoyle}, {James}, {Kay}, {Kuehn}, {Kuropatkin}, {Lahav}, {Lewis}, {Lidman}, {Lima}, {Maia}, {Mann}, {Marshall}, {Martini}, {Melchior}, {Miller}, {Miquel}, {Mohr}, {Nichol}, {Nord}, {Ogando}, {Plazas}, {Reil}, {Sahl{\'e}n}, {Sanchez}, {Santiago}, {Scarpine}, {Schubnell}, {Sevilla-Noarbe}, {Smith}, {Soares-Santos}, {Sobreira}, {Stott}, {Suchyta}, {Swanson}, {Tarle}, {Thomas}, {Tucker}, {Uddin}, {Viana}, {Vikram}, {Walker}, {Zhang}, \& {DES
  Collaboration}}]{2016ApJS..224....1R}
{Rykoff}, E.~S., {Rozo}, E., {Hollowood}, D., {et~al.} 2016, \apjs, 224, 1

\bibitem[{{Sereno}(2015)}]{2015MNRAS.450.3665S}
{Sereno}, M. 2015, \mnras, 450, 3665

\bibitem[{{Sereno} \& {Umetsu}(2011)}]{2011MNRAS.416.3187S}
{Sereno}, M. \& {Umetsu}, K. 2011, \mnras, 416, 3187

\bibitem[{{Sereno} {et~al.}(2020){Sereno}, {Umetsu}, {Ettori}, {Eckert}, {Gastaldello}, {Giles}, {Lieu}, {Maughan}, {Okabe}, {Birkinshaw}, {Chiu}, {Fujita}, {Miyazaki}, {Rapetti}, {Koulouridis}, \& {Pierre}}]{2020MNRAS.492.4528S}
{Sereno}, M., {Umetsu}, K., {Ettori}, S., {et~al.} 2020, \mnras, 492, 4528

\bibitem[{{Sereno} {et~al.}(2018){Sereno}, {Umetsu}, {Ettori}, {Sayers}, {Chiu}, {Meneghetti}, {Vega-Ferrero}, \& {Zitrin}}]{2018ApJ...860L...4Sereno2018}
{Sereno}, M., {Umetsu}, K., {Ettori}, S., {et~al.} 2018, \apjl, 860, L4

\bibitem[{{Singh} {et~al.}(2020){Singh}, {Saro}, {Costanzi}, \& {Dolag}}]{2020MNRAS.494.3728S}
{Singh}, P., {Saro}, A., {Costanzi}, M., \& {Dolag}, K. 2020, \mnras, 494, 3728

\bibitem[{{Springel}(2005)}]{2005MNRAS.364.1105S}
{Springel}, V. 2005, \mnras, 364, 1105

\bibitem[{{Springel} {et~al.}(2005{\natexlab{a}}){Springel}, {Di Matteo}, \& {Hernquist}}]{2005MNRAS.361..776S}
{Springel}, V., {Di Matteo}, T., \& {Hernquist}, L. 2005{\natexlab{a}}, \mnras, 361, 776

\bibitem[{{Springel} {et~al.}(2005{\natexlab{b}}){Springel}, {White}, {Jenkins}, {Frenk}, {Yoshida}, {Gao}, {Navarro}, {Thacker}, {Croton}, {Helly}, {Peacock}, {Cole}, {Thomas}, {Couchman}, {Evrard}, {Colberg}, \& {Pearce}}]{2005Natur.435..629S}
{Springel}, V., {White}, S. D.~M., {Jenkins}, A., {et~al.} 2005{\natexlab{b}}, \nat, 435, 629

\bibitem[{{Springel} {et~al.}(2001){Springel}, {White}, {Tormen}, \& {Kauffmann}}]{2001MNRAS.328..726S}
{Springel}, V., {White}, S. D.~M., {Tormen}, G., \& {Kauffmann}, G. 2001, \mnras, 328, 726

\bibitem[{{Storey-Fisher} {et~al.}(2022){Storey-Fisher}, {Tinker}, {Zhai}, {DeRose}, {Wechsler}, \& {Banerjee}}]{2022arXiv221003203S}
{Storey-Fisher}, K., {Tinker}, J., {Zhai}, Z., {et~al.} 2022, arXiv e-prints, arXiv:2210.03203

\bibitem[{{Tang} {et~al.}(2021){Tang}, {Lin}, {Wang}, \& {Napolitano}}]{2021MNRAS.508.3321Tang}
{Tang}, L., {Lin}, W., {Wang}, Y., \& {Napolitano}, N.~R. 2021, \mnras, 508, 3321

\bibitem[{{Tornatore} {et~al.}(2007){Tornatore}, {Borgani}, {Dolag}, \& {Matteucci}}]{2007MNRAS.382.1050T}
{Tornatore}, L., {Borgani}, S., {Dolag}, K., \& {Matteucci}, F. 2007, \mnras, 382, 1050

\bibitem[{{Umetsu} {et~al.}(2020){Umetsu}, {Sereno}, {Lieu}, {Miyatake}, {Medezinski}, {Nishizawa}, {Giles}, {Gastaldello}, {McCarthy}, {Kilbinger}, {Birkinshaw}, {Ettori}, {Okabe}, {Chiu}, {Coupon}, {Eckert}, {Fujita}, {Higuchi}, {Koulouridis}, {Maughan}, {Miyazaki}, {Oguri}, {Pacaud}, {Pierre}, {Rapetti}, \& {Smith}}]{2020ApJ...890..148U}
{Umetsu}, K., {Sereno}, M., {Lieu}, M., {et~al.} 2020, \apj, 890, 148

\bibitem[{{Umetsu} {et~al.}(2016){Umetsu}, {Zitrin}, {Gruen}, {Merten}, {Donahue}, \& {Postman}}]{2016ApJ...821..116U}
{Umetsu}, K., {Zitrin}, A., {Gruen}, D., {et~al.} 2016, \apj, 821, 116

\bibitem[{{van den Busch} {et~al.}(2022){van den Busch}, {Wright}, {Hildebrandt}, {Bilicki}, {Asgari}, {Joudaki}, {Blake}, {Heymans}, {Kannawadi}, {Shan}, \& {Tr{\"o}ster}}]{2022A&A...664A.170V}
{van den Busch}, J.~L., {Wright}, A.~H., {Hildebrandt}, H., {et~al.} 2022, \aap, 664, A170

\bibitem[{{Vikhlinin} {et~al.}(2009{\natexlab{a}}){Vikhlinin}, {Burenin}, {Ebeling}, {Forman}, {Hornstrup}, {Jones}, {Kravtsov}, {Murray}, {Nagai}, {Quintana}, \& {Voevodkin}}]{2009ApJ...692.1033V}
{Vikhlinin}, A., {Burenin}, R.~A., {Ebeling}, H., {et~al.} 2009{\natexlab{a}}, \apj, 692, 1033

\bibitem[{{Vikhlinin} {et~al.}(2009{\natexlab{b}}){Vikhlinin}, {Kravtsov}, {Burenin}, {Ebeling}, {Forman}, {Hornstrup}, {Jones}, {Murray}, {Nagai}, {Quintana}, \& {Voevodkin}}]{2009ApJ...692.1060V}
{Vikhlinin}, A., {Kravtsov}, A.~V., {Burenin}, R.~A., {et~al.} 2009{\natexlab{b}}, \apj, 692, 1060

\bibitem[{{Villaescusa-Navarro} {et~al.}(2021){Villaescusa-Navarro}, {Angl{\'e}s-Alc{\'a}zar}, {Genel}, {Spergel}, {Somerville}, {Dave}, {Pillepich}, {Hernquist}, {Nelson}, {Torrey}, {Narayanan}, {Li}, {Philcox}, {La Torre}, {Maria Delgado}, {Ho}, {Hassan}, {Burkhart}, {Wadekar}, {Battaglia}, {Contardo}, \& {Bryan}}]{2021ApJ...915...71V}
{Villaescusa-Navarro}, F., {Angl{\'e}s-Alc{\'a}zar}, D., {Genel}, S., {et~al.} 2021, \apj, 915, 71

\bibitem[{{Villaescusa-Navarro} {et~al.}(2022){Villaescusa-Navarro}, {Ding}, {Genel}, {Tonnesen}, {La Torre}, {Spergel}, {Teyssier}, {Li}, {Heneka}, {Lemos}, {Angl{\'e}s-Alc{\'a}zar}, {Nagai}, \& {Vogelsberger}}]{2022ApJ...929..132V}
{Villaescusa-Navarro}, F., {Ding}, J., {Genel}, S., {et~al.} 2022, \apj, 929, 132

\bibitem[{{Vogelsberger} {et~al.}(2014){Vogelsberger}, {Genel}, {Springel}, {Torrey}, {Sijacki}, {Xu}, {Snyder}, {Nelson}, \& {Hernquist}}]{2014MNRAS.444.1518V}
{Vogelsberger}, M., {Genel}, S., {Springel}, V., {et~al.} 2014, \mnras, 444, 1518

\bibitem[{{Wechsler} \& {Tinker}(2018)}]{2018ARA&A..56..435Wechsler}
{Wechsler}, R.~H. \& {Tinker}, J.~L. 2018, \araa, 56, 435

\bibitem[{{Weinberger} {et~al.}(2017){Weinberger}, {Springel}, {Hernquist}, {Pillepich}, {Marinacci}, {Pakmor}, {Nelson}, {Genel}, {Vogelsberger}, {Naiman}, \& {Torrey}}]{2017MNRAS.465.3291W}
{Weinberger}, R., {Springel}, V., {Hernquist}, L., {et~al.} 2017, \mnras, 465, 3291

\bibitem[{{Yan} {et~al.}(2020){Yan}, {Mead}, {Van Waerbeke}, {Hinshaw}, \& {McCarthy}}]{2020MNRAS.499.3445Y}
{Yan}, Z., {Mead}, A.~J., {Van Waerbeke}, L., {Hinshaw}, G., \& {McCarthy}, I.~G. 2020, \mnras, 499, 3445

\end{thebibliography}
\bibliographystyle{aa}

\appendix
\section{Cosmological parameter estimates from classification probability}
\label{sec:appA}
In this Appendix, we summarize the statistical arguments behind using the classification as a starting point to infer cosmology. In particular, we use $n$ cluster observables $O=\{O_1,O_2, .. O_i ... O_n \}$ in a series of cosmological models $M_i$ with corresponding cosmological parameters $\theta_j = (\Omega_m, \sigma_8, h_0, \Omega_b)$.

We can start from the assumption that every single cluster carries the information of the cosmology behind the universe it lives in. The expectation of the cosmological parameters for one observation $O_i$ can be assumed to be 
\begin{align}
    \mu_i = \sum_j P(\theta_j | O_i) \ \theta_j,
\end{align}
where $P(\theta_j | O_i)$ is the conditional distribution related to the individual observable $O_i$.
The corresponding error on this expectation is defined by the variance:
\begin{align}
    \sigma_i^2 = \sum_j (\theta_j - \mu _ i)^2 P(\theta_j | O_i).
\end{align}
The observation $O$ can be thought of as a series of measuring processes, then the expectation is 
\begin{align}
    \mu =  \sum_i^n \mu_i / n,
    \label{eq:mean_par}
\end{align}
and the error is
\begin{align} 
    \sigma = \sqrt{ \sum_i^n \sigma_i^2 / n}.
    \label{eq:err_par}
\end{align}
Eq. \ref{eq:mean_par} and \ref{eq:err_par} represent the main statistics adopted in Sect. \ref{sec:infer para} to estimate the cosmological parameters estimates from independent observations of clusters. 
To fully define them, we need to define the
$P(\theta_j | O_i)$,
i.e., the probability of a single cluster observations $i$ to come from a cosmological model $j$ (see also Sect. \ref{sec:para metric}). In principle, it can be estimated by Bayes probability, as
\begin{align}
    P(\theta_j | O_i) = p(\theta_i) \frac{P(O_i|M_j)}{P(O_i|M)},
\end{align}
where $M = M_1 \bigcup M_2 ... \bigcup M_i ... M_m $ is all your $m$ models, or simulations. The prior here can be $p(\theta_i) = 1/m$, which means a flat distribution in the absence of observations. However, it is difficult to find a smooth probability distribution function for $P(O_i|M_j)$ or $P(O_i|M)$ under small $m$ and high-dimension output simulation data. 

ML technique provides a good way to find out the best fit $P(O_i|M_j)$.
This is possible by training a network with simulation cluster pair $(\theta_j|M_{jk})$, where $M_{jk}$ is one simulation cluster $k$ from a cosmological simulation $j$, and the training label are set to be $P_{ML}(\theta_j|M_{jk}) = 1$. If a series of simulations cover the real observations, ML can provide a good approximation of it via the best likelihood ($P_{ML}$)
\begin{align}
    P(\theta_j|O_{i}) \approx P_{ML}(\theta_j|O_{i}).
\end{align}
Under the statistical viewpoint of ML, this approximation does not have to be as accurate as possible (due to the cluster degeneracy on different simulations with nearly the same parameters), as long as the accuracy is greater than the prior $p(\theta_i)$. The larger the threshold, the more significant the cosmological information carried out by the individual cluster. Of course, the precision of the method increases as a function of the size of the cluster catalog as Eq \ref{eq:err_par} show. {\it This means that we cannot perform cosmology with one cluster.}

\section{Constraints for other cosmologies}
\label{sec:appB}
As a continuation of the results presented in Sect. \ref{sec:infer para}, we extend our analysis to three additional cases, namely M5, M7, and M9, to further demonstrate the parameter prediction power of our machine learning method. 
Our results reveal that the parameter constraints of M7 (Fig. \ref{fig:M7}) are similar to that of M6 (Fig. \ref{fig:M6}). However, M5 (Fig. \ref{fig:M5}) and M9 (Fig. \ref{fig:M9}) are located further away from the center of the parameter space compared to M6 and M7, resulting in a relatively poorer parameter prediction effect for these cases.

\begin{figure}[h]
\centering
\includegraphics[scale=0.32]{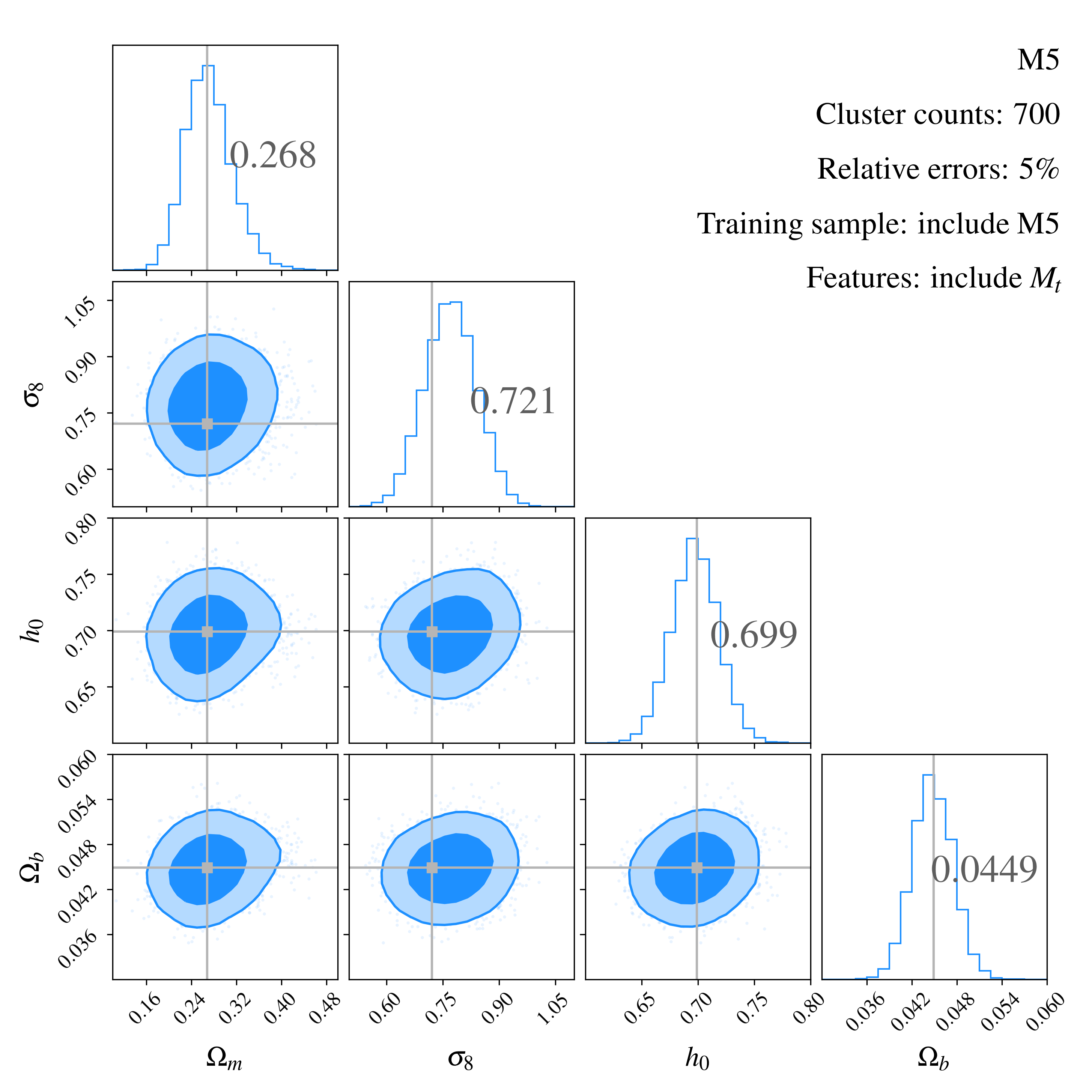}
\caption{Cosmological parameters of M5 inferred by the MLCCA. This graph is the same type as Fig. \ref{fig:M6}. 
The true values of all cosmological parameters are within the 1$\sigma$ confidence interval.}
\label{fig:M5}
\end{figure}

\begin{figure}
\centering
\includegraphics[scale=0.32]{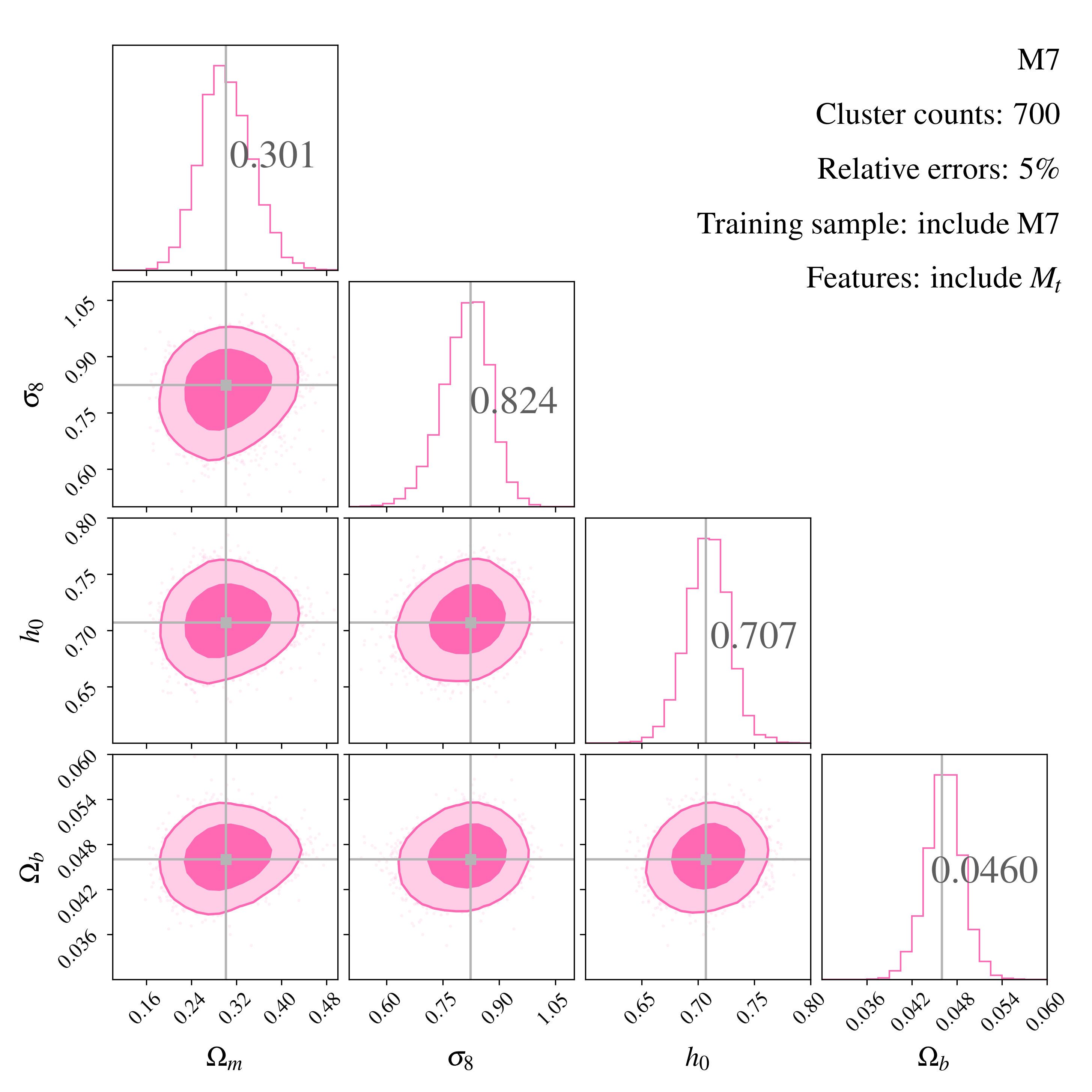}
\caption{Cosmological parameters of M7 inferred by the MLCCA. This graph is the same type as Fig. \ref{fig:M6}. The true values of all cosmological parameters are within the 1$\sigma$ confidence interval.}
\label{fig:M7}
\end{figure}

\begin{figure}
\centering
\includegraphics[scale=0.32,trim=0 0 0 1.4cm]{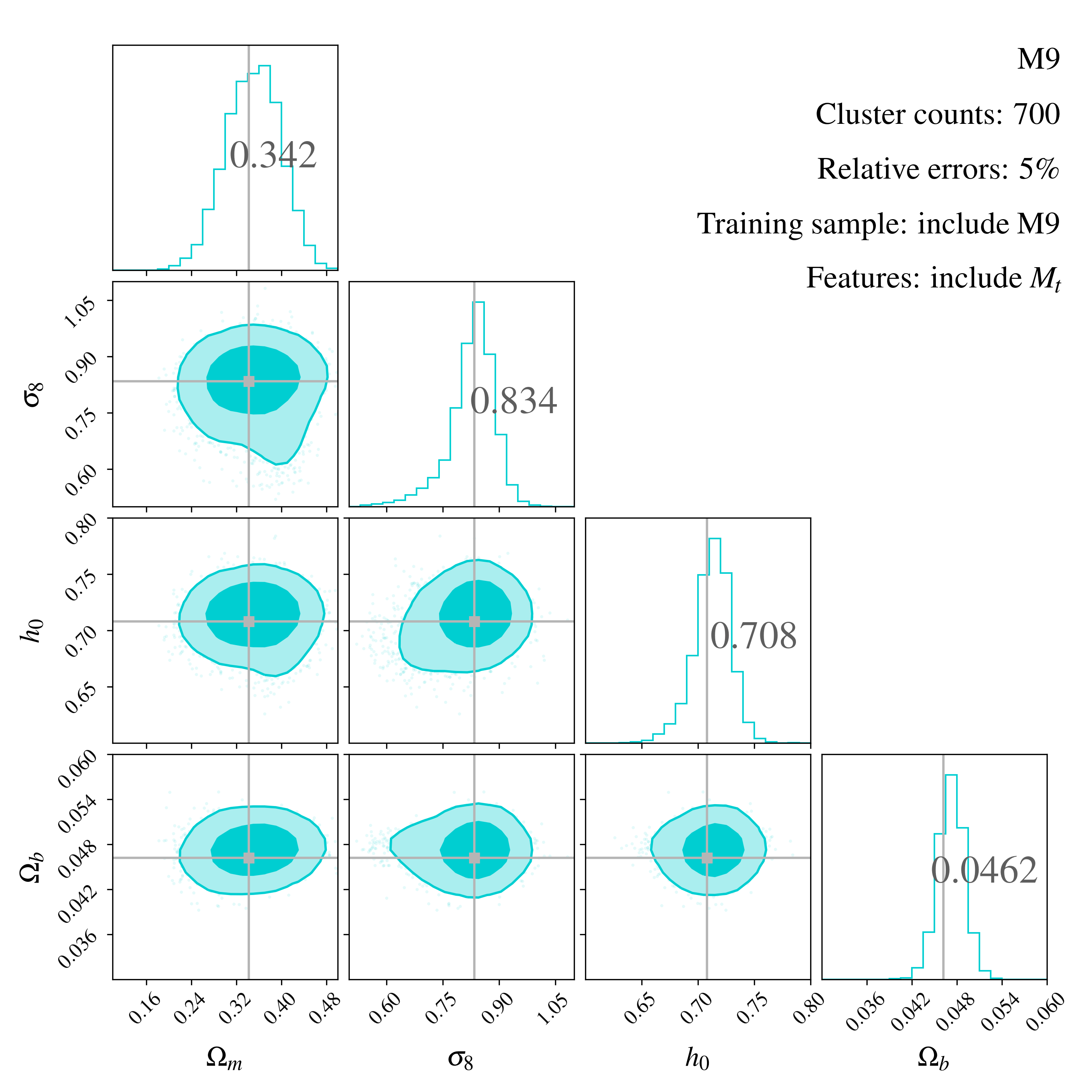}
\caption{Cosmological parameters of M9 inferred by the MLCCA. This graph is the same type as Fig. \ref{fig:M6}. 
The true values of all cosmological parameters are within the 1$\sigma$ confidence interval.}
\label{fig:M9}
\end{figure}

\section{Combination with other methods: effect of training size and error propagation}
\label{sec:appC}
In this section we visually compare the precision of the MLCCA approach with other methods, to check the ability of the new approach to compete with standard approaches and possibly help solve some degeneracies among the cosmological parameters. This check is not meant to be complete, as we use only the weak lensing and CMB results shown in Fig. \ref{fig:para_map}, but is meant to put the results found in Sect. \ref{sec:infer para} in the context of the cosmological parameter tensions. 
In Fig. \ref{fig:error propagation 1} we overlap the confidence contours from weak lensing and CMB as from Fig. \ref{fig:para_map}, which allows us to compare how the degeneracies among the parameters work differently in the different methods. 
In particular, the MLCCA contours are more symmetric around the true values and do not show the classical degeneracy between $\sigma_8$ and $\Omega_m$ parameters found for the weak lensing. 
The size of the 1 and 2$\sigma$ contours are larger than the ones produced by the CMB constraints but compatible with the ones from weak lensing. In principle, we can expect to reduce the size of the MLCCA contours by:
1) increasing the training sample size;
2) assuming a less conservative choice to propagate the errors on the parameters of the individual clusters. 
For the former, we have tested the impact of the training sample selecting from Fig. \ref{fig:number_map} with cluster counts larger than 20k, and repeated the training of the MLCCA using 20k clusters and by still testing on the usual 700 over 20 random extractions, with no overlap with the training sample. We have, thus, excluded cosmologies M1 to M5 and chose to predict the cosmological parameter for the catalog from the most central of the residual cosmology, which was M9. We have then compared the contours with the one obtained from the standard training made on the 7000 cluster sample and shown in Fig. \ref{fig:M9} and found that the accuracy of all cosmological parameters is, in fact, slightly improved. We go from 
$(0.351_{-0.050}^{+0.047}, 0.836_{-0.061}^{+0.046}, 0.714_{-0.016}^{+0.014}, 0.0473_{-0.0018}^{+0.0016})$ 
for $\Omega_m,~ \sigma_8, ~h_0, ~\Omega_b$ for the 7k training case to the 
$(0.355_{-0.043}^{+0.041}, 0.844_{-0.046}^{+0.038}, 0.715_{-0.014}^{+0.013}, 0.0474_{-0.0014}^{+0.0014})$
for the 20k training sample, i.e., with an average $15\pm5$\% improvement. This further motivates the use of larger volumes for future applications. To test the impact of the error propagation, we have designed a comparative experiment for M7 cosmology but bypassed the step of error propagation (see Sect. \ref{sec:para metric}). This is shown in Fig. \ref{fig:error propagation 2}, where we can clearly see that each contour shrinks after canceling the conservative option we have made for the error propagation.

We can finally conclude that the MLCCA has the advantage of fully accounting for the degeneracies among the cosmological parameters, i.e. providing relatively symmetric confidence contours (especially in the presence of a uniform distribution of prior cosmologies), and regardless of the choice of error propagation, it provides comparable precisions on the parameters that can be eventually improved using considerably larger training samples.

\begin{figure}
\centering
\includegraphics[scale=0.266,trim=0 0.2cm 0 1.15cm]{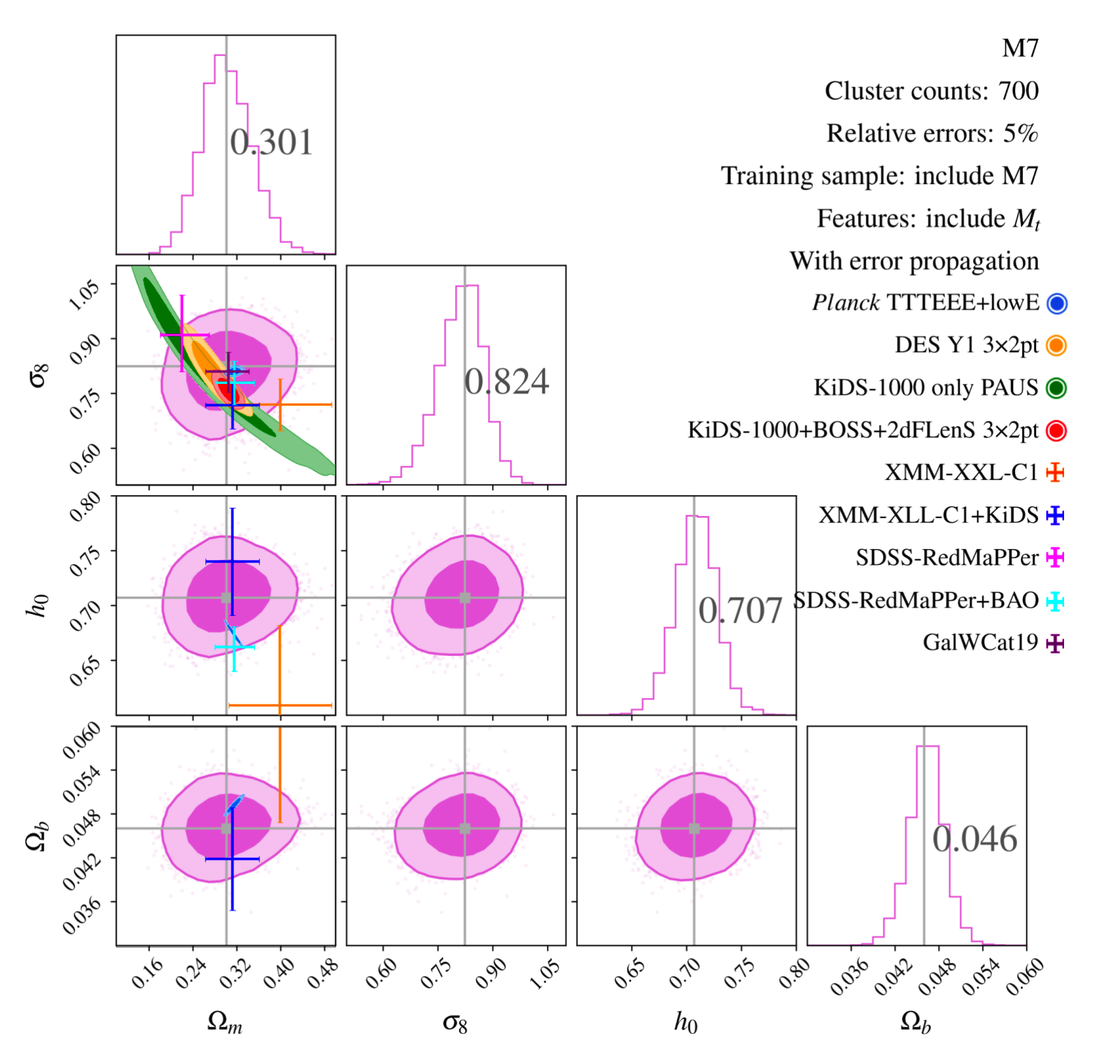}
\caption{Cosmological parameters of M7 inferred by the MLCCA with error propagation. This diagram is overlaid with cosmology constraints as in Fig. \ref{fig:para_map}, with modifications mainly to fit the coordinate range.}
\label{fig:error propagation 1}
\end{figure}

\begin{figure}
\centering
\includegraphics[scale=0.266,trim=0 0.2cm 0 0.3cm]{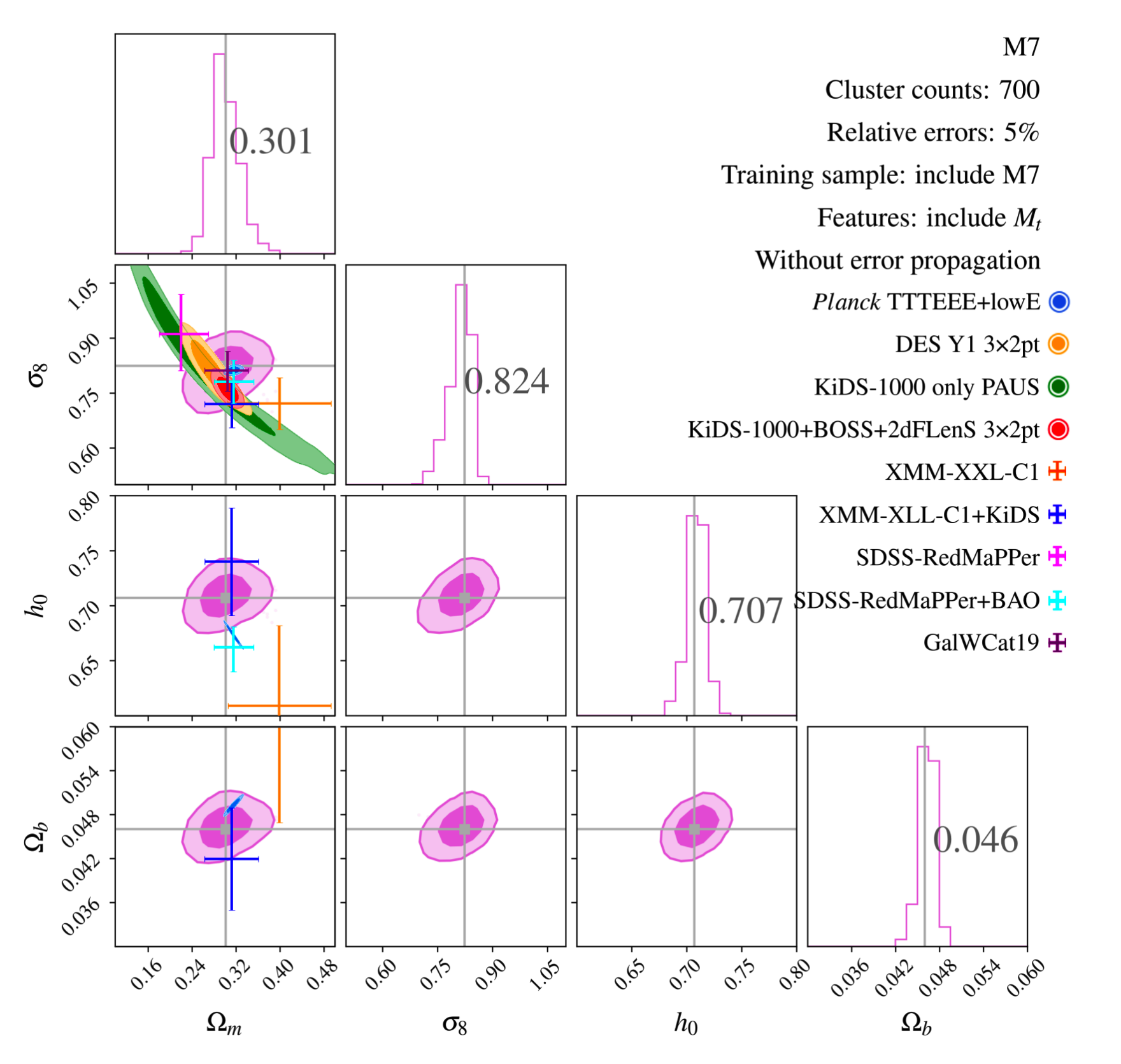}
\caption{Cosmological parameters of M7 inferred by the MLCCA without error propagation. This diagram is overlaid with cosmology constraints as in Fig. \ref{fig:para_map}, with modifications mainly to fit the coordinate range.}
\label{fig:error propagation 2}
\end{figure}

\end{document}